\documentclass[12pt, prd, aps, superscriptaddress, preprint]{revtex4-1}

\usepackage{amsmath}
\usepackage{graphicx}
\usepackage{mathrsfs}
\usepackage{hyperref}
\usepackage{subfigure}
\usepackage{slashed}

\newcommand{\fig}[1]{\mbox{\textcircled{{\scriptsize #1}}}}

\begin{document}

\title{Long distance contribution to the $K_L$-$K_S$ mass difference}

\newcommand\bnl{Brookhaven National Laboratory, Upton, NY 11973, USA}
\newcommand\cu{Physics Department, Columbia University, New York,
      NY 10027, USA}
\newcommand\riken{RIKEN-BNL Research Center, Brookhaven National
      Laboratory, Upton, NY 11973, USA}
\newcommand\soton{School of Physics and Astronomy, University of
  Southampton,  Southampton SO17 1BJ, UK}

\author{N.H.~Christ}\affiliation{\cu}
\author{T.~Izubuchi}\affiliation{\bnl}\affiliation{\riken}
\author{C.T.~Sachrajda}\affiliation{\soton}
\author{A.~Soni}\affiliation{\bnl}
\author{J.~Yu}\affiliation{\cu} 
\collaboration{RBC and UKQCD Collaborations}

\date{June 21, 2012}

\begin{abstract}
We develop and demonstrate techniques needed to compute the long distance contribution to the $K_{L}$-$K_{S}$ mass difference, $\Delta M_K$, in lattice QCD and carry out a first, exploratory calculation of this fundamental quantity. The calculation is performed on 2+1 flavor, domain wall fermion, $16^3\times32$ configurations with a 421 MeV pion mass and an inverse lattice spacing $1/a=1.73$ GeV. We include only current-current operators and drop all disconnected and double penguin diagrams. The short distance part of the mass difference in a 2+1 flavor calculation contains a quadratic divergence cut off by the lattice spacing. Here, this quadratic divergence is eliminated through the GIM mechanism by introducing a valence charm quark.  The inclusion of the charm quark makes the complete calculation accessible to lattice methods provided the discretization errors associated with the charm quark can be controlled.  The long distance effects are discussed for each parity channel separately. While we can see a clear signal in the parity odd channel, the signal to noise ratio in the parity even channel is exponentially decreasing as the separation between the two weak operators increases. We obtain a mass difference $\Delta M_K$ which ranges from $6.58(30)\times 10^{-12}$ MeV to $11.89(81)\times 10^{-12}$ MeV for kaon masses varying from 563 MeV to 839 MeV.  Extensions of these methods are proposed which promise accurate results for both $\Delta M_K$ and $\epsilon_K$, including long distance effects.
\end{abstract}

\maketitle

\section{Introduction}
\label{sec:Introduction}
Lattice QCD provides a first-principles method to compute non-perturbative QCD effects in electroweak processes. In first-order weak interaction processes, the large masses of the $W$ and $Z$ bosons mean that these interactions take place in a very small space-time region, at distances of $O(10^{-18}\,\mathrm{m})$, allowing their effects on the QCD scale to be described by a local four-quark operator, $H_W$.   However, in second-order weak processes, the position of the two $W$ or $Z$ exchanges may be separated by a distance which is much larger than $1/M_W$ and may be as large as $1/\Lambda_{\mathrm{QCD}}$ or $1/m_\pi$.  Such long distance effects contain non-perturbative contributions, making lattice QCD a natural method for their determination. Before this can be achieved however, a number of theoretical and practical problems must be overcome, and the purpose of this paper is to begin tackling these issues.

The $K_{L}$-$K_{S}$ mass difference with a value of $3.483(6)\times10^{-12}$\,MeV~\cite{Nakamura:2010zzi}, is extremely small and is known very accurately.  It is believed to arise from $K^{0}$-$\overline{K}^{0}$ mixing via second-order weak interactions.  However, because of its small size and because it arises from an amplitude in which strangeness changes by two units, this is a promising quantity to reveal phenomena which lie outside the standard model, making the calculation of the standard model contribution to $\Delta M_K$ an important challenge. Conventionally, the standard model contribution to this mass difference is separated into short distance and long distance parts. The short distance part receives contributions from momenta on the order of the charm quark mass, has been evaluated to next-to-next-to-leading order in QCD perturbation theory and represents about $70\%$ of the total mass difference~\cite{Herrlich:1993yv,Brod:2011ty}. If the mass difference can be explained within the standard model, the remaining $30\%$ must come from non-perturbative, long distance effects. 

A further uncertainty associated with this conventional approach is the use of QCD perturbation theory at the scale of the charm quark mass.  As pointed out in the recent next-to-next-to-leading order (NNLO) calculation~\cite{Brod:2011ty} these NNLO order terms are as large as 36\% of the leading order (LO) and next-to-leading order (NLO) terms, raising doubts about the use of QCD perturbation theory at this energy scale.  These uncertainties can be removed if the charm quark is also treated using lattice methods and a connection to perturbation theory attempted as a scale significantly larger than the charm quark mass $m_c$.

It is customary when discussing the $K_{L}$-$K_{S}$ mass difference to follow this convention of referring to distance scales at or below $1/m_c$ as short distance and those larger than $1/m_c$ as long distance.  We will follow this convention here. However, the inverse charm quark mass represents a somewhat large distance to act as boundary between short and long distance regions.  Thus, we should keep in mind that non-perturbative methods may be needed for the proper treatment of a portion of these short distance contributions to the $K_{L}$-$K_{S}$ mass difference and that it may be better to adopt a shorter distance demarcation between short and long distances in the future.

Here we propose a method to compute these long distance effects on a Euclidean lattice~\cite{Christ:2010zz} which also includes a non-perturbative treatment of the charm quark. The method is composed of three parts. First, we devise an Euclidean-space amplitude which can be evaluated in lattice QCD and which contains the second-order mass difference of interest. As explained in the following section, we perform a second-order integration of the product of two first-order weak Hamiltonians in a given space-time volume. The integration sums the contribution to the mass difference from all possible intermediate states. 

Second we must deal with those contributions coming when the separation between the two effective weak operators $H_W$ is at or below the lattice spacing.  In this region, the lattice description is not accurate and for separations approaching $1/M_W$ even the description of the process by a product of two separate four-quark operators breaks down.   In fact, a generic product of two four-quark operators with two pairs of quark fields contracted will diverge quadratically when integrated over the region where the locations of these two operators coincide --- a divergence controlled in reality by the $W$ and $Z$ propagators that are approximated at low energies by the four-quark operator $H_W$.  For such a case, we must introduce sufficient subtractions, themselves represented at low energies by additional $\Delta S=2$ four-quark operators, to make the lattice calculation well-defined and dominated by distances that are large compared to the lattice spacing.  This requires the energy scale $\mu$ at which the subtraction is performed to be smaller than $1/a$: $\mu \ll 1/a$.  If this subtraction is performed in the continuum, it must be arranged so that the subtraction term is infra-red safe, with all internal momenta at a scale where continuum QCD perturbation theory can be applied, {\it i.e.} $\Lambda_{\rm QCD} \ll \mu$.  

Fortunately, for the largest contribution to the mass difference $\Delta M_K$, the GIM mechanism~\cite{Glashow:1970gm} removes this quadratic divergence, leaving a convergent integral involving loop momenta at or below the scale of the charm quark mass.  Naively one might expect the subtraction realized by the GIM mechanism to convert a quadratic into a logarithmic divergence, leaving a short distance part that, in the language of
second-order effective field theory, would need to be removed by adding a new, local counterterm.  However, because of the $V-A$ structure of the standard model, the GIM mechanism effectively results in a double subtraction, leaving a finite amplitude that can be computed without ambiguity in the four-flavor theory.  Thus, for the problem at hand we will simply include the charm quark.  In the approximation that $m_c \ll 1/a$ the complete calculation, including those parts referred to both as long and short distance, can be carried out accurately using lattice methods.

Third, a generalization of the Lellouch-Luscher method~\cite{Lellouch:2000pv} is used to correct potentially large finite-volume effects coming from the two-pion state which can be degenerate with the kaon and the associated principal part appearing in the infinite volume integral over intermediate states~\cite{Christ:2010zz}.  This is an important part of this proposal. However, in the kinematic region studied in this paper, we are unable to resolve the two-pion intermediate state signal from statistical fluctuations, so this last piece cannot be studied numerically in the present work.  We therefore postpone a more complete theoretical discussion of this topic, beyond that presented in Refs.~\cite{Christ:2010zz} and \cite{Christ:2012np}, to a later paper.

An important limitation of the numerical calculation described here is the omission of disconnected diagrams.  Including such diagrams leads to an exponentially falling signal to noise ratio adding serious difficulty to the calculation~\cite{Blum:2011pu}. For this practical reason, we omit this type of diagram from this first study of long distance effects in lattice QCD.  The problem of how best to calculate disconnected diagrams with good precision is more general than the present calculation and is a subject of very active research.  The techniques currently being developed will be applied at the next stage to the calculation of the mass difference.  The main aim of this paper is to show that the other issues, special to this second order weak calculation, can be resolved. 

The $K_L-K_S$ mass difference considered in this paper is a particularly interesting example of a class of rare, second order weak processes in which long distance effects ({\it i.e.} effects arising when the two exchanged $W$ or $Z$ bosons are separated by distances at the QCD scale) play an important role.  Closely related to $\Delta M_K$ is the somewhat smaller long distance contribution to $\epsilon_K$, the parameter describing indirect CP violation in the kaon system~\cite{{Buras:2010pza}}, which is discussed here in Appendix~\ref{sec:epsilon}.  Also related to the calculation performed here are various rare kaon decays in which pairs of $W$ and $Z$ bosons and photons are exchanged as a $K$ meson decays into a pion and lepton or neutrino pair~\cite{Isidori:2005tv}.

In this paper, we perform a first study of long distance effects using a $2+1$ flavor, domain wall fermion, $16^3\times32\times16$ lattice ensemble with a $421$\,MeV pion mass and an inverse lattice spacing $1/a=1.73$ GeV. This paper is organized as follows. In  Sec.~\ref{sec:amplitude} we introduce the four point, second-order weak amplitude computed in this work and explain how to extract the finite volume approximation to the mass difference from this amplitude.  (In Appendix~\ref{sec:kinematics} we apply standard perturbation theory to the time development operator generated by the combined strong and weak interactions to motivate this construction and compare it with less favorable alternative approaches.) In Sec.~\ref{sec:simulation} we describe the setup of this calculation including the ensemble used and kinematic regions explored.  Section~\ref{sec:contractions} gives the details of the effective operators used and contractions evaluated. In Sec.~\ref{sec:short}, we discuss the quadratic divergence arising from short distance effects when only three flavors of valence quarks are present and explain how to use the GIM mechanism to remove it. 

In Sec.~\ref{sec:long}, the long distance portion of the amplitude is separated into two parts according to the parity of the intermediate state and each part is discussed separately. In Sec.~\ref{sec:mass}, we renormalize the lattice operators and present the resulting $K_L-K_S$ mass differences.  Section~\ref{sec:NLO_compare} contains a comparison of our lattice calculation with the corresponding NLO perturbation theory result. Since both calculations are limited to the same box graphs and the NLO formulae can be evaluated at the kinematics used in the lattice calculation, this provides a meaningful comparison of these two approaches, with the lattice result approximately twice as large as that found in NLO perturbation theory for our relatively heavy, 421 MeV pion.   Further discussion and our conclusions are given in Sec.~\ref{sec:conclusion}.  Appendix~\ref{sec:epsilon} contains a brief review of the different terms which contribute to the mass difference $\Delta M_K$ and to the indirect CP violation parameter $\epsilon_K$ in the standard model.  We discuss their relative sizes, the distance scales involved and propose a strategy, following the methods presented here, to compute each of these pieces using a combination of perturbation theory and lattice methods, with well controlled errors.  This paper provides a complete account of a calculation reported in preliminary form in Ref.~\cite{Yu:2011gk}.
\newpage

\section{Second order weak amplitude}
\label{sec:amplitude}
If we neglect CP violating effects, which are at the $0.1\%$ level, the standard model contribution to the  $K_{L}$-$K_{S}$ mass difference is given by:
\begin{equation}
\Delta M_K = 2M_{\overline{0}0} =  2\mathscr{P}\sum_{n}\frac{\langle\overline{K}^0|H_W|n\rangle \langle n|H_W|K^0\rangle}{M_K-E_n}.
\label{eq:delta_mk_cont}
\end{equation}
where $H_W$ is the $\Delta S=1$ effective weak Hamiltonian.  The operator $H_W$ represents the effects of the exchange of a $W$ boson at energies much less than the $W$ boson's mass and can be written as a sum of four-quark operators multiplied by Wilson coefficients.  This operator is described thoroughly in Ref.~\cite{Buchalla:1995vs} and the details of the particular choice for $H_W$ used in this calculation are given in Section~\ref{sec:mass}. In Eq.~\eqref{eq:delta_mk_cont} and elsewhere (except Appendix~\ref{sec:epsilon}), we will neglect CP violating effects, treating the off-diagonal mass mixing matrix element $M_{\overline{0}0}$ as real and omitting CP violating terms from the effective weak Hamiltonian $H_W$.  In Eq.~\eqref{eq:delta_mk_cont} we are summing over all possible intermediate states $|n\rangle$ with energy $E_{n}$. This generalized sum includes an integral over intermediate-state energies and the $\mathscr{P}$ indicates that the principal part should be taken when evaluating the integral over the $E_{n}=M_{K}$ singularity.   This formula does not correctly treat intermediate states whose energies are on the order of the masses of the $W$ and $Z$ bosons.  As will be discussed below, such states are unimportant for $\Delta M_K$ in the standard model.

Since the second order mass difference given in Eq.~\eqref{eq:delta_mk_cont} must appear in the Dyson-Wick expansion for the time evolution operator at second order in the $\Delta S=1$ effective weak Hamiltonian $H_W$, we should expect a similar expression to enter the following four-point correlator which can be directly studied using lattice methods:
\begin{equation}
G(t_f,t_2,t_1,t_i)=\langle0|T\left\{\overline{K^0}(t_f)H_W(t_2)H_W(t_1)\overline{K^0}(t_i)\right\}|0\rangle,
\label{eq:unintegrated_correlator}
\end{equation}
where $T$ is the usual time ordering operator.  Here the initial $K^0$ state is generated by the kaon source $\overline{K^0}(t_i)$ at the time $t_i$ and the final $\overline K_{0}$ state is destroyed by the kaon sink $\overline{K}^{0}(t_f)$ at time $t_f$ and we assume $t_f \gg t_i$. The two effective Hamiltonians act at the times $t_2$ and $t_1$.  Assuming that the time separations $t_f-t_k$ and $t_k-t_i$ for $k=1$ and 2  are sufficiently large that the time development operator will project onto the $K^0$ and $\overline{K^0}$ initial and final states and inserting a complete set of energy eigenstates $|n\rangle$, we find:
\begin{equation}
G(t_f,t_2,t_1,t_i)=N_K^2 e^{-M_K(t_f-t_i)}\sum_{n}\langle\overline{K^0}|H_W|n\rangle\langle n|H_W|K^0\rangle e^{-(E_n-M_K)|t_2-t_1|},
\label{eq:unintcorr}
\end{equation}
where $N_K$ is the normalization factor for the kaon interpolating operator.  If we fix the times $t_i$ and $t_f$, then this correlator depends only on the time separation between the two Hamiltonians $|t_2-t_1|$.  We will refer to $G(t_f,t_2,t_1,t_i)$ as the unintegrated correlator. The unintegrated correlator receives contributions from all possible intermediate states. The terms in this sum over intermediate states show exponentially decreasing or increasing behavior with increasing  $|t_2-t_1|$ depending on whether $E_n$ lies above or below $M_K$. 

\begin{figure}[!htp]
\centering
\includegraphics[width=0.7\textwidth]{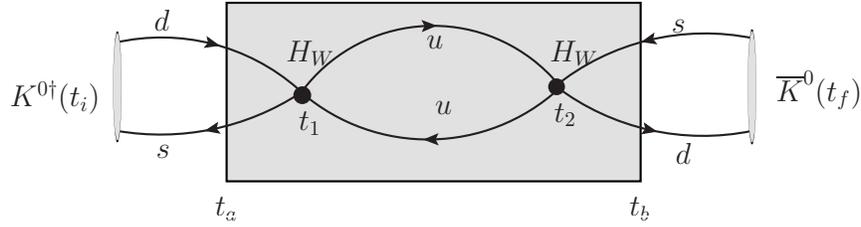}
\caption{One type of diagram contributing to $\mathscr{A}$ in Eq.~\eqref{eq:integrated_correlator}. Here $t_2$ and $t_1$ are integrated over the time interval $[t_a,t_b]$, represented by the shaded region.}
\label{fig:demo}
\end{figure}

We can integrate the times $t_1$ and $t_2$ in the unintegrated correlator over a time interval $[t_a,t_b]$ and obtain:
\begin{equation}
\mathscr{A}=\frac{1}{2}\sum_{t_2=t_a}^{t_b}\sum_{t_1=t_a}^{t_b}\langle0|T\left\{\overline{K^0}(t_f)H_W(t_2)H_W(t_1)\overline{K^0}(t_i)\right\}|0\rangle.
\label{eq:integrated_correlator}
\end{equation}
We call this amplitude the integrated correlator. The integrated correlator is represented schematically in Fig.~\ref{fig:demo}.  After inserting a sum over  intermediate states and summing explicitly over $t_2$ and $t_1$ 
in the interval $[t_a,t_b]$ one obtains:
\begin{equation}
\begin{split}
\mathscr{A}   =& N_K^2e^{-M_K(t_f-t_i)} \Bigg\{ \sum_{n\ne n_0}\frac{\langle\overline{K}^0|H_W|n\rangle\langle n|H_W|K^0\rangle}{M_K-E_n}\left( -T + \frac{e^{(M_K-E_n)T}-1}{M_K-E_n}\right)\\
			&+ \frac{1}{2}\langle\overline{K}^0|H_W|n_0\rangle\langle n_0|H_W|K^0\rangle T^2\Bigg\}.
\end{split}
\label{eq:integration_result}
\end{equation}
Here $T=t_b-t_a+1$ and the sum includes all possible intermediate states except a possible state $|n_0\rangle$ which is degenerate with the kaon, $E_{n_0}=M_K$.   (In this discussion and in the remainder of this paper, we express all dimensionful quantities in lattice units unless otherwise specified.)  The contribution from such a degenerate state appears separately as the final term on the right hand side of this equation.   The method proposed in Ref.~\cite{Christ:2010zz} to control finite volume errors requires that the spatial volume be adjusted to create such a degenerate $\pi-\pi$ state and that this state be omitted from the finite volume expression used as an approximation to the infinite volume quantity $\Delta M_K$.~\footnote{Together with G.~Martinelli, we are also exploring the possibility of controlling finite volume effects without requiring a two-pion state to be degenerate with the kaon, generalizing the approach to finite-volume effects in $K\to\pi\pi$ decays developed in \cite{Lin:2001ek,Kim:2005gf}.}  The expression on the right-hand side of Eq.~\eqref{eq:integration_result} has been made easier to recognize by replacing the quantity $1-\exp{(M_K-E_n)a}$, which results from the sum over the discrete times $t_1$ and $t_2$, by its value in the continuum limit, $\it i.e.$ by either zero or $(E_n-M_K)a$ as appropriate.

The coefficient of the term which is proportional to $T$ in Eq.~\eqref{eq:integration_result} gives the finite-volume approximation to $\Delta M_K$ up to some normalization factors:
\begin{equation}
\Delta M_K^{FV} =  2 \sum_{n\neq n_0} \frac{\langle\overline{K}^0|H_W|n\rangle\langle n|H_W|K^0\rangle}{M_K-E_n}
\label{eq:fvmassdiff}
\end{equation}
The other terms in Eq.~\eqref{eq:integration_result} can be classified into four categories according to their dependence on $T$: 
\begin{enumerate}
\item[i)] 
The term independent of $T$ within the large parentheses. This constant does not affect our determination of the mass difference from $\mathscr{A}$. 
\item[ii)] Terms exponentially decreasing as $T$ increases coming from states $|n\rangle$ with $E_n>M_K$.  These terms are negligible for sufficiently large $T$. 
\item[iii)] Terms exponentially increasing as $T$ increases coming from states $|n\rangle$ with $E_n<M_K$. These will be the largest contributions when $T$ is large and must be removed as discussed in the paragraph below. 
\item[iv)] The final term proportional to $T^2$ coming from states degenerate with the kaon.  As discussed below, this term must be identified and removed in order to relate the finite- and infinite-volume expressions for $\Delta M_K$ following the method of Ref.~\cite{Christ:2010zz}.
\end{enumerate}
This behavior of the integrated correlator is interpreted in Appendix~\ref{sec:kinematics} by using standard perturbation theory to analyze the time development generated by the sum of the QCD and weak Hamiltonian.  This provides insight into Eq.~\eqref{eq:integration_result} and allows other alternative choices of correlation function to be easily discussed.

The exponentially growing terms, introduced in item iii) above, pose a significant challenge.  Fortunately, the two leading terms corresponding to the vacuum and single pion states can be computed separately and subtracted. In this work, since no disconnected diagrams are included, there is no contribution from the vacuum state. The matrix element $\langle \pi^0 |H_W|K^0\rangle$ can be obtained from three-point correlation functions which allows the exponentially growing single-pion term to be determined and removed. 

A second approach to remove these two unwanted exponentially growing terms exploits the chiral Ward identities to add to $H_W$ terms proportional to the scalar and pseudo-scalar densities, $\overline{s}d$ and $\overline{s}\gamma^5d$ with coefficients chosen to eliminate the two matrix elements $\langle \pi|H_W|K^0\rangle$ and $\langle 0|H_W|K^0\rangle$.  Since these two densities can be written as the divergence of the vector and axial currents respectively, they cannot contribute to an on-shell matrix element such as that given in Eq.~\eqref{eq:unintegrated_correlator}. (Note, this statement remains valid when the effective, weak interaction, current-current and QCD penguin operators are added to the action so no contact terms are needed for this approach to be applied to the second-order processes considered here.)   This approach is similar to the subtraction that we carry out in this paper but instead of removing only the exponentially growing term in Eq.~\eqref{eq:integration_result}, such an addition will remove all single pion and vacuum contributions from that equation, including their appearance in the sum over intermediate states $|n\rangle$.  We have not explored this approach here because our omission of disconnected diagrams has already removed possible vacuum intermediate state contributions and the density $\overline{s}d$ will contribute only to type 3 and type 4 amplitudes (see Figs.~\ref{fig:type3} and \ref{fig:type4} below) which are not included in the present calculation.

Two-pion states with energies below $M_K$ may also exist and, if present, must be explicitly identified and removed. For our current kinematics, the only $\pi-\pi$ state with an energy possibly below $M_K$ is the threshold state with two pions essentially at rest.  In the following we study the contribution of this state as the kaon mass is varied.  In a future, more physical calculation the difficulty of two-pion states with energy below $M_K$ can be avoided if we introduce G-parity boundary conditions to force each pion to have a non-zero momentum and then tune the energy of lightest $\pi-\pi$ state to be degenerate with that of the kaon, following Ref.~\cite{Christ:2010zz}.

The approach developed in Ref.~\cite{Christ:2010zz} to control finite-volume errors requires a choice of spatial box and boundary conditions which results in a $\pi-\pi$ state, $|n_0\rangle$ whose energy approximates that of the kaon. If this degeneracy is precise with an accuracy $|M_K-E_{n_0}| \ll 1/T$ then this state will contribute the term proportional to $T^2$ term in Eq.~\eqref{eq:integration_result} above.  Since such degeneracy needs only to be achieved at a level of $|M_K-E_{n_0}| \ll \Lambda_{\rm QCD}$
(the scale at which the result will depend on $M_K-E_{n_0}$) to properly control finite-volume errors, we may instead need to identify this nearly degenerate state as a term showing very slow exponential increase or decrease with $T$.   This is the final, important piece of a lattice calculation of $\Delta M_K$.   However, in this work we have been unable to distinguish a clear signal from the $\pi-\pi$ intermediate state.  Thus, this last step cannot be studied in the current work.  We will discuss this issue further in Sec.~\ref{sec:long}.

\section{Simulation details}
\label{sec:simulation}
The calculation is performed on a lattice ensemble generated with the Iwasaki gauge action and 2+1 flavors of domain wall fermions at a coupling $\beta=2.13$. The space-time volume is $16^3\times32$ and the inverse lattice spacing $a^{-1}=1.729(28)$GeV. The fifth-dimensional extent is $L_s=16$ and the residual mass is $m_{\rm res}=0.00308(4)$ in lattice units.  (In the following all quantities will be expressed in lattice units unless otherwise stated.  Occasionally, an explicit factor of the lattice spacing $a$ may be added for clarity.)   The sea light and strange quark masses are $m_l=0.01$ and $m_s=0.032$ respectively, corresponding to a pion mass $M_{\pi}=421$\,MeV and a kaon mass $M_{K}=563$\,MeV. We use 800 configurations, each separated by 10 time units.  This ensemble is described in greater detail in Ref.~\cite{Blum:2011pu} and is also similar to the earlier ensembles described and analyzed in Ref.~\cite{Allton:2007hx}, except that the current ensemble has a more physical value for the sea quark mass and was generated with a better RHMC algorithm.

We will use Fig.~\ref{fig:demo} to explain the set up of this calculation. Two Coulomb gauge-fixed kaon sources are located at time slices $t_i=0$ and $t_f=27$ respectively. The two effective weak operators $H_W(t_i)_{i=1,2}$ are introduced in the interval $4 \le t_1,t_2 \le 23$. We calculate the four-point function defined in Eq.~\eqref{eq:unintegrated_correlator} for all possible choices of $t_1$ and $t_2$.  Note that the diagram given in Fig.~\ref{fig:demo} is only one type of  possible contraction. We will discuss the contractions in detail in Sec.~\ref{sec:contractions}. 

For given values of $t_1$ and $t_2$, each of the two effective operators should be integrated over the whole spatial volume since these two volume averages would result in reduced statistical noise. However, there is no easy way to do this because of two difficulties. First, we are not able to compute all of the light-quark propagators connecting the two operators. It is impractical to use point source propagators since there will be $16^3$ point sources on each time slice.   In simpler cases, this difficulty can be avoided by the use of a stochastic source distributed over the time slice.  However, an attempt to use this technique in the present case failed to give a signal that could be recognized above the noise.  Even if this first difficulty of generating the multitude of needed point source propagators could be overcome, we would still face a second difficulty:  the number of operations needed to calculate all the contractions would be $O(V^2)$, where $V$ is the space-time volume of the lattice.  This also would be too time consuming. Thus, we sum the location of only one of the two operators over the spatial volume and, relying on the translational symmetry of the other ingredients in the calculation, fix the spatial location of other operator at the origin $(0,0,0)$.  For each of the contractions in our calculation, these two weak operators enter in distinct ways and we average the two cases where one operator is fixed at the origin and the other integrated over the spatial volume to
improve the statistics.

We use periodic boundary conditions in the spatial directions for the Dirac operator when computing the propagators. In the temporal direction, we calculate propagators for both periodic and anti-periodic boundary conditions and take their average for the propagator that we use. This effectively doubles the temporal extent of the lattice and suppresses around-the-world effects to a negligible level.  (This approach is equivalent to working on a lattice of size $16^3\times 64$ with gauge fields invariant under a translation of 32 sites in the time direction.)  The most expensive part of this simulation is solving for the light quark propagators. There are 2 wall source light quark propagators and 20 point source light quark propagators, one on each time slice between $t_a=4$ and $t_b=23$. So in total we need to calculate $(20+2)\times2=44$ propagators, where the factor of two comes from our two choices of temporal boundary conditions.  Further each propagator requires 12 Dirac operator inversions, one for each spin and color.  This large number of light-quark Dirac operator inversions makes this calculation a good candidate for the use of the EigCG technique~\cite{Stathopoulos:2007zi,Liu:2011jp}.  We collect the lowest $100$ eigenvectors and use them to accelerate the light-quark Dirac operator inversions. The overhead associated with collecting these low modes is amortized over many inversions and the number of conjugate gradient iterations is reduced by a factor of $6$. 

\begin{figure}[!htp]
\centering
\includegraphics[width=0.7\textwidth]{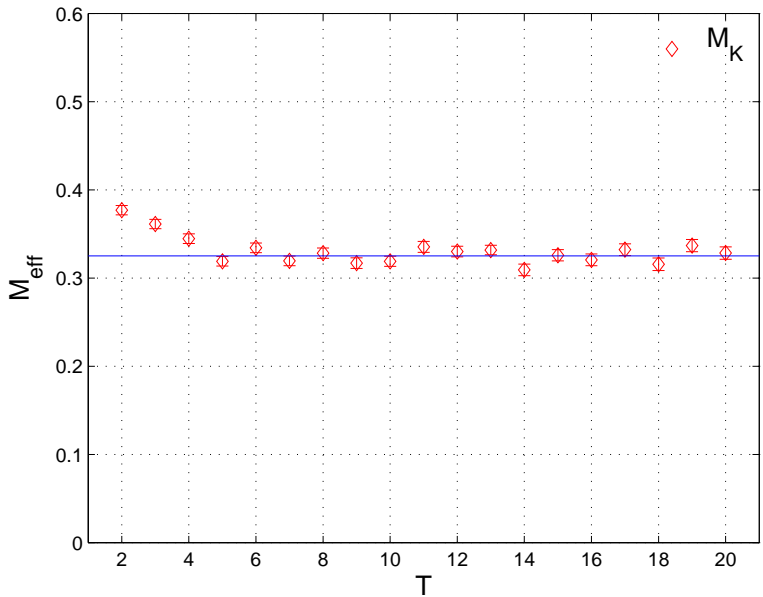}
\caption{A plot of the kaon effective mass found from the two point correlator between a wall source and a wall-sink using $m_l=0.01$ and $m_s=0.032$.  The blue line shows the result of the fit.}
\label{fig:kaon_eff_mass}
\end{figure}

We mentioned in Sec.~\ref{sec:amplitude} that the time separation between the kaon wall source and the $\Delta S=1$ weak operators should be large enough to project onto kaon states. In the set up of this calculation, the two operators can be located at any time slice between $[4,23]$. So the time separation between the kaon source or sink and either effective weak operator is guaranteed to be equal or larger than $4$. In Fig.~\ref{fig:kaon_eff_mass} we give a sample kaon effective mass plot for $m_l=0.01$ and $m_s=0.032$.
This plot suggests that the effects of excited kaon states will be negligible when the separation between source and sink is 5 or larger.  We therefore use the restricted range $[5,22]$ for $t_k$ in the following analysis, discarding the results when either operator is at the location $t_k=4$ or 23 for $k=1$ and 2.

In order to reduce short distance effects to a level which can be accurately controlled using lattice methods, we introduce a valence charm quark into our calculation.  In Sec.~\ref{sec:short}, we investigate the resulting GIM cancellation for different charm quark masses. These masses are given in Table.~\ref{tab:charm_mass}, where we use mass renormalization factor $Z^{\overline{MS}}_m$(2 Gev)=1.498~\cite{Aoki:2010dy}. When we discuss the long distance effects in Sec.~\ref{sec:long}, we choose a $863$\,MeV valence charm quark mass and several different valence strange quark masses. The strange quark masses and corresponding kaon masses are given in Table.~\ref{tab:kaon_mass}.  The up and down quark masses are kept at their unitary value, equal to the 0.01 mass of the sea quark.

\begin{table}[!htp]
\centering
\caption{Valence charm quark masses used to implement the GIM cancellation. The upper row gives the bare masses in lattice units. The lower row contains the $\overline{\rm MS}$ masses at a scale of 2 GeV.}
\begin{ruledtabular}
\begin{tabular}{c | c c c c c c}
$m_c$  & 0.132 & 0.165 & 0.198 & 0.231 & 0.264 & 0.330\\
\hline
$m_c$ (MeV) & 350 & 435 & 521 & 606 & 692 & 863 \\
\end{tabular}
\end{ruledtabular}
\label{tab:charm_mass}
\end{table}

\begin{table}[!htp]
\centering
\caption{Valence strange quark mass (upper row) and kaon mass (lower row), both in lattice units.}
\begin{ruledtabular}
\begin{tabular}{c | c c c c c c c c}
$m_s$ & 0.01 & 0.032 & 0.06 & 0.075 & 0.09 & 0.11 & 0.13 & 0.18\\
\hline
$M_K$ & 0.2431(8)& 0.3252(7) & 0.4087(7) & 0.4480(7) & 0.4848(8) & 0.5307(8) & 0.5738(8) & 0.6721(10)\\
\end{tabular}
\end{ruledtabular}
\label{tab:kaon_mass}
\end{table}

\section{Operators and contractions}
\label{sec:contractions}
In this section we will describe the $\Delta S=1$ effective weak operators and the contractions that are evaluated in this calculation.  The first-order, $\Delta S=1$ effective weak Hamiltonian including four flavors can be written as:
\begin{equation}
H_W=\frac{G_F}{\sqrt{2}}\sum_{q,q^{\prime}=u,c}V_{qd}V^{*}_{q^{\prime}s}(C_1Q_1^{qq^{\prime}}+C_2Q_2^{qq^{\prime}})
\label{eq:H_W}
\end{equation}
where $q$ and $q'$ are each one of the two charge 2/3 quarks in the four flavor theory ($u$ and $c$), $V_{qd}$ and $V_{q's}$ are Cabibbo-Kobayashi-Maskawa (CKM) matrix elements, $C_1$ and $C_2$ are Wilson coefficients and we include only the current-current operators, which are defined as:
\begin{equation}
\begin{split}
Q_1^{qq{\prime}}&=(\bar{s}_id_i)_{V-A}(\bar{q}_jq^{\prime}_j)_{V-A}\\
Q_2^{qq{\prime}}&=(\bar{s}_id_j)_{V-A}(\bar{q}_jq^{\prime}_i)_{V-A}\,,
\end{split}
\label{eq:operator}
\end{equation}
where $i,j$ are color indices and the spinor indices are contracted within each pair of brackets.  The subscript $V-A$ on each fermion bilinear indicates the usual difference of vector and axial currents with the four-vector index on the currents appearing in each of the two bilinear factors contracted.  We neglect the penguin operators in the effective Hamiltonian. This is a good approximation since these operators are suppressed by a factor $\tau = -V_{td}V^{*}_{ts}/V_{ud}V^{*}_{us} = 0.0016$ in a four flavor theory.  (Such operators will be discussed in Sec.~\ref{sec:epsilon} when a possible calculation of the long distance contribution to $\epsilon_K$ is considered.) 

\begin{figure}[!htp]
\begin{ruledtabular}
\begin{tabular}{cc}
\includegraphics[width=0.4\textwidth]{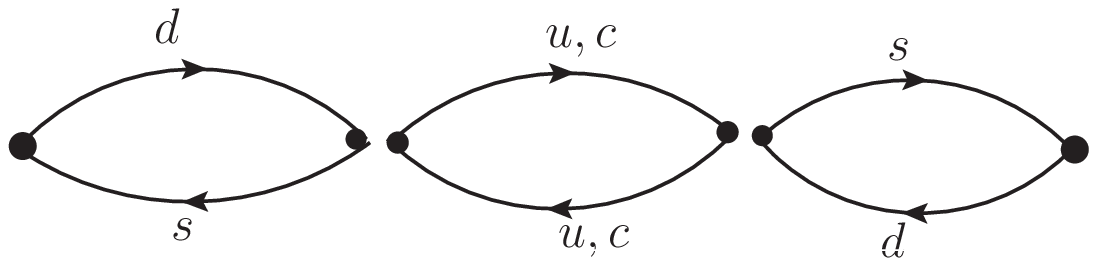} & \includegraphics[width=0.4\textwidth]{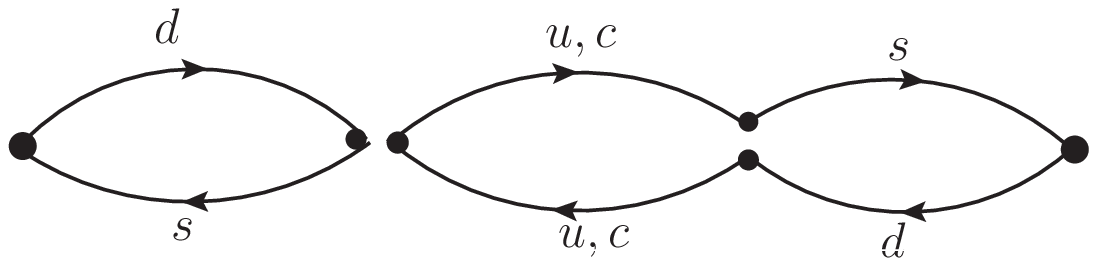} \\
\fig{1} & \fig{2} \\
\hline
\includegraphics[width=0.4\textwidth]{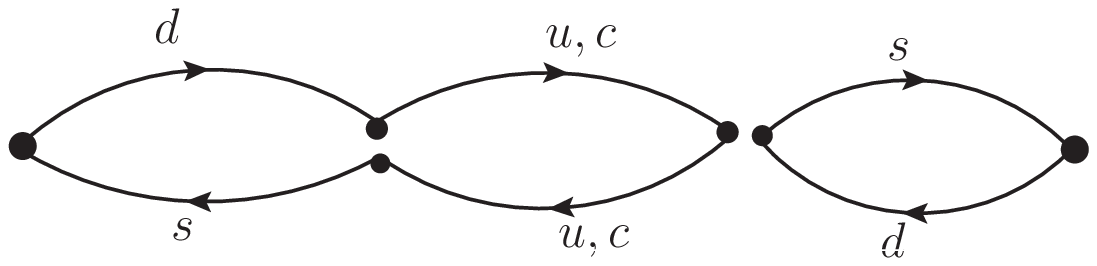} & \includegraphics[width=0.4\textwidth]{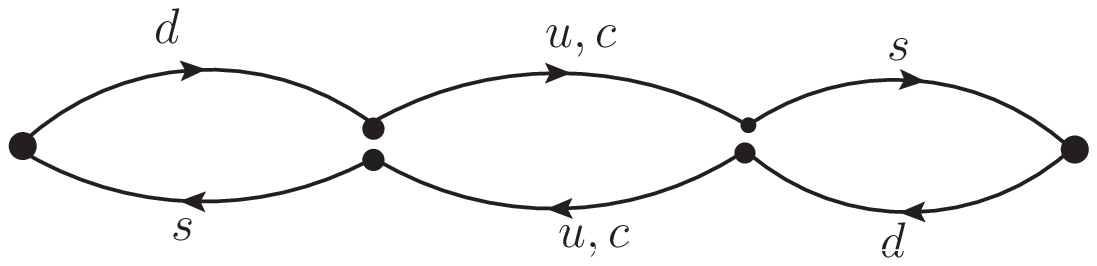}\\
\fig{3} & \fig{4}\\
\end{tabular}
\end{ruledtabular}
\caption{Diagrams for type 1 contractions. The two two-quark vertices associated with the kaon sources correspond to a spinor product including a $\gamma_5$ matrix. Each of the four two-quark vertices associated with four quark operators correspond to a contraction of color indices.  The spinor products, which include the matrix $\gamma_{\mu}(1-\gamma_5)$, connect incoming and outgoing quark lines which carry the same electric charge.  Vertices where the quark lines are joined in this fashion then have the color and spin contracted in the same pattern and correspond to the operator $Q_1$.  Where the quark lines and corresponding color contractions for quarks with different electric charges are joined, the operator $Q_2$ appears.}
\label{fig:type1}
\end{figure}

\begin{figure}[!htp]
\begin{ruledtabular}
\begin{tabular}{cc}
\includegraphics[width=0.4\textwidth]{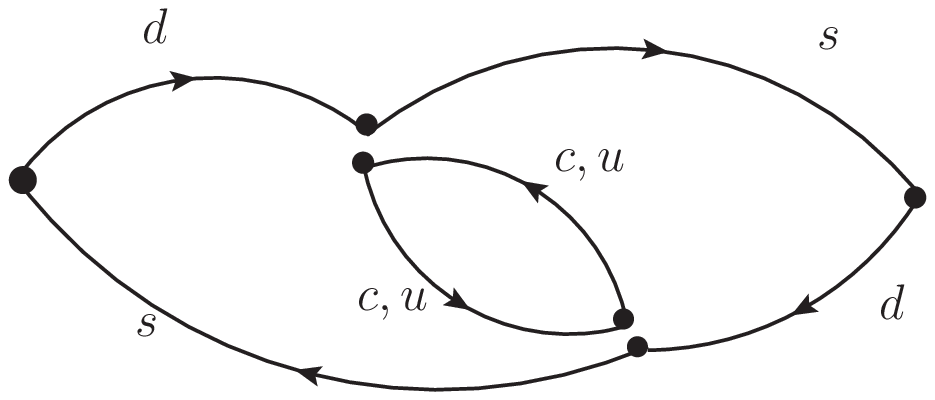} & \includegraphics[width=0.4\textwidth]{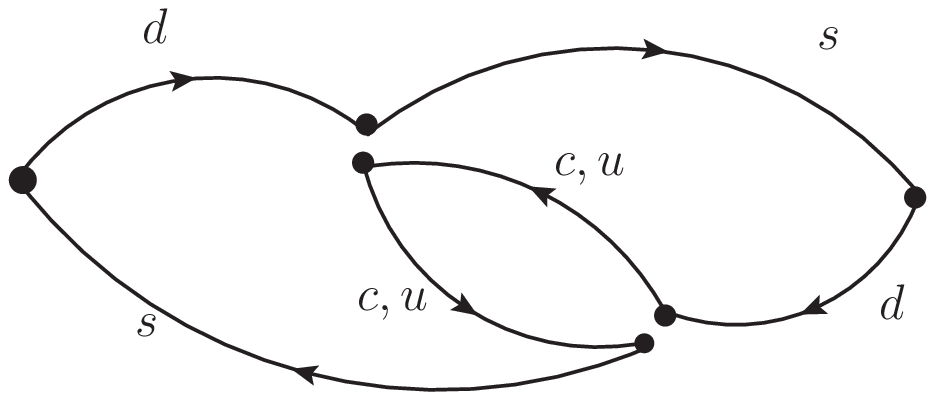} \\
\fig{5} & \fig{6} \\
\hline
\includegraphics[width=0.4\textwidth]{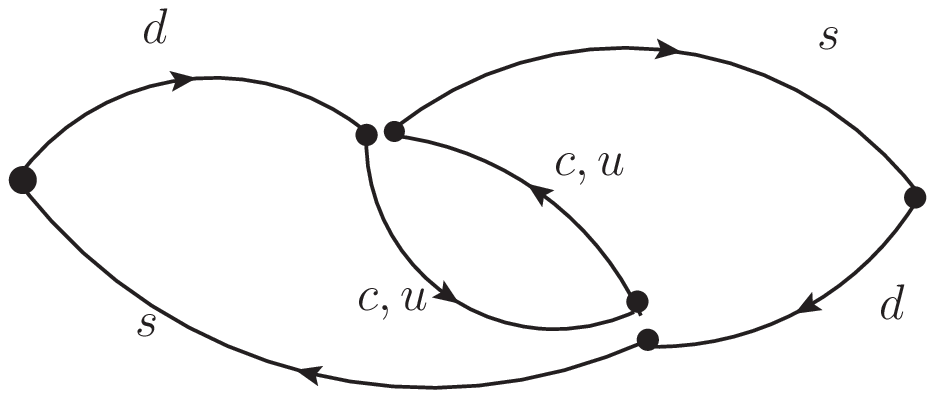} & \includegraphics[width=0.4\textwidth]{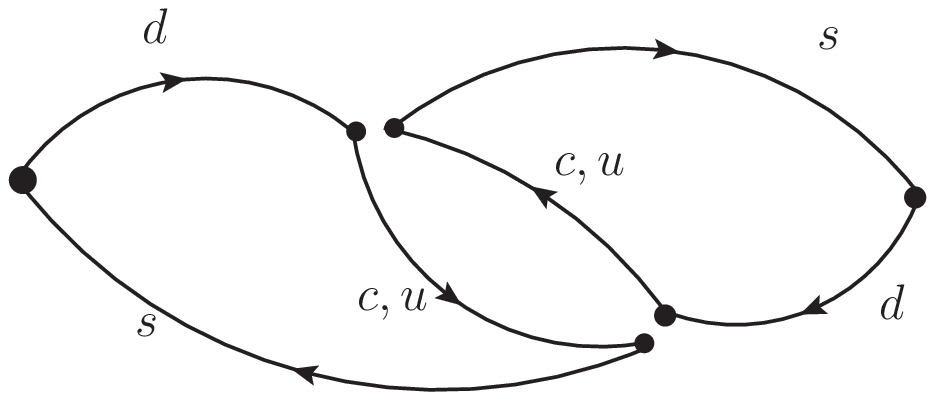}\\
\fig{7} & \fig{8}\\
\end{tabular}
\end{ruledtabular}
\caption{Diagrams for type 2 contractions.  The conventions used here are the same as those explained in the caption to Fig.~\ref{fig:type1}.}
\label{fig:type2}
\end{figure}

\begin{figure}[!htp]
\begin{ruledtabular}
\begin{tabular}{cc}
\includegraphics[width=0.4\textwidth]{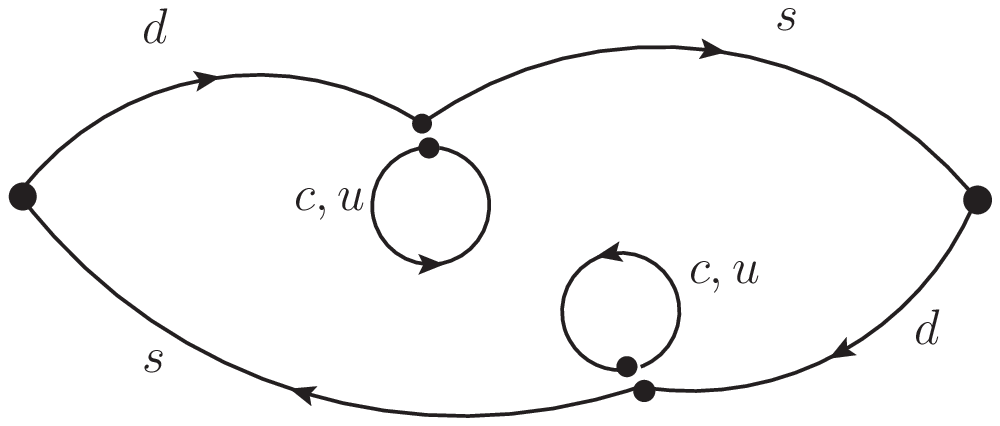} & \includegraphics[width=0.4\textwidth]{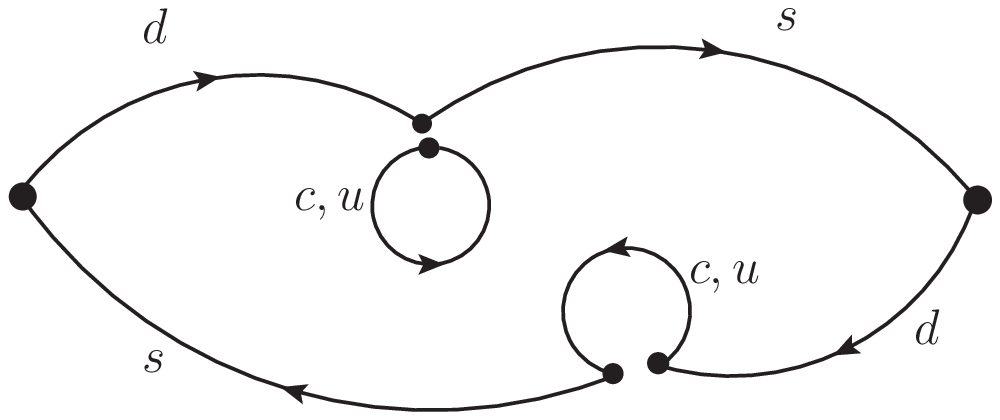} \\
\fig{13} & \fig{14} \\
\hline
\includegraphics[width=0.4\textwidth]{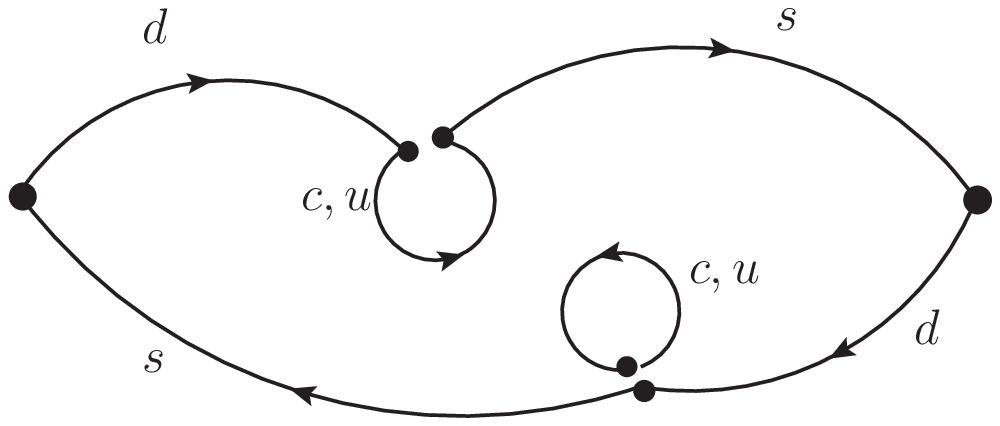} & \includegraphics[width=0.4\textwidth]{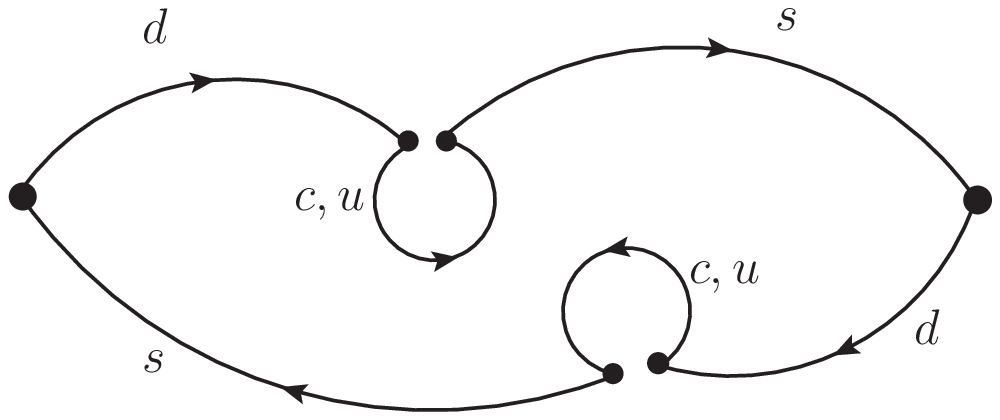}\\
\fig{15} & \fig{12}\\
\end{tabular}
\end{ruledtabular}
\caption{Diagrams for type 3 contractions, which are not included in this calculation.}
\label{fig:type3}
\end{figure}

\begin{figure}[!htp]
\begin{ruledtabular}
\begin{tabular}{cc}
\includegraphics[width=0.4\textwidth]{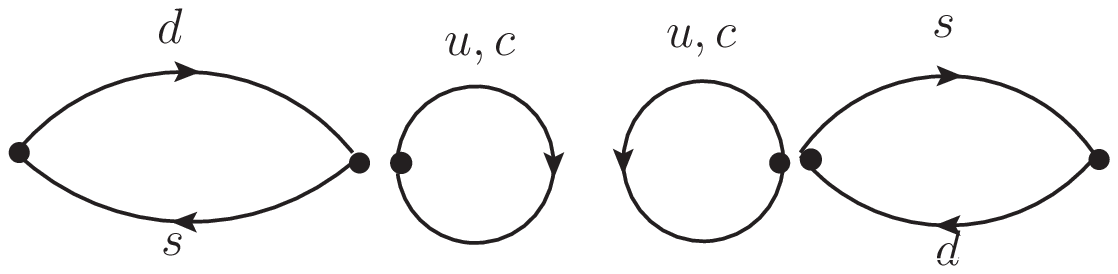} & \includegraphics[width=0.4\textwidth]{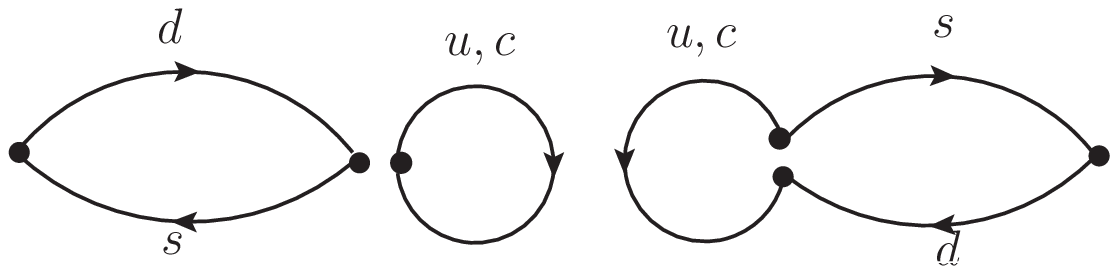} \\
\fig{9} & \fig{10} \\
\hline
\includegraphics[width=0.4\textwidth]{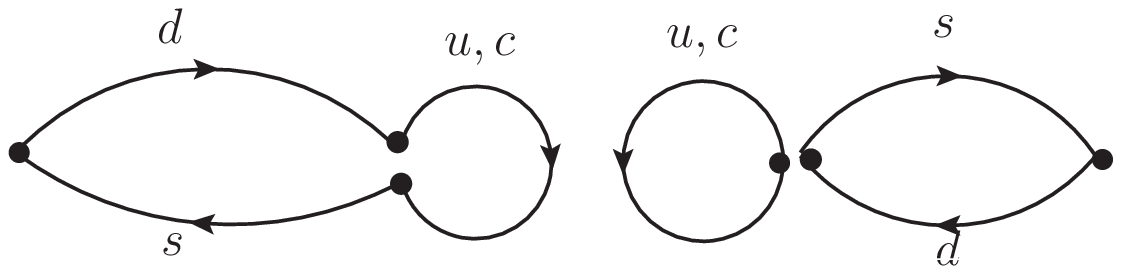} & \includegraphics[width=0.4\textwidth]{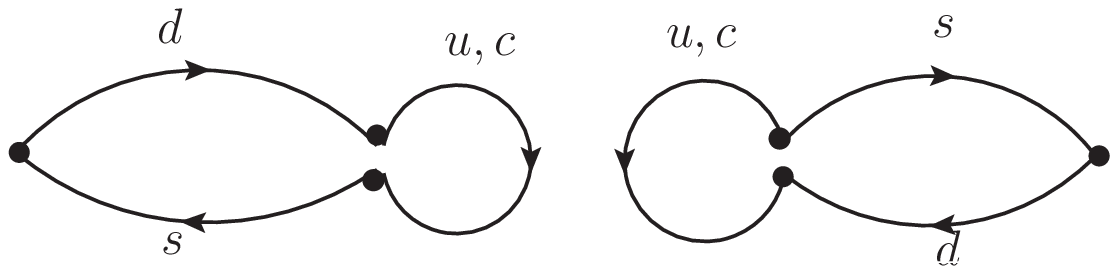}\\
\fig{11} & \fig{12}\\
\end{tabular}
\end{ruledtabular}
\caption{Diagrams for type 4 contractions, which are not included in this calculation.}
\label{fig:type4}
\end{figure}

We list all the possible contractions contributing to the four point correlators in Figs.~\ref{fig:type1}-\ref{fig:type4}. There are in total 16 diagrams which are labeled by circled numbers and we categorize them into four types according to their topology. There are six quark propagators in each diagram. Four of these propagators are connected to the kaon wall sources while two propagators connect one of the weak operators to the other or each weak operator to itself.  We call these two quark propagators internal propagators. In a four flavor theory, the flavor of the internal quark propagators can be either up or charm.  We therefore have four different combinations for each diagram: $uu$, $cc$, $uc$ and $cu$. We use these labels in a subscript to denote the flavor of the two internal quark propagators. For example, the first diagram with two internal up quark propagators is represented by $\fig{1}_{uu}$, and the GIM cancellation occurs in the combination:
\begin{equation}
\fig{1}_{\rm GIM} = \fig{1}_{uu} + \fig{1}_{cc} - \fig{1}_{uc} - \fig{1}_{cu}.
\label{eq:GIM_cancel}
\end{equation}

Because of the arrangement of quark flavors and spin contractions in  the operators   $Q_1^{qq'}$ and $Q_2^{qq'}$ the spin indices on quark fields which carry the same charge are always contracted with an interposed $\gamma_{\mu}(1-\gamma_5)$ spin matrix.  Therefore, the pattern of spin contractions need not be represented in Figs.~\ref{fig:type1}-\ref{fig:type4}.  Instead, the separation of each four-quark vertex into two pairs of two quark vertices shown in those figures indicates the pattern of color contractions.  Thus, when two quark lines carrying the same charge are joined in those figures that arrangement of spin and color contractions is the same and the operator $Q_1^{qq'}$ appears at that vertex.  If lines with different charge are joined, it is the operator $Q_2^{qq'}$ that appears.

All the correlation functions are given by combinations of these contractions. For example, 
\begin{equation}
\langle\overline{K^0}(t_f)Q_1^{uu}(t_2)Q_1^{uu}(t_1)\overline{K^0}(t_i)\rangle=\fig{1}_{uu}-\fig{5}_{uu}-\fig{9}_{uu}+\fig{13}_{uu},
\label{eq:complete}
\end{equation}
where the contractions identified by circled numbers do not carry the minus sign coming from the number of fermion loops.  Instead these minus signs appear explicitly in Eq.~\eqref{eq:complete}.  Since our definition of the kaon interpolation operators is $K^{0}=i(\bar{d}\gamma_{5}s)$, there will be a minus sign, $i^2=-1$, coming from two kaon sources. This minus sign is also not included in the contractions.

In a unitary calculation, we need to include all types of diagrams. However, we do not include type 3 and type 4 diagrams in this calculation for two reasons.  The first reason is practical.   We would need to compute an additional stochastic wall source for each time slice to evaluate the new loop graphs which appear in the type 3 and 4 contractions. This would approximately double the computation time.  More importantly, type 4 diagrams are disconnected diagrams which are extremely noisy and would require a far larger statistical sample than is being used here~\cite{Blum:2011pu}.  The second reason is phenomenological.  There is some empirical evidence suggesting that the contribution from type 3 and type 4 diagrams may be small.  For example, disconnected graphs similar to those of type 4 are often small when contributing to other processes where they are said to be ``OZI suppressed''~\cite{Zweig:1964jf,*Okubo:1963fa,*Iizuka:1966fk}.  The omission of such diagrams is also consistent with the results of the recent study of $\Delta I = 1/2$ $K\to\pi\pi$ decays~\cite{Blum:2011pu} in which the contribution of disconnected diagrams was found to be zero within rather large errors.  (Note, in the 2+1 flavor calculation of Ref.~\cite{Blum:2011pu}, diagrams containing a closed loop formed from a single quark line did give large $O(1/a^2)$ contributions from off-shell states and required careful treatment.  However, in the case of four flavors, GIM cancellation renders such loops convergent, reducing them in size by a factor of $(m_c a)^2$.  As a result, such disconnected diagrams may require less complex treatment in the four-flavor theory considered here.)  Of course in a complete calculation these diagrams must be calculated explicitly after which the precision of the Zweig suppression will be known.

Neglecting type 3 and type 4 diagrams, Eq.~\eqref{eq:complete} reduces to:
\begin{equation}
\langle\overline{K^0}(t_f)Q_1^{uu}(t_2)Q_1^{uu}(t_1)\overline{K^0}(t_i)\rangle=\fig{1}_{uu}-\fig{5}_{uu}.
\label{eq:q1q1}
\end{equation}
There are two other possible operator combinations in this calculation:
\begin{equation}
\begin{split}
\langle\overline{K^0}(t_f)Q_2^{uu}(t_2)Q_2^{uu}(t_1)\overline{K^0}(t_i)\rangle&=\fig{4}_{uu}-\fig{8}_{uu}\\
\langle\overline{K^0}(t_f)(Q_1^{uu}(t_2)Q_2^{uu}(t_1)+Q_2^{uu}(t_2)Q_1^{uu}(t_1))\overline{K^0}(t_i)\rangle&=-\fig{2}_{uu}-\fig{3}_{uu}+\fig{6}_{uu}+\fig{7}_{uu}.
\end{split}
\end{equation}
After GIM cancellation, these become:
\begin{equation}
\begin{split}
\langle\overline{K^0}(t_f)Q_{11}^{\rm GIM}(t_2,t_1)\overline{K^0}(t_i)\rangle&=\fig{1}_{\rm GIM}-\fig{5}_{\rm GIM}\\
\langle\overline{K^0}(t_f)Q_{22}^{\rm GIM}(t_2,t_1)\overline{K^0}(t_i)\rangle&=\fig{4}_{\rm GIM}-\fig{8}_{\rm GIM}\\
\langle\overline{K^0}(t_f)\left(Q_{12}^{\rm GIM}(t_2,t_1)+Q_{21}^{\rm GIM}(t_2,t_1)\right)\overline{K^0}(t_i)\rangle&=-\fig{2}_{\rm GIM}-\fig{3}_{\rm GIM}+\fig{6}_{\rm GIM}+\fig{7}_{\rm GIM}.
\end{split}
\label{eq:GIM_contractions}
\end{equation}
Here the subscript ``GIM'' under the circles indicates the same combination of internal quark line flavors as is given in Eq.~\eqref{eq:GIM_cancel}. The four operator products $Q_{ij}^{\rm GIM}(t_2,t_1)$ appearing on the left-hand side of Eq.~\eqref{eq:GIM_contractions} are each the appropriate sum of all four combinations of intermediate charm and up quarks:
\begin{equation}
	\begin{split}
		Q_{ij}^{\rm GIM}(t_2,t_1) &= Q_i^{uu}(t_2)Q_j^{uu}(t_1)+Q_i^{cc}(t_2)Q_j^{cc}(t_1)\\
															&-Q_i^{uc}(t_2)Q_j^{cu}(t_1)-Q_i^{cu}(t_2)Q_j^{uc}(t_1)\qquad i,j=1,2.
	\end{split}
\end{equation}

\begin{figure}[!htp]
\begin{ruledtabular}
\begin{tabular}{cc}
\includegraphics[width=0.4\textwidth]{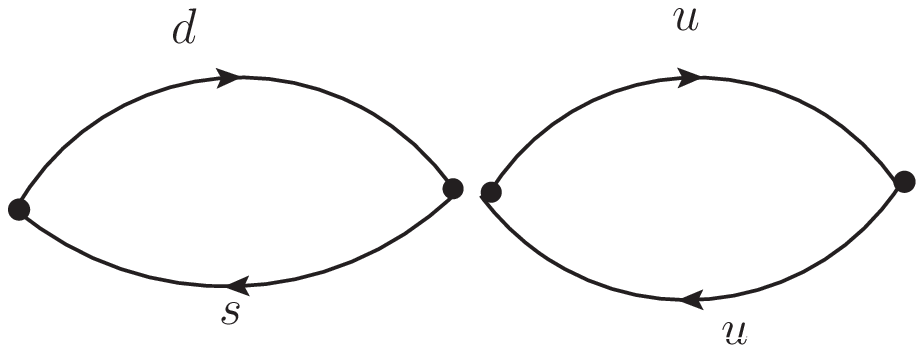} &
\includegraphics[width=0.4\textwidth]{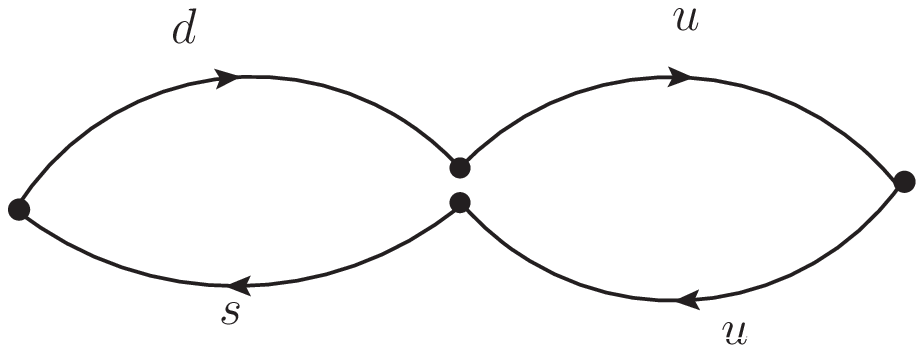}
\end{tabular}
\end{ruledtabular}
\caption{Diagram for the $\langle\pi^0 | H_W | K^0\rangle$ contractions. The vertex at each of the two kaon sources includes a $\gamma_5$ matrix.   The meaning of the vertices is the same as those in the previous figures and is explained in the caption to Fig.~\ref{fig:type1}}
\label{fig:ktopi}
\end{figure}

As discussed in Sec.~\ref{sec:amplitude}, we need to calculate the matrix element $\langle\pi^0 | H_W | K^0\rangle$ in order to remove the exponentially growing term in the second order correlator. However, the definition of the $\pi^0$ intermediate state must be reconsidered in this non-unitary calculation. In a unitary theory, $\bar{u}u$, $\bar{d}d$ and $\bar{s}s$ will mix with each other through disconnected diagrams. Then the resulting energy eigenstates are $\pi^0$, $\eta$ and $\eta^{\prime}$, where $\pi^0$ is defined as $i(\bar{u}\gamma_5u-\bar{d}\gamma_5d)/\sqrt{2}$. However, in our non-unitary calculation, all disconnected diagrams are neglected and correlators of the operators $i(\bar{u}\gamma_5u\pm\bar{d}\gamma_5d)$ will reveal independent but symmetrical ``states'' with the same mass.   Since only up quarks can appear in our intermediate state, we must use the interpolating operator $i\bar{u}\gamma_5u$ to create our $\pi^0$ state and can neglect the effects of the symmetrical state created by $\bar{d}\gamma_5d$.  Thus, in our calculation of $\langle\pi^0 | H_W | K^0\rangle$, we use $\pi^0=i\bar{u}\gamma_5u$ (with no $1/\sqrt{2}$ factor) and only include the pair of contractions shown Fig.~\ref{fig:ktopi}.

\section{Short distance contribution}
\label{sec:short}
In this section, we discuss the short distance contribution to our calculation of $\Delta M_K$ in detail.   We begin by discussing results without a charm quark and their dependence on a short distance, position-space cutoff.  We then introduce a charm quark and examine the resulting GIM cancellation.  

All the results presented in this section are for integrated correlators composed of the operator combination $Q_1 \cdot Q_1$, {\it i.e.} both four quark operators are $Q_1$ operators.  (This case is presented for illustration since it is for this combination of operators that we have data which includes a short distance, position-space cutoff.)  The results are the average of $600$ configurations separated by $10$ time units, with valence quark masses $m_l=0.01$ and $m_s=0.032$. The resulting pion and kaon masses are $m_{\pi}=0.2431(8)$ and $m_K=0.3252(7)$ respectively. The $\pi^0$ state is the only intermediate state lying below the kaon mass for these kinematics.

\subsection{Quadratic divergence at short distance}
\begin{figure}[!htp]
\begin{tabular}{cc}
\includegraphics[width=0.5\textwidth]{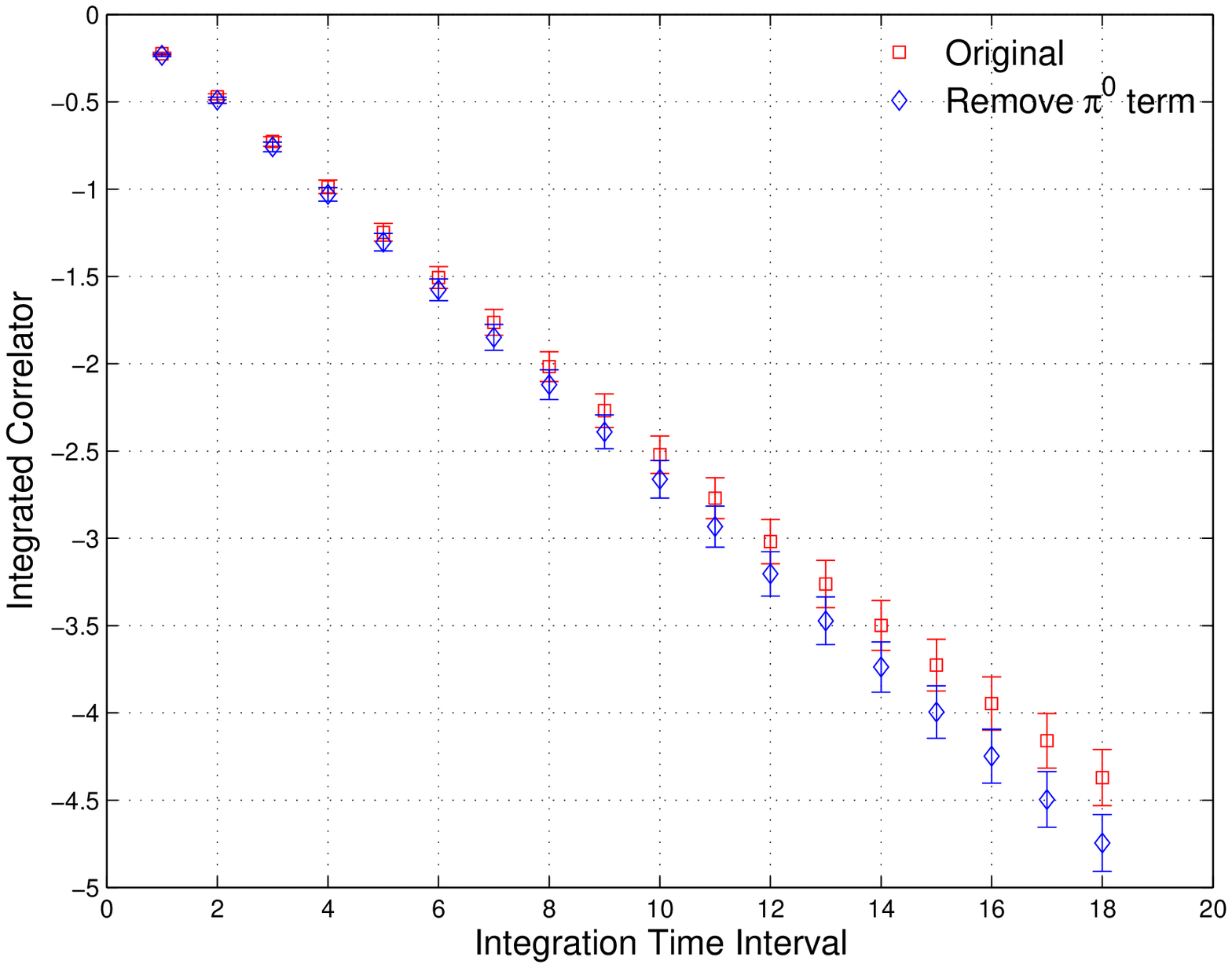} & \includegraphics[width=0.5\textwidth]{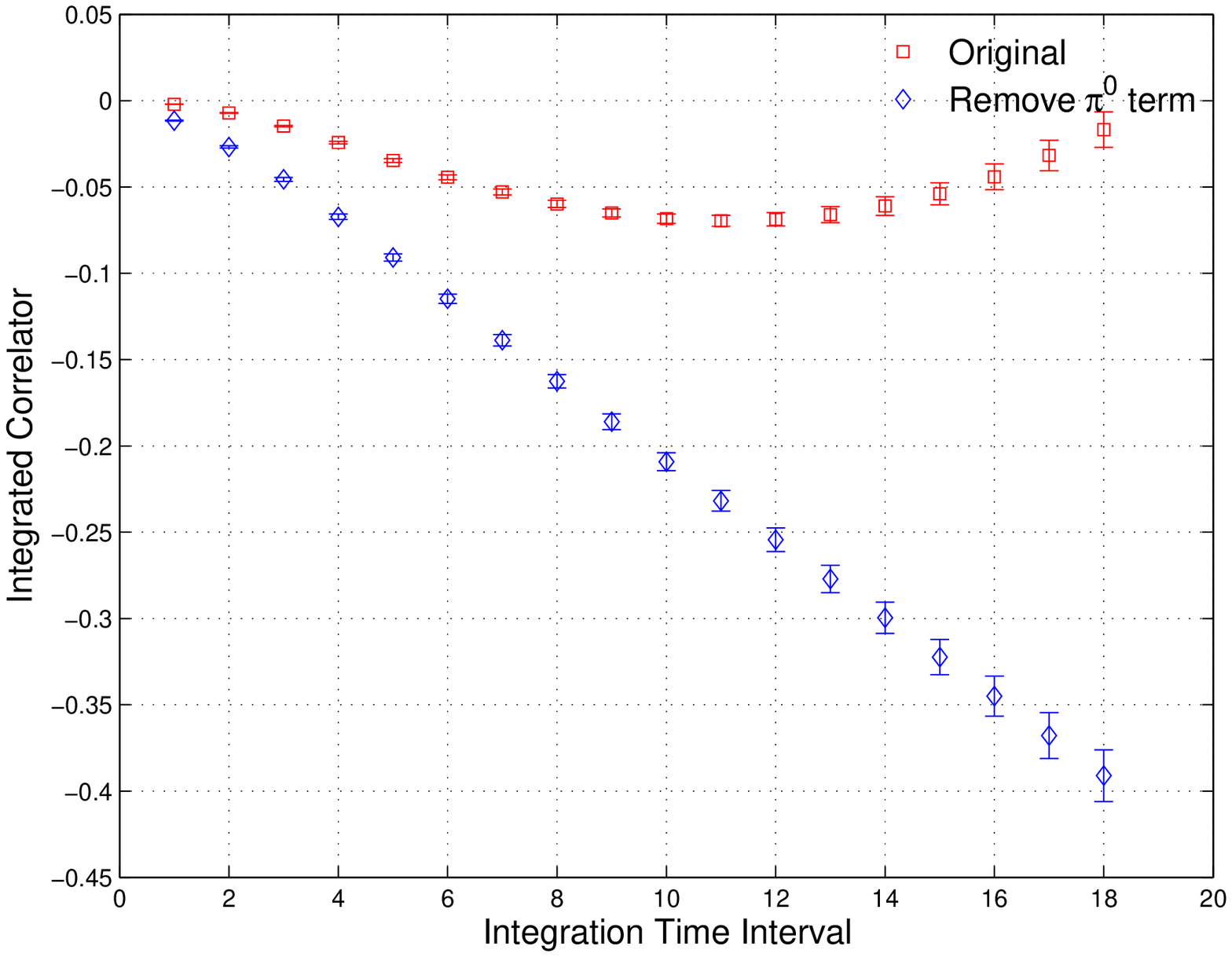}\\
(a) & (b)
\end{tabular}
\caption{The integrated correlator as a function of integration time interval $T$. (a) The original result without any artificial position-space cutoff; (b) The result with a cutoff radius of 5.  The red squares and blue diamonds are the results before and after the subtraction of the exponentially growing $\pi^0$ term, respectively. For both plots we include only the operator combination $Q_1 \cdot Q_1$.}
\label{fig:cutoff}
\end{figure}

In Eq.~\eqref{eq:integration_result}, we can see that the integrated correlator depends only on the separation between $t_a$ and $t_b$ which we defined earlier as $T=t_b-t_a+1$, the number of discrete times lying in the interval $[t_a,t_b]$. For a given value of $T$, all $(t_a,t_a+T-1)$ pairs which lie in the range $[5,22]$ are possible choices of this integration interval. We calculate all of them and use the averaged result after normalization as the final definition of integrated correlator:
\begin{equation}
\overline{\mathscr{A}}(T;t_i,t_f) = \frac{1}{19-T}\frac{e^{M_K(t_f-t_i)}}{N_K^2}\sum_{t_a=5}^{23-T}\mathscr{A}(t_a,t_b=t_a+T-1;t_i,t_f).
\label{eq:avgintcorr}
\end{equation}
In the left panel of Fig.~\ref{fig:cutoff}, we plot the integrated correlator as a function of the integration time interval $T$. Here the valence charm quark is not included, so there is no GIM cancellation. There are two curves in this plot: the red squares correspond to the integrated correlator defined in Eq.~\eqref{eq:avgintcorr}, the blue diamonds represent the results after the exponentially growing $\pi^0$ term is removed. The $\pi^0$ contribution to the integrated correlator can be determined using Eq.~\eqref{eq:integration_result}, where the $\langle\pi^0|H_W|K^0\rangle$ matrix element is determined from a three point correlator calculation. Note that only the exponentially growing $\pi^0$ term and a constant term coming from the $\pi^0$ are removed; the $\pi^0$ contribution to the term proportional to $T$ is retained as required by Eq.~\eqref{eq:integration_result}. The left-hand plot suggests that the exponentially growing $\pi^0$ term is only a small part of the result. This can be explained as follows. The integrated correlator receives contributions from all possible intermediate states. The short distance part, which comes from heavy intermediate states, is expected to be power divergent. The $\pi^0$ contribution, which is long distance physics, contains no such divergence and is small compared to the divergent short distance part even though it is exponentially growing with $T$. 

\begin{figure}[!htp]
\centering
\includegraphics[width=0.7\textwidth]{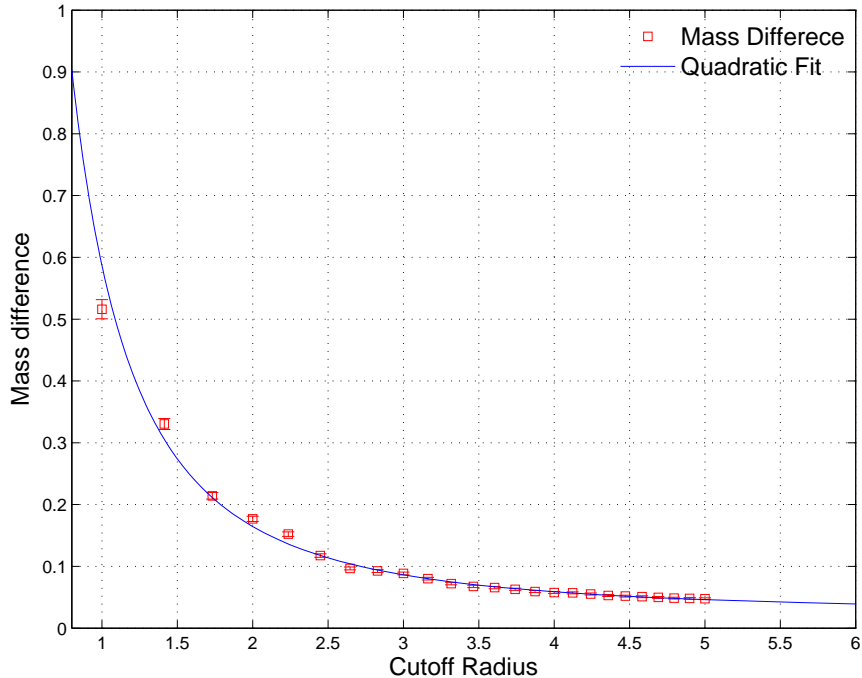}
\caption{The mass difference $\Delta M_K^{11}$ defined in Eq.~\eqref{eq:dmkq1q1} for different values of the cutoff radius $R$.  The blue curve is the two parameter fit to a $1/R^2$ behavior defined in Eq.~\eqref{eq:InverseR2_fit}}
\label{fig:fitcutoff}
\end{figure}

To investigate the divergent character of short distance part in detail, we introduce an artificial position-space cutoff radius $R$.  When we perform the double integration, we require the space-time separation between the positions of the two operators to be larger than or equal to this cutoff radius:
\begin{equation}
\sqrt{(t_2-t_1)^2+(\vec{x}_2-\vec{x}_1)^2}\geq R
\end{equation}
The right-hand plot in Fig.~\ref{fig:cutoff} presents the result with a cutoff radius of 5. Comparing this plot with the left plot, we can see that the amplitude of the integrated correlator is reduced by a factor of approximately $10$ and the exponentially growing $\pi^0$ term is now a very important part of the result which significantly changes the behavior of the correlator at long distance. All these observations suggest that the short distance contribution is substantially reduced after we impose the cutoff.  We can also plot the mass difference $\Delta M_K$ as a function of this cutoff radius $R$. The mass difference on a finite lattice is defined in Eq.~\eqref{eq:fvmassdiff}. However, we consider only the operator $Q_1$ here, so we define:
\begin{equation}
\Delta M_K^{11} =  2 \sum_{n\neq n_0} \frac{\langle\overline{K}^0|Q_1^{uu}|n\rangle\langle n|Q_1^{uu}|K^0\rangle}{M_K-E_n},
\label{eq:dmkq1q1}
\end{equation}
where the superscript $11$ means both operators are $Q_1$. This quantity is given by the slope of the coefficient of linear term in Eq.~\eqref{eq:integration_result} when $T$ is sufficiently large that the exponentially falling terms can be neglected.  We choose to fit the slope of the integrated correlator in the range $9 \le T \le 18$.  In Fig.~\ref{fig:fitcutoff} we show the dependence of $\Delta M_K^{11}$  on the cutoff radius $R$. The blue curve is a naive uncorrelated two parameter fit:
\begin{equation}
\Delta M_K^{11} (R)= \frac{b}{R^2} + c\,,
\label{eq:InverseR2_fit}
\end{equation}
where $b$ and $c$ are constants.  The fitting result shows a  convincing, power divergent short distance contribution.

\subsection{Valence charm quark and GIM cancellation}
\begin{figure}[!htp]
\centering
\includegraphics[width=0.7\textwidth]{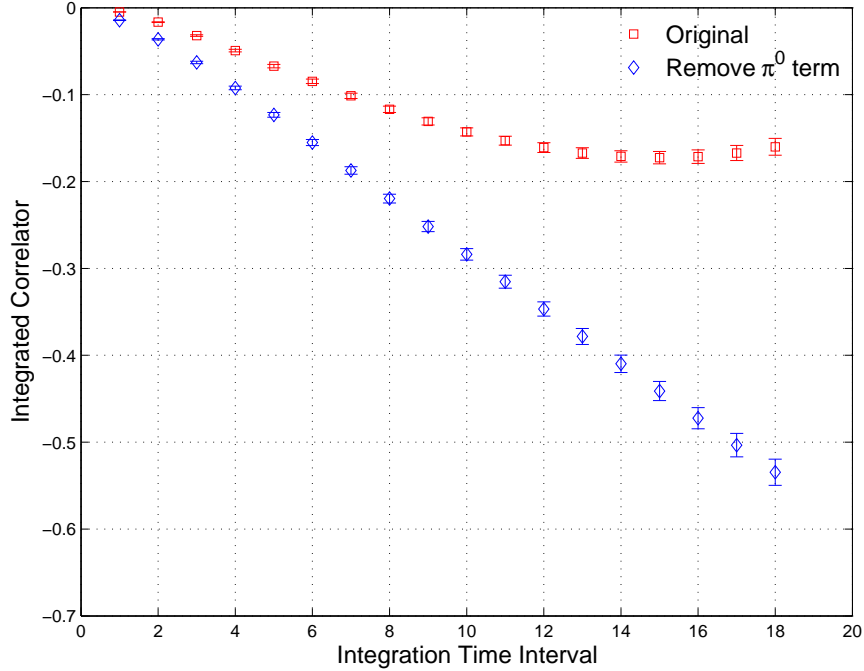}
\caption{The integrated correlator after GIM cancellation with a 0.863 GeV valence charm quark. The red squares and blue diamonds are the results before and after the subtraction of the exponentially increasing $\pi^0$ term respectively. We include only the $Q_1 \cdot Q_1$ operator combination in this plot.}
\label{fig:1000mevcharm}
\end{figure}

The short distance contribution in a lattice calculation is necessarily unphysical, principally determined by the lattice cutoff.  To control these short distance effects, we introduce a valence charm quark.  The resulting GIM mechanism will then substantially reduce the short distance contribution. The implementation of the GIM cancellation in this calculation is quite straightforward.   We simply replace the two internal up quark propagators in the contractions with the appropriate difference between up quark and charm quark propagators. We use six different valence charm quark masses which are given in Tab.~\ref{tab:charm_mass}. In Fig.~\ref{fig:1000mevcharm} we plot the integrated $Q_1\cdot Q_1$ correlator after GIM cancellation with a $863$ MeV valence charm quark mass.  We can compare this plot with those in Fig.~\ref{fig:cutoff}. The behavior of the integrated correlator after GIM cancellation is quite similar to the result after introducing the artificial position-space cutoff. The GIM cancellation reduces the amplitude by approximately a factor of $10$. Thus, as expected, the short distance contribution is substantially reduced by the GIM mechanism.

\begin{figure}[!htp]
\centering
\includegraphics[width=0.7\textwidth]{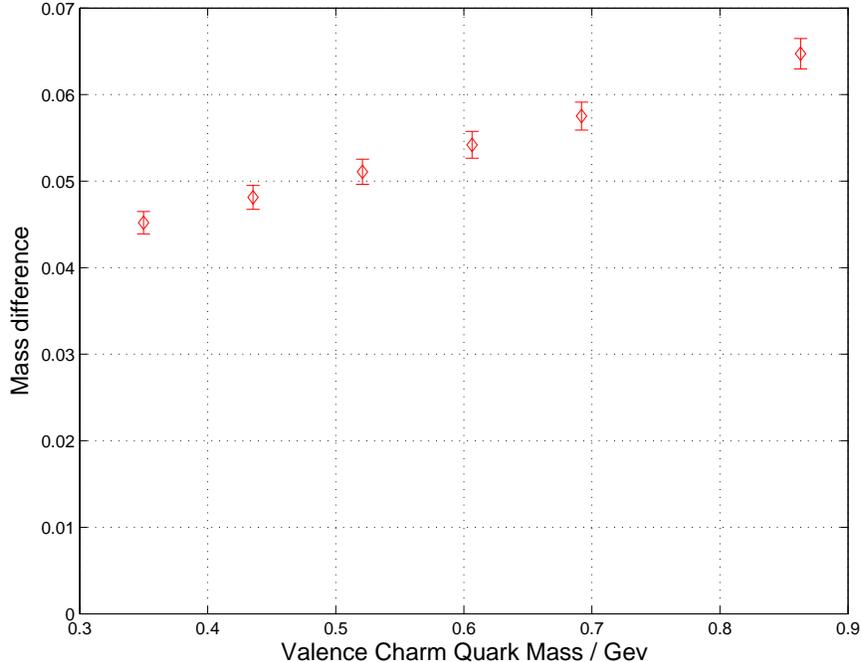}
\caption{The mass difference $\Delta M_K^{11}$, defined in Eq.~\eqref{eq:dmkq1q1} after GIM cancellation as a function of the valence charm quark mass.  }
\label{fig:dmvscharm}
\end{figure}

In Fig.~\ref{fig:dmvscharm}, we plot the mass difference for different valence charm masses. The definition of the mass difference $\Delta M_K^{11}$ is similar to that given in Eq.~\eqref{eq:dmkq1q1}, but the GIM cancellation is now included. The mass difference is obtained from the slope in $T$ of the integrated correlator using the fitting range $T\in[9,18]$. The values of $\Delta M_K^{11}$ are listed in the Tab.~\ref{tab:dmkq1q1mc}. The plot shows that the mass difference increases as the charm quark mass increases. This is expected since the cancellation between the up and charm quark propagators will be more complete for a lighter charm quark.

\begin{table}[!htp]
	\caption{The mass difference $\Delta M_K^{11}$, defined in Eq.~\eqref{eq:dmkq1q1} after GIM cancellation, evaluated for different charm quark masses.  These results were obtained  from 600 configurations, use a kaon mass of 563 MeV and are the matrix elements of bare lattice operators without Wilson coefficients or renormalization factors.}
	\begin{ruledtabular}
		\begin{tabular}{c|cccccc}
			$m_c$ (MeV) & 350 & 435 & 521 & 606 & 692 & 863 \\
			\hline
			$\Delta M^{11,GIM}_{K}$ & 0.0452(13) & 0.0481(14) & 0.0511(15) & 0.0542(15) & 0.0575(16) & 0.0647(18) \\
		\end{tabular}
	\end{ruledtabular}
	\label{tab:dmkq1q1mc}
\end{table}

\subsection{Short distance subtraction}
\label{subsec:short_distance_subtraction}
One might expect that the GIM cancellation would reduce the quadratic divergence present in a 2+1 flavor lattice calculation of $\Delta M_K$ to a milder logarithmic divergence leaving an unphysical, short distance artifact of the form $\mathrm{ln}(m_c a)$ reflecting a physical $\mathrm{ln}(m_c/M_W)$ short distance contribution, inaccessible to a lattice calculation.  However, because of the $V-A$ structure of the weak vertices in the standard model, the $u$ and $c$ quark masses appear only quadratically in the internal quark lines so that the difference of those propagators introduces a factor of $m_c^2-m_u^2$ for each of the two internal quark lines, reducing the overall degree of divergence by four units.  It should be noted that in our calculation this $(V-A)$-induced cancellation reduces the result by approximately an order of magnitude, a reduction that depends critically upon the use of chiral lattice fermions. Thus, the GIM cancellation is complete, leaving only convergent integrals in a theory built from the effective four-quark operator $H_W$, with all ``short distance'' contributions coming from distances on the order of $1/m_c$.  Thus, if potential lattice artifacts associated with the large value of $m_c a$ can be neglected and the role of the omitted charmed sea quarks is small, then the present calculation (or one in which diagrams with all possible topologies have been included) will capture all important aspects of $\Delta M_K$.  If $m_c a$ is sufficiently large that it cannot be neglected or charmed sea quarks need to be included, then these difficulties can be systematically addressed in a later, 2+1+1 flavor calculation at a smaller lattice spacing.  In contrast, the corresponding continuum calculation cannot properly treat the long distance region where the two weak operators are separated by a distance of a few tenths of a Fermi and even the short distance part with momenta on the order of $m_c$ may suffer from poor convergence of the perturbation expansion.

In a generic calculation of a second order weak quantity using an effective four-Fermi form for each of the weak operators, divergences will be encountered because of the singular behavior that results as the two operators approach each other in space time.  In the most difficult case, this divergence will be quadratic and two subtractions will be required before the calculation using the effective theory becomes well defined.  If the GIM subtraction reduces the divergence to one which is only logarithmic, then the needed single subtraction can be carried out explicitly using a Rome-Southampton style~\cite{Martinelli:1995ty}, RI/MOM subtraction.  Indeed, in an earlier version of this work~\cite{Yu:2011gk}, we were unaware of the absence of a $\mathrm{ln}(m_c a)$ term in the lattice calculation with valence charm and we performed such an explicit RI/MOM subtraction to remove it.  This subtraction was determined from the bi-local operator formed from the product of the two effective weak Hamiltonians by requiring that a particular, spin- and color-projected, four external quark, Landau gauge-fixed vertex vanish at a specific kinematic point.  This subtraction was chosen in such a way that it could be both easily applied in the lattice calculation and also computed perturbatively so that the correct subtraction term could be restored to the lattice result.  In fact, the subtraction term reported in Ref.~\cite{Yu:2011gk}, performed at a scale $\mu=2$ GeV, was zero within errors, consistent with the absence of a true short distance, $\mathrm{ln}(m_c/M_W)$, contribution to  $\Delta M_K$.

\section{Long distance contribution}
\label{sec:long}
In this section we will examine the long distance contribution to our calculation of $\Delta M_K$ in detail.  As we have discussed in Sec.~\ref{sec:amplitude}, the intermediate states lying below the kaon mass will contribute terms which grow exponentially as the time interval $T$, over which the bi-local, second order weak interaction operators are integrated, is increased.  These terms do not contribute to the physical mass difference $\Delta M_K$ and must be identified and removed.   For physical quark masses such states include the vacuum, $\pi^0$, $\pi$-$\pi$ and three $\pi$ states. There is no vacuum state contribution in this work and for our kinematics the kaon mass is below the three-pion threshold.  Thus, in the present calculation we are most interested in the $\pi^0$ and $\pi$-$\pi$ intermediate states. The different parity of these two states allows us to study their contributions separately.  Each left-left, $\Delta S=1$ four quark operator can be separated into parity conserving and violating parts:
\begin{equation}
LL = (VV+AA) - (VA+AV).
\end{equation}
The product of the two left-left operators can then be written as the sum of four terms:
\begin{equation}\label{eq:llxll}
\begin{split}
LL\otimes LL&=(VV+AA)\otimes(VV+AA) + (VA+AV)\otimes(VA+AV)\\
&-(VV+AA)\otimes(VA+AV) - (VA+AV)\otimes(VV+AA).
\end{split}
\end{equation}
The third and fourth terms of Eq.~\eqref{eq:llxll} change the parity and hence cannot contribute to the matrix element between $K^0$ and $\overline{K^0}$ states.  In the first term on the right-hand side of Eq.~\eqref{eq:llxll} both operators are parity conserving, which implies that the intermediate state must have odd parity.  In the second term, both operators are parity violating, so the intermediate states have even parity. We can distinguish these two contributions and investigate the $\pi^0$ (parity odd) and $\pi$-$\pi$ (parity even) intermediate states separately.

The integrated correlator receives contributions from both short and long distances. Therefore, in this section we examine the unintegrated correlators in Eq.~\eqref{eq:unintegrated_correlator}, where we can explicitly study the case of large time separation between the two $\Delta S=1$ operators. The results presented in this section are for an average of 800 configurations separated by 10 time units, with a valence light quark mass $m_l=0.01$ which corresponds to a pion mass $m_\pi=0.2431(8)$ and eight valence strange quark masses whose values together with the corresponding kaon masses are given in Tab.~\ref{tab:kaon_mass}. 

\subsection{Parity-odd channel}
For this case, corresponding to the contribution of the first term in Eq.~\eqref{eq:llxll},  both operators are parity conserving which implies that all intermediate states have odd parity.    As can be seen from Eq.~\eqref{eq:unintcorr}, in the limit of large time separation $|t_2-t_1|$ the contribution from heavier states will decrease exponentially and only the lightest states will survive.  For the parity-odd case this lightest state is the $\pi^0$ so that the unintegrated correlator becomes:
\begin{equation}
G(t_f,t_2,t_1,t_i) = N_K^2e^{-M_K(t_f-t_i)}\langle\overline{K^0}|H_W|\pi^0\rangle\langle \pi^0|H_W|K^0\rangle e^{-(M_{\pi}-M_K)|t_2-t_1|}.
\label{eq:unintegrated_pi0}
\end{equation}
The unintegrated correlator only depends on the time separation $T_H=t_2-t_1$ at given $t_i$ and $t_f$.  For a given value of $T_H$, all $(t_1,t_1+T_H)$ pairs in the range $[5,22]$ are possible choices. We compute all of them, take their average and remove the normalization factor $N_K^2$.  The result is the unintegrated correlator $\overline{G}(T_H;t_f,t_i)$:
\begin{equation}
\overline{G}(T_H;t_f,t_i)=\frac{1}{t_f-t_i-9-T_H}\frac{e^{M_K(t_f-t_i)}}{N^2_K}\sum_{t_1=t_i+5}^{t_f-5-T_H}G(t_f,t_2=t_1+T,t_1,t_i)\,,
\label{eq:unintegrated_def}
\end{equation}
where we have adopted the order $t_2>t_1$ and imposed the restriction $t_1 \ge t_i+5$ and
$t_f-5 \ge t_2$.

We also compute the three point correlator needed to extract the matrix element $\langle\pi^0 | Q_i | K^0\rangle$. We can then compare our lattice result for the unintegrated correlator given in Eq.~\eqref{eq:unintegrated_def} for large $T_H$ with the contribution of a single $\pi^0$ shown in Eq.~\eqref{eq:unintegrated_pi0}. The single-pion matrix elements are given in Tab.~\ref{tab:ktopi} for the set of 8 kaon masses. As we have explained in Sec.~\ref{sec:simulation}, we use $\pi^0=i\bar{u}\gamma_5u$ and only compute the diagrams shown in Fig.~\ref{fig:ktopi}. 

\begin{table}[!htp]
\caption{Results for single-pion matrix elements, $\langle\pi^0 | Q_i | K^0 \rangle$, at various kaon masses. We use $\pi^0=i\bar{u}\gamma_5u$ and only include the diagrams in Fig.~\ref{fig:ktopi}.}
\label{tab:ktopi}
\begin{ruledtabular}
\begin{tabular}{ccc}
$M_K$ & $\langle \pi^0 | Q_1^{uu} | K^0\rangle$ & $\langle \pi^0 | Q_2^{uu} | K^0\rangle$\\
\hline
0.2431(8) & 0.02107(29) & -0.00779(26) \\
0.3252(7) & 0.02729(30) & -0.00954(23) \\   
0.4087(7) & 0.03300(33) & -0.01067(22) \\
0.4480(7) & 0.03550(35) & -0.01103(22) \\
0.4848(8) & 0.03773(36) & -0.01128(22)  \\
0.5307(8) & 0.04037(39) & -0.01149(22) \\
0.5738(8) & 0.04271(42) & -0.01160(23)  \\
0.6721(10) & 0.04753(49) & -0.01156(25) \\
\end{tabular}
\end{ruledtabular}
\end{table}

\begin{figure}[!htp]
\begin{tabular}{cc}
\includegraphics[width=0.5\textwidth]{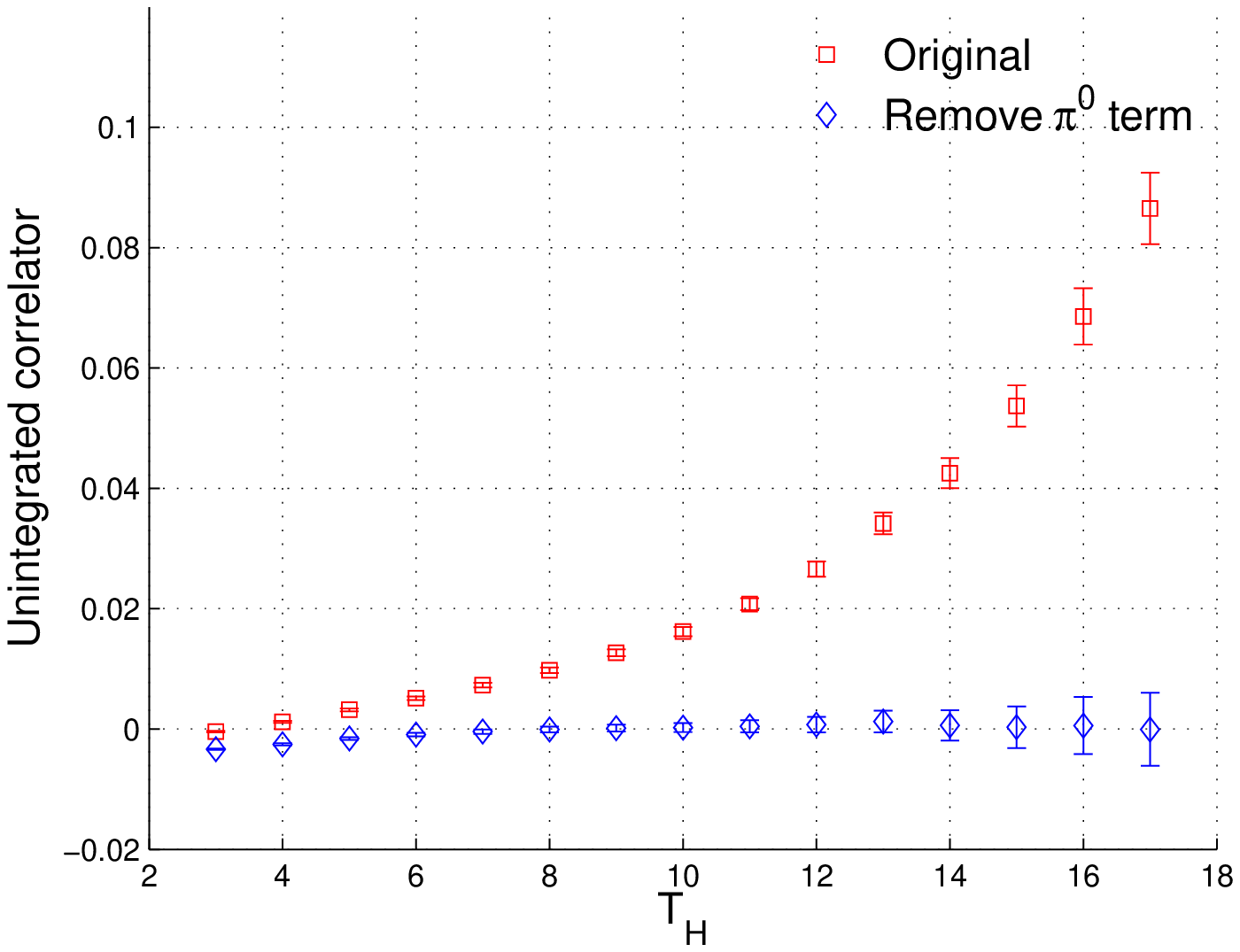} &
\includegraphics[width=0.5\textwidth]{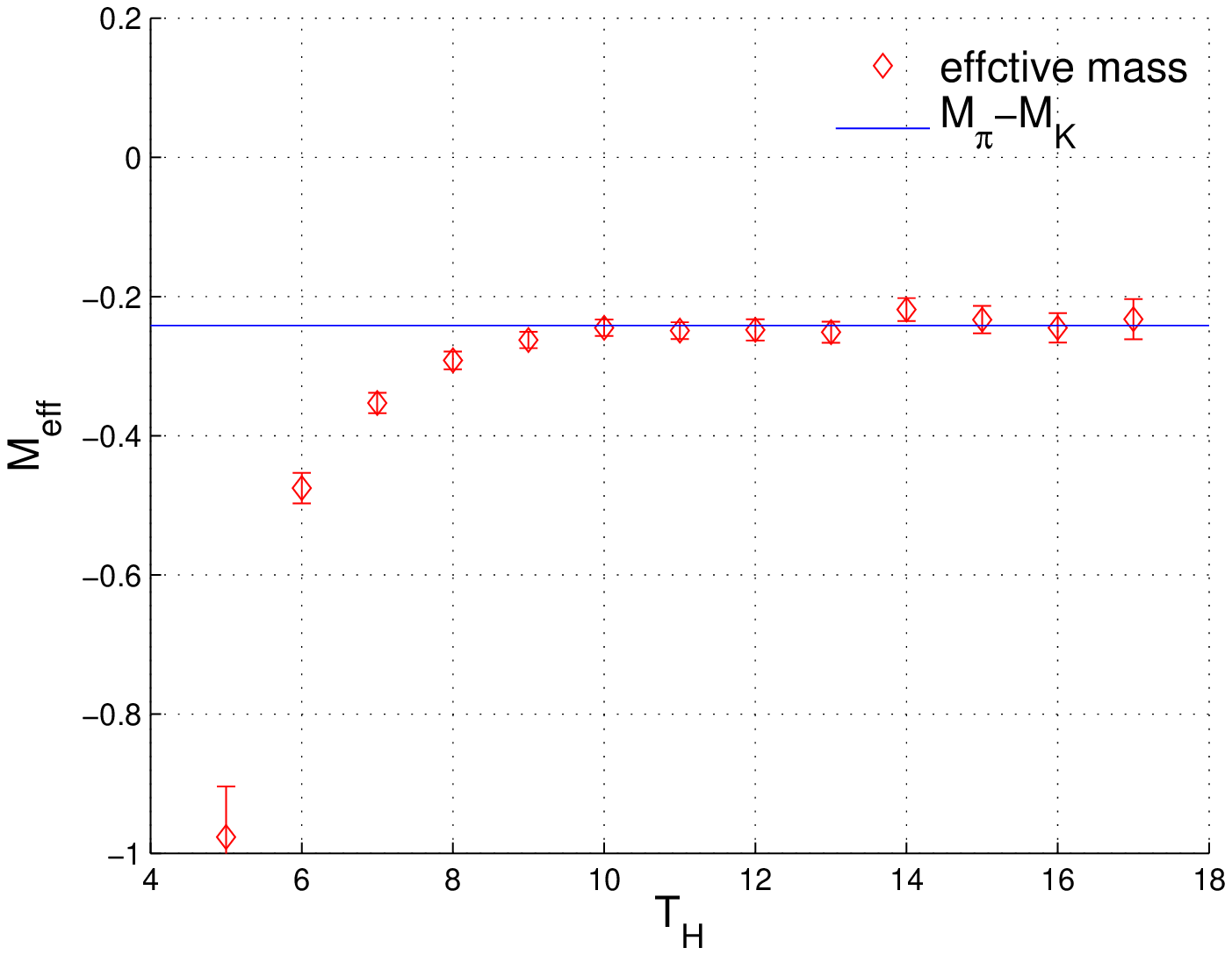} 
\end{tabular}
\caption{A plot of unintegrated correlator $\overline{G}$ and resulting effective mass for the combination of operators  $Q_1 \cdot Q_1$ and a kaon mass $M_K=0.4848(8)$.  Only the product of the parity even components of the two operators is included. In the left-hand plot, the red diamonds and blue squares show the result before and after subtraction of the $\pi^0$ term. In the right-hand plot, the red diamonds are effective masses obtained from the unintegrated correlator. The blue horizontal line shows the ``exact'' value of $M_{\pi}-M_K$ obtained from the two point correlator calculation.}
\label{fig:q1q1unintcorr}
\end{figure}

\begin{figure}[!htp]
\begin{tabular}{cc}
\includegraphics[width=0.5\textwidth]{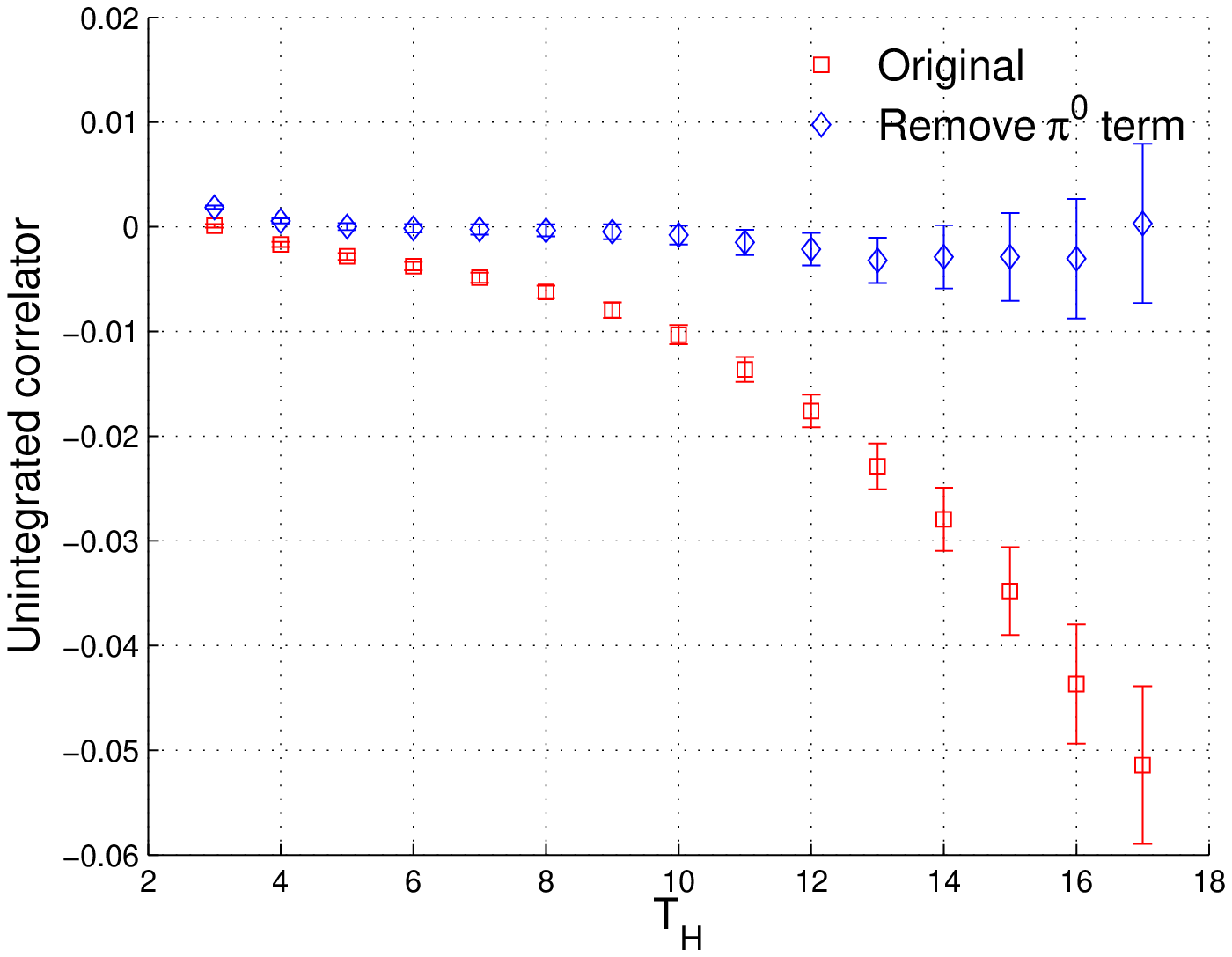} &
\includegraphics[width=0.5\textwidth]{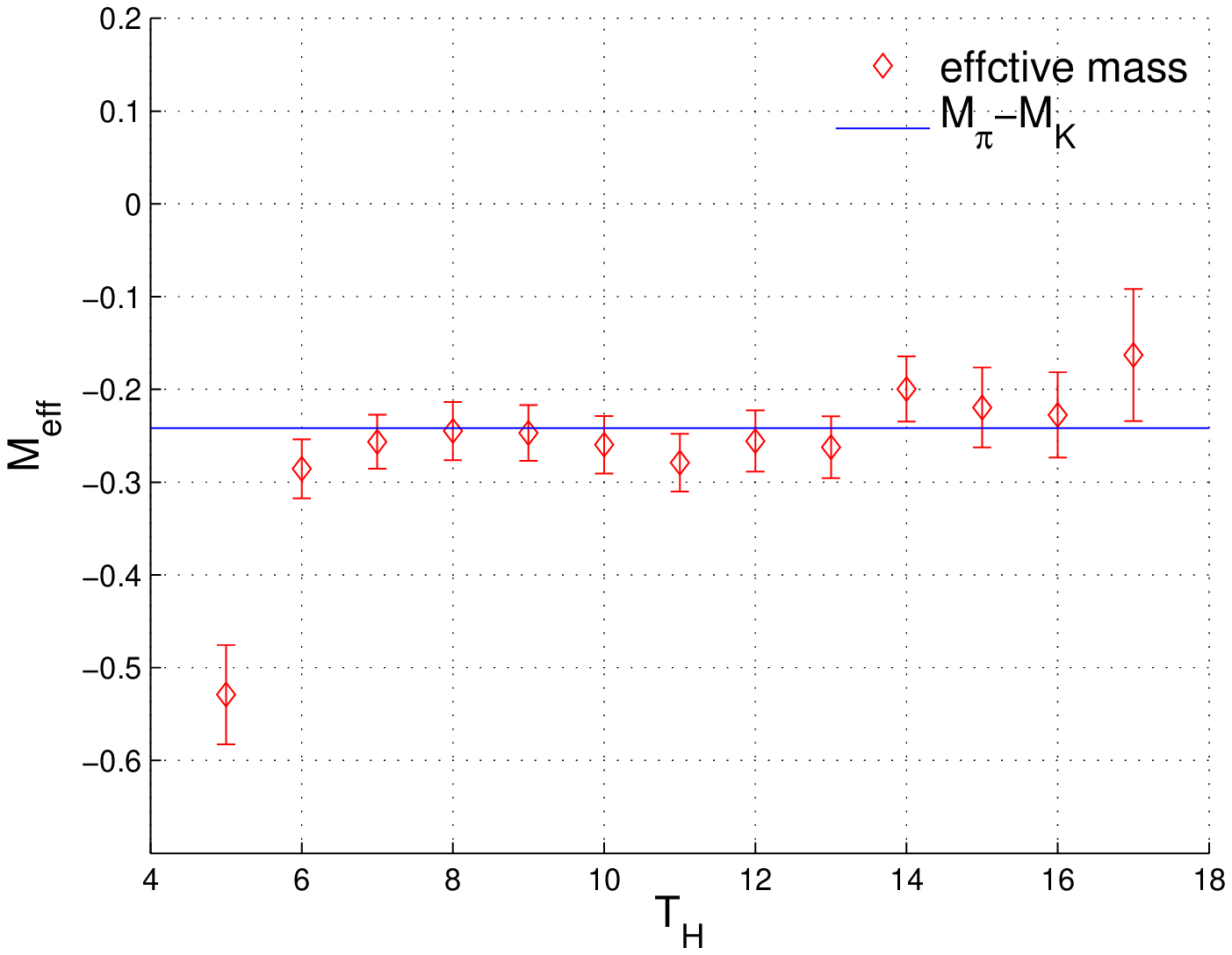} 
\end{tabular}
\caption{Plots of the unintegrated correlato r$\overline{G}$ and corresponding effective mass for the operator combination $Q_1 \cdot Q_2$ at a kaon mass $M_K=0.4848(8)$.  Only the product of the parity even components of the two operators is included.}
\label{fig:q1q2unintcorr}
\end{figure}

\begin{figure}[!htp]
\begin{tabular}{cc}
\includegraphics[width=0.5\textwidth]{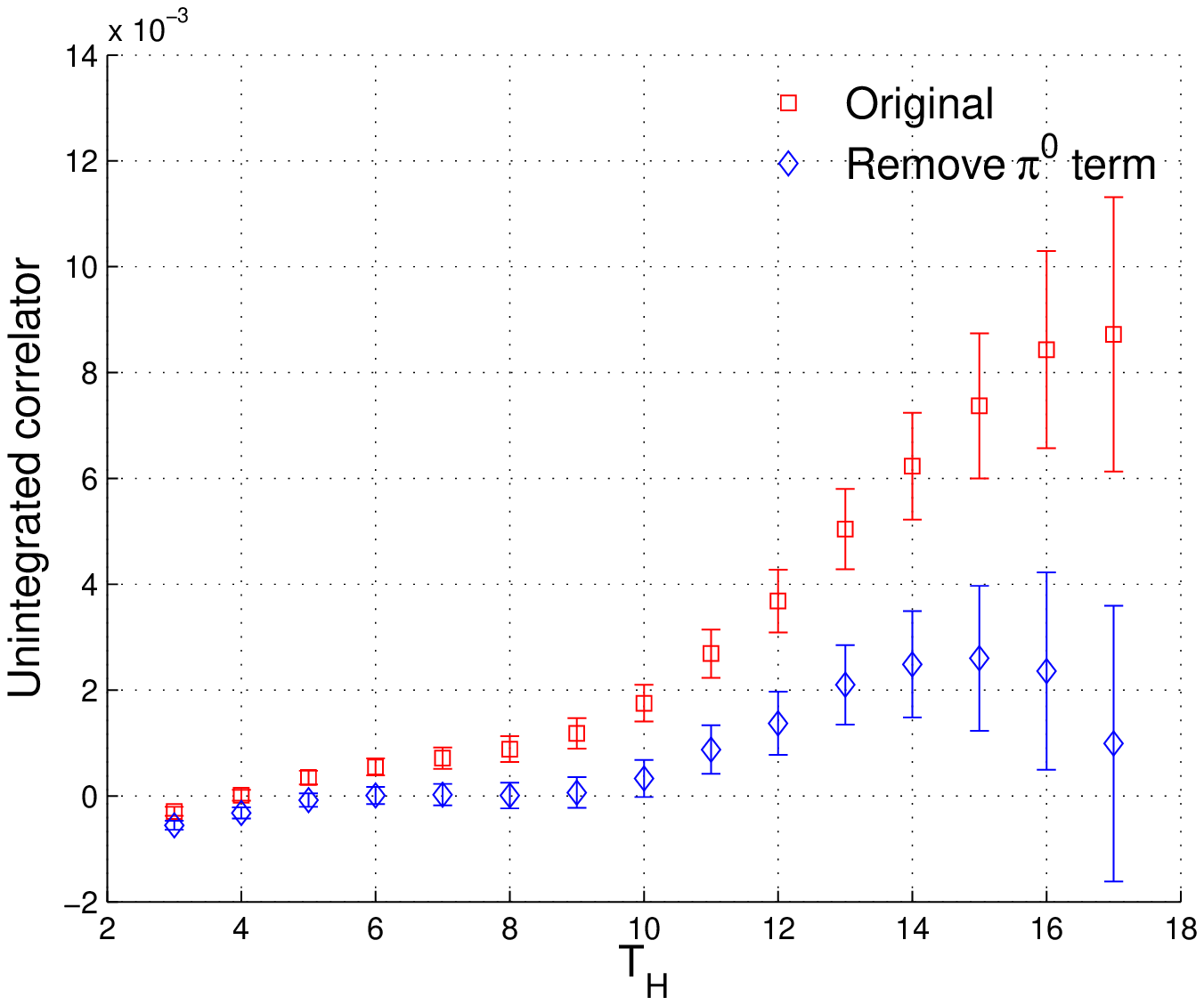} &
\includegraphics[width=0.5\textwidth]{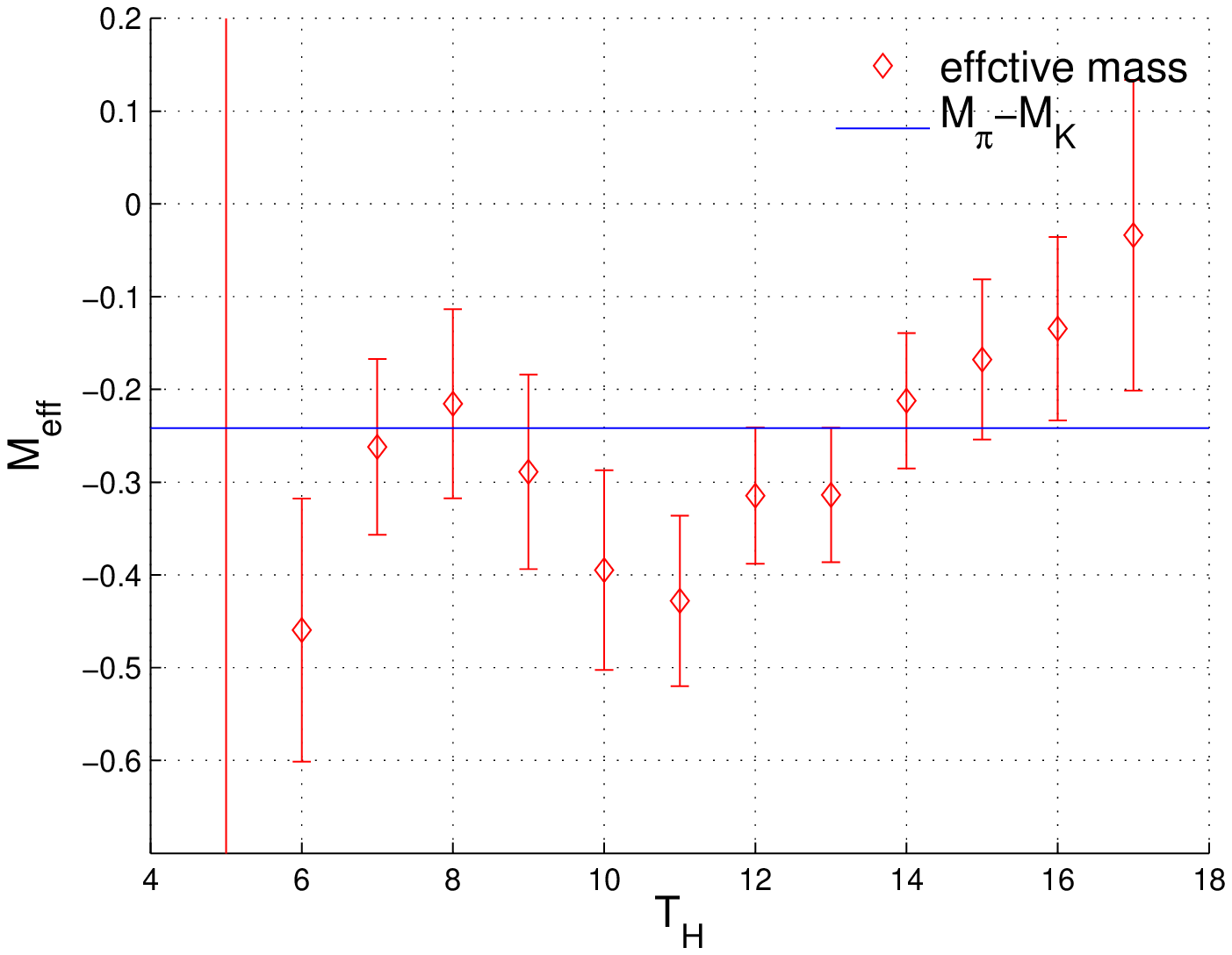} 
\end{tabular}
\caption{Plots of the unintegrated correlator and corresponding effective mass for the operator combination $Q_2 \cdot Q_2$ at a kaon mass $M_K=0.4848(8)$.   Only the product of the parity even components of the two operators is included.}
\label{fig:q2q2unintcorr}
\end{figure}

In Figs.~\ref{fig:q1q1unintcorr}-\ref{fig:q2q2unintcorr}, we plot the unintegrated correlators and resulting effective masses for the kaon mass $M_K=0.4848(8)$. The three figures correspond to the different operator combinations: $Q_1 \cdot Q_1$, $Q_1 \cdot Q_2$ and $Q_2 \cdot Q_2$, respectively. In the plots of the unintegrated correlators we show both original results and the results after the subtraction of the $\pi^0$ contribution. This subtraction is done using the numerical results in Tab.~\ref{tab:ktopi}. Since only the $\pi^0$ term should be present for large time separations, we expect that the results after subtraction should be consistent with zero for large $T_H$.  In the effective mass plots, we calculate the effective mass $M_X-M_K$ from the unintegrated correlators, here $M_X$ is the mass of the intermediate state. For this parity conserving case, the lightest state is the pion. The ``exact'' $M_{\pi}-M_K$ mass obtained from two point correlator calculation is shown in the plots as a blue horizontal line which agrees well with the computed effective mass.  Although all three figures show the expected behavior, we find that the statistical errors seen for the different operator combinations are quite different. The operator combination $Q_1 \cdot Q_1$ has the smallest errors while $Q_2 \cdot Q_2$ has the largest. 

\begin{figure}[!htp]
\centering
\includegraphics[width=0.7\textwidth]{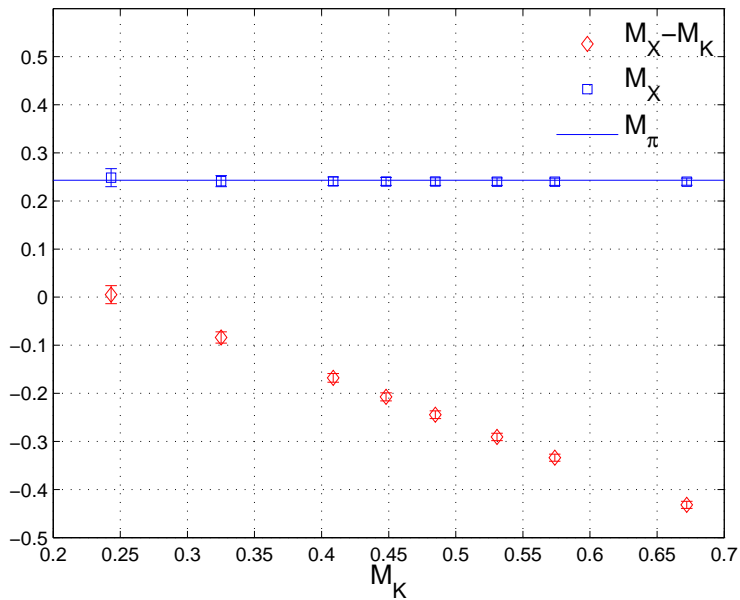}
\caption{Intermediate state masses determined for all eight kaon masses from the unintegrated correlators of the parity even portion of the operators $Q_1 \cdot Q_1$. The red diamonds are the fitting results and should correspond to the difference $M_X-M_K$.   The blue squares are obtained from the results for $M_X-M_K$ by adding the result for $M_K$ obtained from the two-point kaon correlators.  The blue horizontal line is the ``exact'' pion mass given by the two point function calculation.}
\label{fig:pionmass}
\end{figure}

In Fig.~\ref{fig:pionmass}, we plot the intermediate state masses obtained from unintegrated correlators at eight different kaon masses for the $Q_1 \cdot Q_1$ case. The mass $M_X-M_K$ is obtained from a two parameter exponential fit and compared with the difference of  $M_K$ and $M_{\pi}$ obtained directly from the two point correlators.  The intermediate state mass agrees very well with the single pion mass for all choices of kaon mass.

\subsection{Parity even channel}
In this section, we examine the case where parity violating operators appear at both vertices. This requires that the intermediate states have even parity. The long distance behavior is expected to be dominated by the two-pion intermediate state, which is the lightest parity-even state.

\begin{figure}[!htp]
\begin{tabular}{cc}
$Q_1 \cdot Q_1$ & $Q_1 \cdot Q_2$ \\
\includegraphics[width=0.5\textwidth]{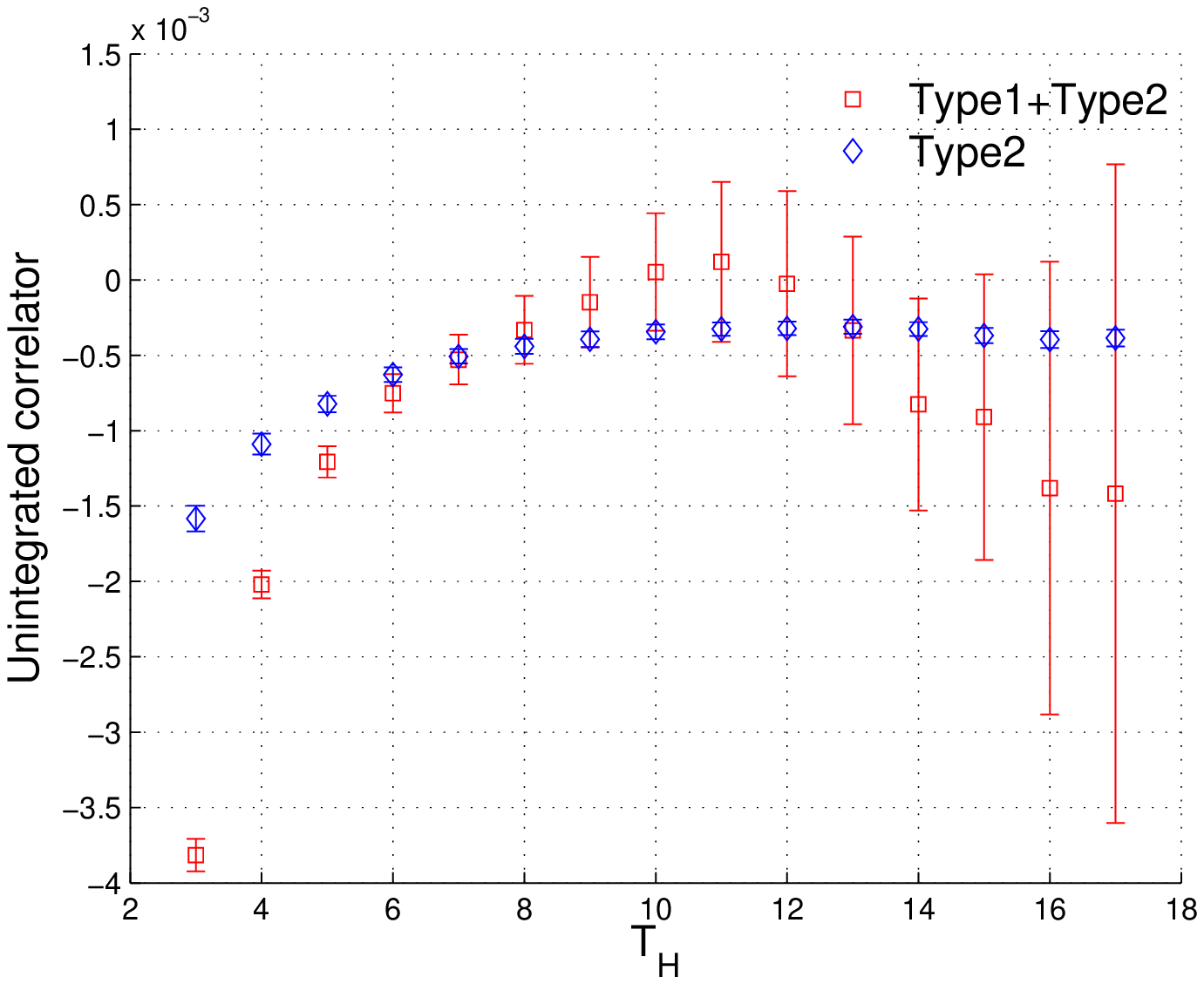}&
\includegraphics[width=0.5\textwidth]{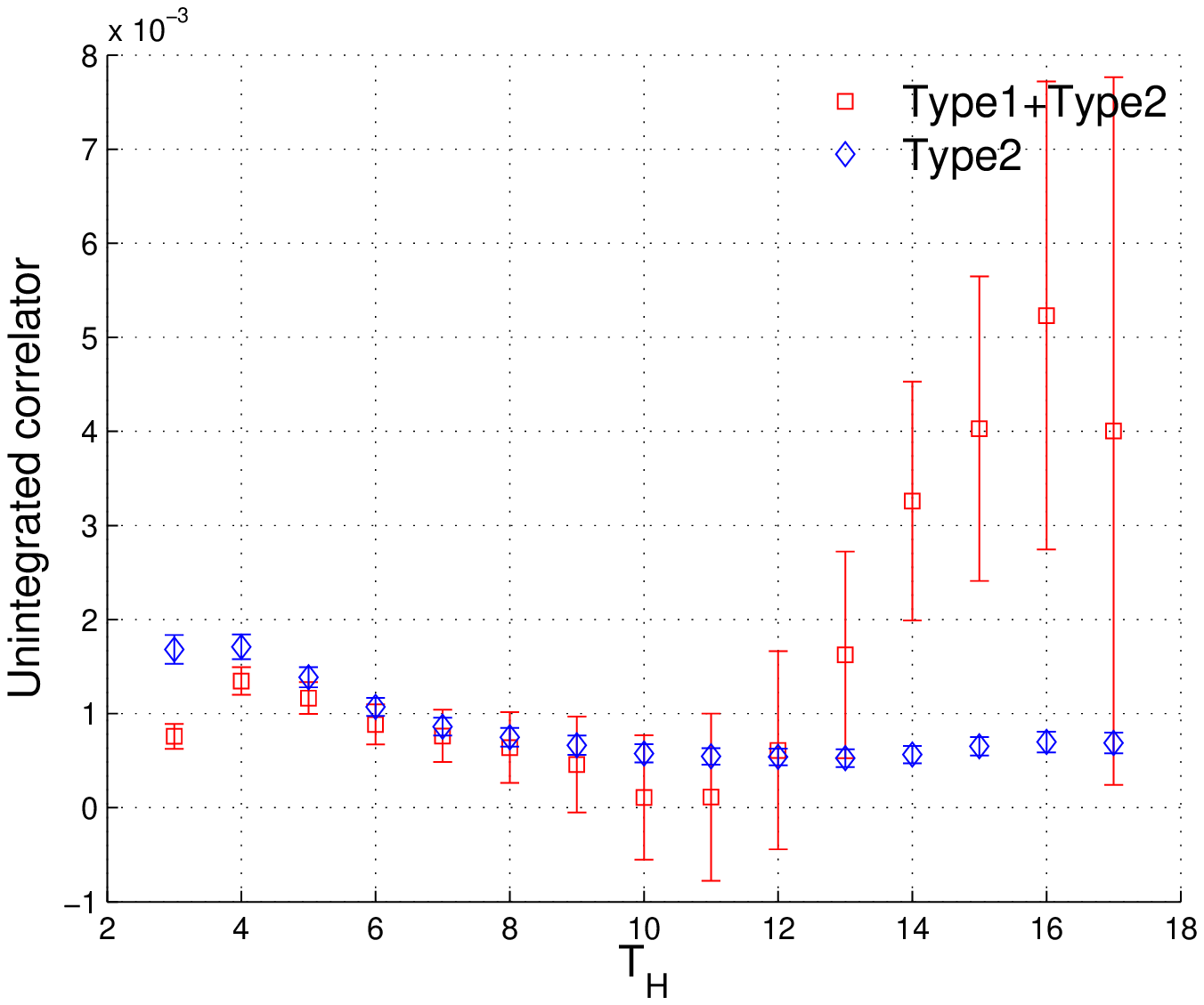}\\
$Q_2 \cdot Q_2$ & Intermediate state mass\\
\includegraphics[width=0.5\textwidth]{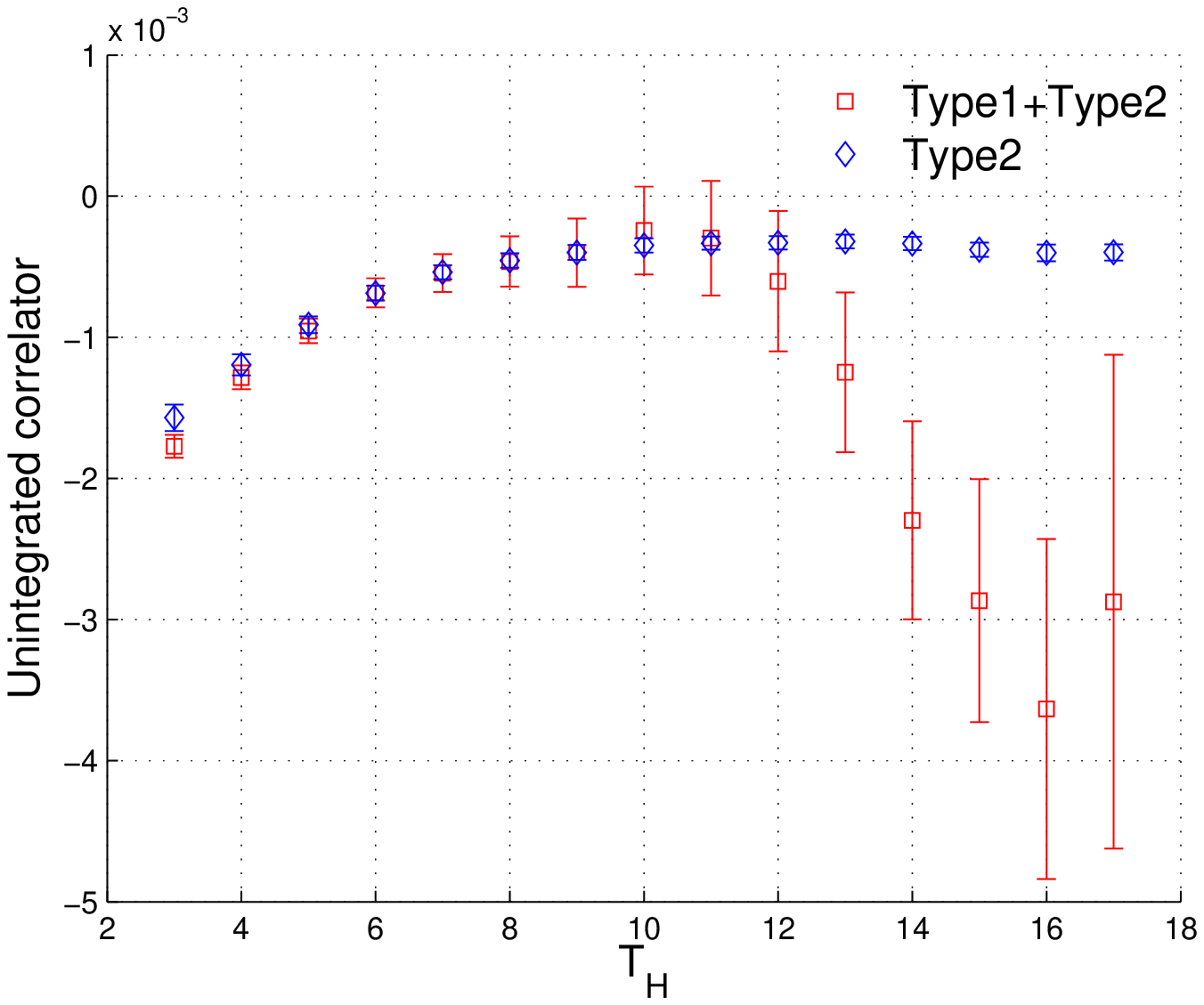}&
\includegraphics[width=0.5\textwidth]{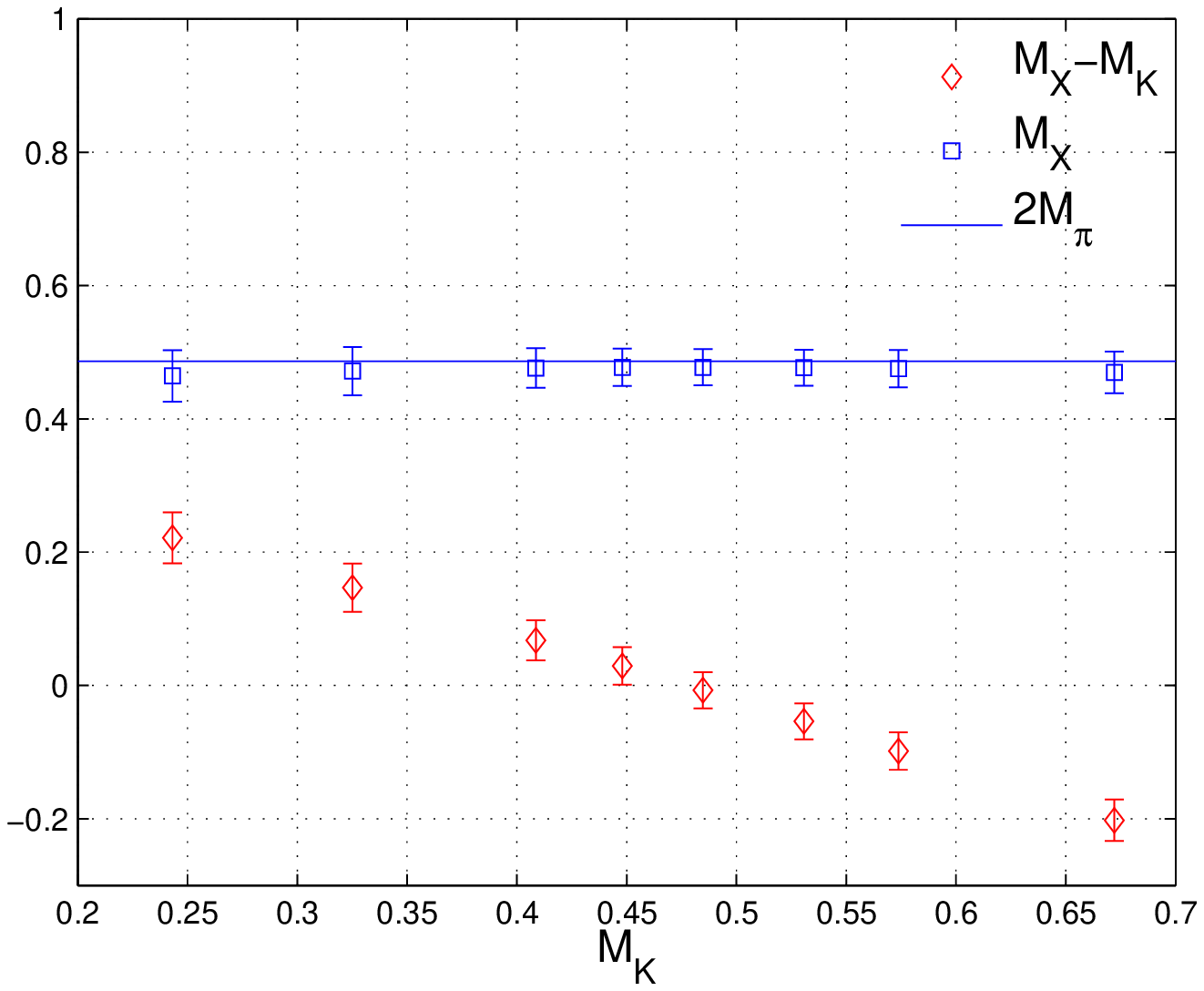}
\end{tabular}
\caption{The unintegrated correlators for different, parity-odd operator combinations at a kaon mass $M_K=0.4848(8)$. We plot both the full results and the results from type 2 diagrams only. The last plot is the fitted intermediate state mass (red diamonds) and the sum of that mass and the kaon mass (blue squares) for all choices of kaon masses. The results shown in this last plot are obtained from fitting the type 2 diagrams alone.  Because the type 2 diagrams shown in the last plot are only a subset of those needed for a physical calculation, we do not expect the effective mass shown in this last plot to be either the $I=0$ or $I=2$ finite volume $\pi-\pi$ energy.  We view the agreement with $2m_\pi$ as coincidental.}
\label{fig:pvunintcorr}
\end{figure}

In Fig.~\ref{fig:pvunintcorr}, we present the unintegrated correlators for the three different products of parity violating operators evaluated at a kaon mass $M_K=0.4848(4)$. This kaon mass is very close to the energy of two pions at rest, so we expect to get a plateau at large time separation $T_H$.  However, our results are extremely noisy at long distance and we are not able to identify such a plateau. This large noise can be explained as follows.  Although the signal should come from two-pion intermediate states, we will also have noise, whose size can be estimated from the square of the Green's functions being studied.  In this squared Green's function the source and sink are composed of the product of two parity-violating operators and two kaon sources and sinks.  Such a Green's function will receive a contribution from a two-pion intermediate state.  The noise will fall with increasing separation $|t_2-t_1|$ between the weak operators as the square root of this Green's function, implying that this noise will behave as $e^{- |t_2-t_1| m_\pi}$, dominating the two-pion signal which falls more rapidly as $e^{- |t_2-t_1| 2m_\pi}$.  Thus, the signal to noise ratio will fall exponentially for large time separation. The situation here is very similar to what is found for disconnected diagrams.  This argument is consistent with our observation that most of the noise comes from type 1 diagrams, shown in Fig.~\ref{fig:type1}, because the topology of type 2 diagrams does not allow a single-pion contribution to their noise.

This argument is confirmed by plotting the results from type 2 contractions only.  If we analyze the type 2 diagrams alone, and fit the resulting intermediate state masses the results agree with the two-pion mass very well, as seen in the lower right panel of Fig.~\ref{fig:pvunintcorr}.

\section{The $K_L-K_S$ mass difference}
\label{sec:mass}
In order to use the numerical results presented in the previous sections to calculate the physical  $K_L-K_S$ mass difference, we must connect our four-quark lattice operators with the physical $\Delta S=1$ effective weak Hamiltonian $H_W$ given in Eq.~\eqref{eq:H_W}.  Thus, we must determine the Wilson coefficients and normalize the lattice operators in the same scheme in which the Wilson coefficients are computed.  We will follow the same procedure used in previous work~\cite{Blum:2011pu,Blum:2001xb}.  The Wilson coefficients are evaluated in the $\overline{\rm MS}$ NDR scheme using the formulae in Ref.~\cite{Buchalla:1995vs}.  The lattice operators are first non-perturbatively normalized in the RI/MOM scheme and then converted into the $\overline{\rm MS}$ NDR scheme using formulae provided by Lehner and Sturm, extending to our four-flavor case the results given in Ref.~\cite{Lehner:2011fz}.

We will consider only  the current-current operators defined in Eq.~\eqref{eq:operator} which enter the present calculation.  In particular, we are only interested in the operators:
\begin{equation}
\begin{split}
\widetilde{Q}_1&=(\bar{s}_iu_j)_{V-A}(\bar{u}_jd_i)_{V-A}-(\bar{s}_ic_j)_{V-A}(\bar{c}_jd_i)_{V-A}\\
\widetilde{Q}_2&=(\bar{s}_iu_i)_{V-A}(\bar{u}_jd_j)_{V-A}-(\bar{s}_ic_i)_{V-A}(\bar{c}_jd_j)_{V-A}\\
Q_1^{cu}&=(\bar{s}_iu_j)_{V-A}(\bar{c}_jd_i)_{V-A}\\
Q_2^{cu}&=(\bar{s}_iu_i)_{V-A}(\bar{c}_jd_j)_{V-A}\\
Q_1^{uc}&=(\bar{s}_ic_j)_{V-A}(\bar{u}_jd_i)_{V-A}\\
Q_2^{uc}&=(\bar{s}_ic_i)_{V-A}(\bar{u}_jd_j)_{V-A} .
\end{split}
\label{eq:basis1}
\end{equation}
These six operators can be categorized into three groups according to their different flavor structure.  Operator mixing will take place within each group.  The discussion of operator mixing is simplified if we define a second, equivalent basis:
\begin{equation}
\begin{split}
Q_+^X &=  Q_1^X + Q_2^X \\
Q_- ^X&= Q_1^X - Q_2^X ,
\end{split}
\label{eq:basis2}
\end{equation}
where the label $X$ takes on the three values `$\,\widetilde{\ }\,$',  $cu$, $uc$ appearing in Eq.~\eqref{eq:basis1}.  Thus, we have three groups of operators $\widetilde{Q}_\pm$, $Q^{cu}_\pm$ and $Q^{uc}_\pm$. The advantage of this basis is that $Q_+$ belongs to the (84,1) irreducible representation of $SU(4)_L\times SU(4)_R$, while $Q_-$ belongs to the (20,1) representation~\cite{Giusti:2004an}.  Since the renormalization will be carried out in the $SU(4)_L\times SU(4)_R$ symmetric limit of vanishing $u$, $d$, $s$ and $c$ quark masses, the operators $Q_+$  and $Q_-$ will not mix with each other or any other dimension 6 operator.   Finally $SU(4)_L\times SU(4)_R$ symmetry requires that the renormalization factors for all operators in the same representation will be identical.

Although the basis in Eq.~\eqref{eq:basis2} is favored theoretically, we choose to use the basis in Eq.~\eqref{eq:basis1} for our actual calculation since it is those operators whose matrix elements are obtained from the explicit contractions which we evaluate.  The effects of the Wilson coefficients and all operator renormalization and mixing can then be summarized by:
\begin{equation}
\begin{split}
H_W&=\frac{G_F}{\sqrt{2}}\sum_{q,q^{\prime}=u,c}V_{qd}V^{*}_{q^{\prime}s}\sum_{i=1,2}C_i^{\overline{\rm MS}}(\mu)(1+\Delta r^{{\rm RI}\rightarrow\overline{\rm MS}})_{ij}(Z^{{\rm lat}\rightarrow RI})_{jk}Q_k^{qq^{\prime},{\rm lat}}(\mu)\\
&=\frac{G_F}{\sqrt{2}}\sum_{q,q^{\prime}=u,c}V_{qd}V^{*}_{q^{\prime}s}\sum_{i=1,2}C_i^{\rm lat}(\mu)Q_i^{qq^{\prime},{\rm lat}}(\mu)
\label{eq:Heff_DS=1} .
\end{split}
\end{equation}
All the operator renormalization and mixing are performed at a scale $\mu=2.15$ GeV. 
As summarized above, the Wilson coefficients $C_i^{\overline{\rm MS}}(\mu)$ are calculated following equation (5.8) - (5.21) in Ref.~\cite{Buchalla:1995vs} using the parameters $\alpha_s(M_Z)=0.1184$, $M_Z=91.1876$ GeV, $M_W=80.399$ GeV and $m_b(m_b)=4.19$ GeV ~\cite{Nakamura:2010zzi}. 

We use formulae provided by Lehner and Sturm which extend their earlier, 2+1 flavor  results~\cite{Lehner:2011fz} for the  matching matrix $\Delta r^{{\rm RI} \to \overline{\rm MS}}$ to the four-flavor case being studied here.  Their $2\times 2$ matching matrix is given by:
\begin{equation}
	\Delta r = \frac{\alpha_s(\mu)}{4\pi}
	\begin{pmatrix} 
		-4\ln(2)   &-8+12\ln(2) \\ 
		-8+12\ln(2)&-4\ln(2) 
	\end{pmatrix} .
\end{equation}
Here $\alpha_s(\mu)$ is calculated using the two-loop formula given by equation (3.19) in Ref.~\cite{Buchalla:1995vs}.  For $\mu=2.15$ GeV, $\alpha_s = 0.2974$.

The lattice operators are related non-perturbatively to operators renormalized in a regularization independent, Rome-Southampton~\cite{Martinelli:1995ty} scheme following the method developed in Ref.~\cite{Blum:2001xb} but using non-exceptional momenta~\cite{Sturm:2009kb} at a scale of $\mu=2.15$ GeV. Specifically, we use the RI/SMOM($\gamma_{\mu}$,$\slashed{q}$) scheme ~\cite{Lehner:2011fz}. Here the first $\gamma_{\mu}$ means that the projectors are constructed from $\gamma$ matrices. The second $\slashed{q}$ identifies the wave function renormalization scheme. We take the value $Z_q^{\slashed{q}}=0.8016(3)$ from ~\cite{Blum:2011pu}. Combining all three ingredients we obtain the final coefficients $C^{\rm lat}_i,\, i=1,2$ that must be applied to the bare lattice operators to construct the complete $\Delta S=1$ effective weak Hamiltonian given in Eq.~\eqref{eq:Heff_DS=1}.  The results for these coefficients and the ingredient from which they are constructed are given in Tab.~\ref{tab:npr}.  Note the diagonal character of the renormalization for the operator basis $Q^X_\pm$ can be seen from the structure of the $2\times2$ matrices given in this table, with equal diagonal and equal off-diagonal elements in our $Q^X_{i},\, i=1,2$ basis.

\begin{table}[!htp]
\caption{The Wilson coefficients, the ${\rm RI} \to \overline{\rm MS}$ matching matrix, the non-perturbative ${\rm lat} \to {\rm RI}$ operator renormalization matrix and their final product, all at a scale $\mu=2.15$ GeV shown in columns one through four respectively.}
\begin{ruledtabular}
\begin{tabular}{cc|cc|cc|cc}
$C_1^{\overline{\rm MS}}$ & $C_2^{\overline{\rm MS}}$ & $\Delta r_{11} = \Delta r_{22}$ &  $\Delta r_{12} = \Delta r_{21}$ & $Z_{11}=Z_{22}$ & $Z_{12}=Z_{21}$ & $C_1^{\rm lat}$ & $C_2^{\rm lat}$ \\
\hline
-0.2967 & 1.1385 & -$6.562\times 10^{-2}$ & $7.521\times 10^{-3}$ & 0.5916 & -0.05901 & -0.2216 & 0.6439
\end{tabular}
\end{ruledtabular}
\label{tab:npr}
\end{table}

\begin{figure}[!htp]
\includegraphics[width=0.7\textwidth]{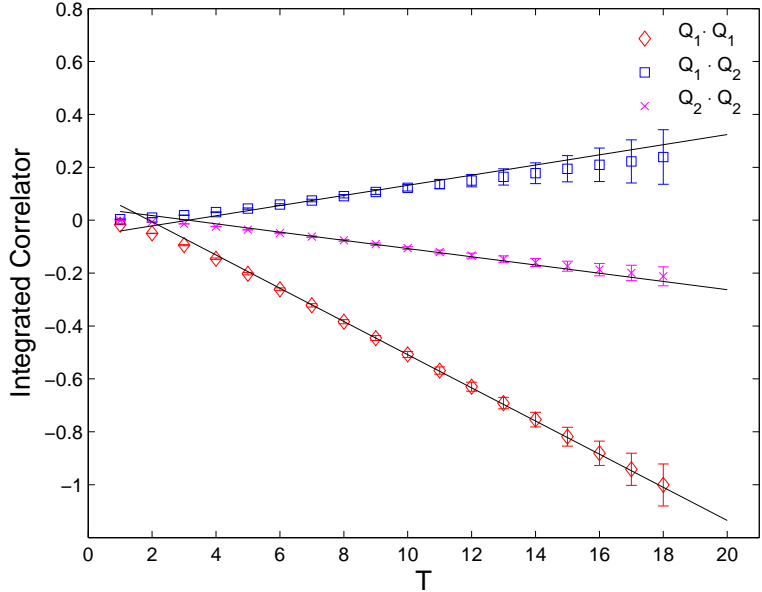}
\caption{Lattice results for the integrated correlator given in Eq.~\eqref{eq:integration_result} for the three operator products $Q_1 \cdot Q_1$,  $Q_1 \cdot Q_2$ and  $Q_2 \cdot Q_2$ and the case $M_K=834$ MeV.  The three lines give the linear fits to the data in the time interval [9,18] used to extract the corresponding values given in Tab.~\ref{tab:mass}.  (Note the slope of the integrated correlator
as a function of the time $T$ given in Eq.~\eqref{eq:integration_result} must be multiplied by $-2$ to obtain the corresponding contribution to $\Delta M_K$.) }
\label{fig:fit_result}
\end{figure}

We now combine all these ingredients and determine the mass difference $\Delta M_K$ in physical units.  The mass difference $\Delta M_K$ can be obtained by fitting the integrated correlator in the limit that the integration region $[t_a,t_b]$ become large.   We fit the dependence of the integrated correlator on $T=t_b-t_a+1$ to a linear function over the range $9 \le t_b-t_a \le 18$.  Figure~\ref{fig:fit_result} shows the computed values for the integrated correlator as a function of $T$ and the corresponding linear fit for each of the three operator products $Q_1 \cdot Q_1$,  $Q_1 \cdot Q_2$ and  $Q_2 \cdot Q_2$, for the case $M_K=834$ MeV. The results are given in Tab.~\ref{tab:mass}.  The lattice mass differences given in this table have a common factor $10^{-2}$ which is not shown.  The errors given in the table are statistical only.  

Although we have data for eight different kaon masses, we present results for only the seven kaon masses ranging from 563 MeV to 1162 MeV.  We do not give results for the lightest kaon because it is degenerate with the pion while the standard formula for $\Delta M_K$, which we are using, assumes that the $K^0$ and $\overline{K^0}$ are the only coupled, single-particle, degenerate states.  While listed for completeness, the three heavier kaon masses of 918, 993 and 1162 MeV are more massive than the threshold two-pion intermediate state and will therefore contain an unknown, exponentially growing contamination which we have been unable to identify and remove. 

Given our pion mass of 421 MeV, the two-pion intermediate state will be close to degenerate with kaon for the $M_K=839$ MeV case.   Were we to follow the prescription proposed in Ref.~\cite{Christ:2010zz} to control finite volume effects, we should choose this degenerate case and then remove completely the contribution of the degenerate, two-pion intermediate state, which should appear in the integrated correlator with the time dependence $(t_b-t_a)^2$.  However, as explained earlier, we are not able to identify the two-pion intermediate state within errors. This implies that the approximately on-shell, two-pion intermediate state contributes only a small part to the mass difference in our calculation and should have a small effect, at least on the results for 563 MeV $\le M_K \le 834$ MeV.

\begin{table}[!htp]
\caption{The contribution of the three operator products evaluated here to the mass difference $\Delta M_K$ for the seven different choices of the kaon mass listed in the first column in MeV.  The quantities in columns two through four are the simple lattice matrix elements of the operator products $Q_i^{qq'}Q_j^{q'q}$ for each $i,j = 1,2$, summed over the four values of $q,q' = u,c$, without Wilson coefficients or renomalization factors and have been scaled to remove a factor $10^{-2}$. These results are obtained from a fitting range [9,18]. The final column gives the complete contribution to $\Delta M_K$, expressed in physical units.  The results for the three largest values of the kaon mass are contaminated by an unknown, exponentially growing two-pion contribution which we have been unable to identify and subtract but are given here for completeness.  These results come from 800 configurations and use a charm quark mass of 863 MeV.}
\begin{ruledtabular}
\begin{tabular}{c|ccc|c}
$M_K$  (MeV)    & $\Delta M_K^{11}$ 
& $\Delta M_K^{12}$ 
& $\Delta M_K^{22}$
& $\Delta M_{K}$ ($\times 10^{-12}$ MeV) \\
\hline
563 & 6.42(15)  & -2.77(16)   & 1.56(9)    & 6.58(30)\\
707 & 8.94(23) & -3.16(27)   & 2.26(14)  & 8.85(48)\\
775 & 10.65(29) &  -3.49(35)  & 2.67(18)  & 10.32(62)\\
834 & 12.55(37) &  -3.84(46)  & 3.11(24)  & 11.89(81)\\
918 & 15.36(50) & -4.34(66)  & 3.75(34) & 14.20(115) \\
993 & 18.51(69) & -4.91(93) & 4.49(48) & 16.83(164) \\
1162 & 28.23(154)& -6.97(220) & 6.99(112) & 25.58(382) \\ 
\end{tabular}
\end{ruledtabular}
\label{tab:mass}
\end{table}

\section{Comparison with NLO perturbative calculation}
\label{sec:NLO_compare}

A direct comparison between our results and the experimental  value of $\Delta M_K$ has limited value because our kaon and pion masses are far from physical and we have not included all diagrams. However, we can learn something about the degree to which the present perturbative calculations describe $\Delta M_K$ for our unphysical kinematics by comparing our result with that obtained perturbatively by evaluating the perturbative formula at the kaon and pion masses used in our present calculation.   While there are now results for $\Delta M_K$ computed at NNLO given in Ref.~\cite{Brod:2011ty}, complete expressions for the results are not given in that brief letter.  Therefore, we choose to compare with the NLO result of Herrlich and Nierste~\cite{Herrlich:1993yv} for which complete information is available in published form.  Since the full results at NLO and NNLO orders differ by 36\% at the physical point, the agreement with our result should be only approximate and this use of the NLO result adequate for our purpose.   This comparison with NLO perturbation theory may also lessen the significance of our omission of disconnected diagrams, which do not appear at NLO.  We will compare this NLO result, evaluated at our kinematics, with our lattice calculation carried out using 600 configurations at the unitary quark masses $m_l=0.01$ and $m_s=0.032$ ($M_\pi=421$ MeV and $M_K=563$ MeV) for a series of valence charm quark masses.

The mass difference in the perturbative calculation is given by:
\begin{equation}
\Delta M_K = \frac{G_F^2}{6\pi^2}f_K^2\hat{B}_KM_K(\lambda-\frac{\lambda^3}{2})^2\eta_1(\mu_c,m_c(\mu_c))m_c^2(\mu_c),
\end{equation}
which can be obtained, for example, from Eq.~(12.1) in Ref.~\cite{Buchalla:1995vs}.  Here $\lambda$ is the sine of the Cabibbo angle, one of the four Wolfenstein parameters entering the CKM matrix, $\mu_c$ is the scale at which the four-flavor theory is matched to that with three flavors and the kaon decay constant $f_K$ is defined using conventions which make its physical value equal to 155 MeV.  The two non-perturbative parameters, the kaon decay constant $f_K$ and the kaon bag parameter $\hat B_K$, evaluated in the renormalization group invariant (RGI) scheme, can also be computed for the unphysical values of $m_l$ and $m_s$ listed above.  For the present calculation we find it convenient to directly compute the matrix element of left-left operator:
\begin{equation}
 \langle O_{LL} \rangle
                 = \langle \bar{K}^0 | (\bar{s}d)_{V-A}(\bar{s}d)_{V-A} | K^0 \rangle,
\label{eq:OLL}
\end{equation}
obtaining the value  0.00462(5) for $m_l=0.01$ and $m_s=0.032$.  Here we use non-relativitic normalization for the kaon states: $\langle K(\vec p)|K(\vec p\,^\prime)\rangle = \delta^3(\vec p - \vec p\,^\prime)$. This lattice result can be converted to the RGI scheme by multiplying by the factor:
\begin{equation}
Z^{RGI}_{VV+AA} = Z^{RGI}_{BK}Z^2_A,
\end{equation}
where $Z^{RGI}_{BK}=1.27$ and $Z_A=0.7161$ are taken from Ref.~\cite{Allton:2008pn}.

The expression for the mass difference then becomes:
\begin{equation}
\Delta M_K = \frac{G_F^2}{8\pi^2}Z^{RGI}_{BK}Z^2_A\langle O_{LL} \rangle(\lambda-\frac{\lambda^3}{2})^2\eta_1\left(\mu_c,m_c(\mu_c)\right)m_c^2(\mu_c) .
\label{eq:pt_result}
\end{equation}
Here the factors $Z^{RGI}_{BK}Z^2_A\langle O_{LL}\rangle$ are lattice quantities determined for the kinematics studied here while $\eta_1$ is determined from the NLO perturbation theory calculation of Ref.~\cite{Herrlich:1993yv}, summarized in Ref.~\cite{Buchalla:1995vs}.  Specifically, Eq.~\eqref{eq:pt_result} corresponds to the term in Eq.~(12.1) of Ref.~\cite{Buchalla:1995vs} containing $\eta_1$.  Note, the two right-most factors in Eq.~(12.1) do not appear in our Eq.~\eqref{eq:pt_result} since they have been incorporated in $\hat B_K$, changing it to the RGI scheme.  We evaluate $\eta_1$ using Eq.~(12.31) of Ref.~\cite{Buchalla:1995vs}.  We now compare this perturbative result with our non-perturbative, lattice calculation of the same box topology and for the same quark masses.

\begin{figure}[!htp]
\includegraphics[width=0.7\textwidth]{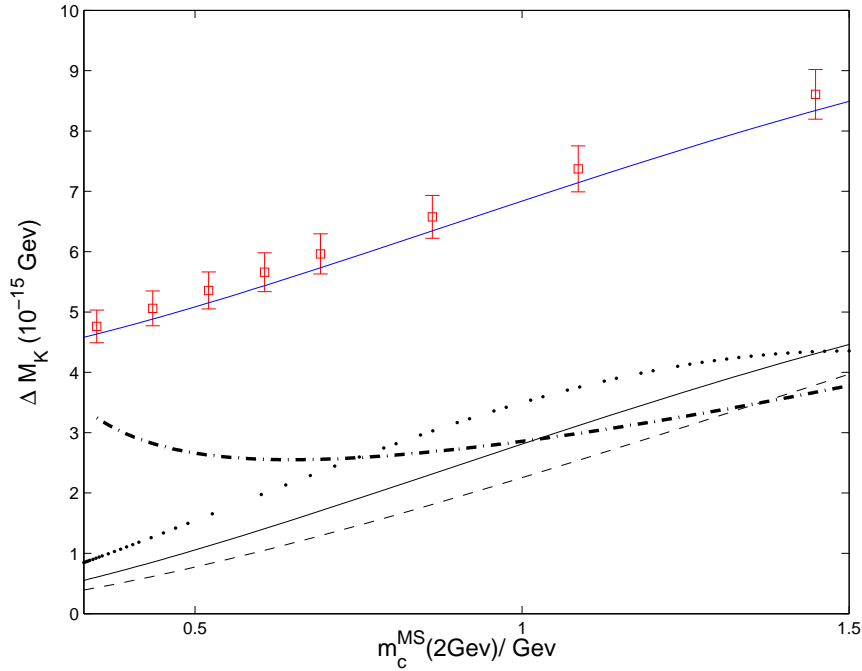}
\caption{The lattice results for $\Delta M_K$ plotted as a function of $m_c$ for the single kaon mass $M_K=563$. Here the charm quark mass is defined in the $\overline{MS}$ scheme at a scale $\mu = 2$ GeV.  The top solid curve is the result of a correlated fit to the ansatz given in Eq.~\eqref{eq:fit_ansatz}.  The same result but with the long distance constant $a$ omitted gives the lowest, solid curve.  The dotted and dashed lines give the perturbative result for the choices of matching scale $\mu_c=1$ and 1.5 GeV respectively. Finally the dash-dot curve corresponds to the choice $\mu=m_c$.}
\label{fig:compare_with_nlo}
\end{figure}

\begin{table}[!htp]
\caption{The quantity $\Delta M_K$ for various charm quark masses and $M_K=563$ MeV. Here the charm quark mass is given in the $\overline{MS}$ scheme at a scale $\mu=2$ GeV.  The third and fourth rows give the lattice results and NLO perturbation result respectively. For the perturbative result, the matching between four and three flavors is done at $\mu_c=m_c(m_c).$  The first row contains the values of $(m_c a)^2$ as an indication of the size of finite lattice spacing errors which may corrupt the comparison between the lattice and NLO perturbative results.}
\begin{ruledtabular}
\begin{tabular}{c|cccccccc}
	$m_c$ (MeV) & 350 & 435 & 521 & 606 & 692 & 863 & 1086 & 1449 \\
\hline
$(m_ca)^2$ & 0.04 & 0.06 & 0.09 & 0.12 & 0.16 & 0.25 & 0.39 & 0.70 \\
$\Delta M_K$ ($10^{-15}$ GeV)& 4.76(27) & 5.06(29) & 5.36(31) & 5.66(32) & 5.96(33) & 6.58(35) & 7.37(38) & 8.61(41) \\
$\Delta M_K^\mathrm{NLO}$ ($10^{-15}$ GeV) & 3.24 & 2.82 & 2.63 & 2.56 & 2.56 & 2.68 & 2.99 & 3.67
\end{tabular}
\end{ruledtabular}
\label{tab:compare_with_nlo}
\end{table}

In our lattice calculation, we determine $\Delta M_K$ for a series of charm quark masses.  We can exploit this mass dependence to attempt to separate the complete lattice result into short and long distance parts as follows.  As is discussed in Appendix~\ref{sec:epsilon}, the dominant contribution to $\Delta M_K$ is proportional to the CKM matrix element product $|V_{cd} V_{cs}^*|^2$ and for large $m_c$ grows as $m_c^2$ as is suggested by the perturbative result in Eq.~\eqref{eq:pt_result}.   As is also implied by that equation, additional factors of $\ln(m_c^2)$ will appear in higher order perturbation theory.  If $\Delta M_K$ is examined for $m_c \gg \Lambda_\mathrm{QCD}$, in addition to such $m_c^2\ln^n(m_c^2)$ terms, we should also expect a constant piece, coming from long distance effects in which the charm quark mass plays a negligible role, with the remaining mass dependence behaving as $1/m_c^2$ for large $m_c$.  As explained in the discussion of the GIM subtraction in Sec~\ref{subsec:short_distance_subtraction}, the charm quark mass enters only as $m_c^2$ which implies there are no terms behaving as $m_c$ or $1/m_c$.  Note, the non-zero density of Dirac eigenvalues, $\rho(\lambda)$ at zero eigenvalue $\lambda=0$ would induce a non-perturbative, chiral symmetry breaking $m_c$ term in the limit of small $m_c$, but has no effect on the large $m_c$ limit being considered here.  This limit is determined only by the large $\lambda$ behavior of $\rho$.

We use this large $m_c$ expansion to parameterize the dependence of $\Delta M_K$ on $m_c$ by adopting the ansatz:
\begin{equation}
\Delta M_K (m_c) = a + b\, m_c^2 + c\, m_c^2 \ln(m_c),
\label{eq:fit_ansatz}
\end{equation}
where we drop the possible $1/m_c^2$ term.  The quadratic plus quadratic times logarithmic form of the terms with coefficients $b$ and $c$ can be found in the NLO perturbative expansion Eq.~\eqref{eq:pt_result} if we use a fixed value of  $\mu_c$ as $m_c$ varies.  Thus, the constants $b$ and $c$  are determined by short distance physics, arising from length scales of order $1/m_c$ and should be accessible to a perturbation theory calculation.  In contrast the $a$ term involves non-perturbative phenomena and long distances.  The perturbative calculation also contains a long distance part which contributes to the constant $a$.  However, this is suppressed by a factor of $(m_{ud}/m_c)^2$ which is at most 0.5\% for our lightest charm quark mass.

In Fig.~\ref{fig:compare_with_nlo} we plot our results for $\Delta M_K$ as a function of the charm quark mass as well as the result from the fit to the ansatz given in Eq.~\eqref{eq:fit_ansatz}.  The upper solid curve shows the entire fitting function given in  Eq.~\eqref{eq:fit_ansatz} while the lower solid curve has the non-perturbative terms proportional to $a$ removed.   A comparison of these two solid curves in Fig.~\ref{fig:compare_with_nlo} suggests that for unphysically massive $M_\pi=421$ MeV and $M_K = 563$ MeV and a charm quark mass of 1.2 GeV, approximately 50\% of $\Delta M_K$ comes from long-distance effects.

Also shown in Fig.~\ref{fig:compare_with_nlo} are the perturbative results for a number of different choices of the matching scale $\mu_c$.  Numerical values for the lattice and perturbative calculations are given in Tab.~\ref{tab:compare_with_nlo} for the case in which the matching scale is taken to be the charm quark mass, $\mu_c=m_c$. The errors given in the table are statistical only. Since there is no degenerate two-pion channel for these kinematics and the disconnected diagrams are neglected in both the lattice and NLO perturbation theory calculation, we expect that the discretization error is the most important systematic error affecting this comparison.  We list the values of $(m_ca)^2$ in Tab.~\ref{tab:compare_with_nlo} to give an estimate for the size of the discretization error.  From Fig.~\ref{fig:compare_with_nlo} we see that the NLO perturbative results depend dramatically on the choice of $\mu_c$, a well known difficulty with the NLO calculation.   This dependence should be small if both $m_c$ and $\mu_c$ are sufficiently large that NLO perturbation theory is a good approximation. In fact, the NNLO result computed by Brod {\it et al.} gives a large, 36\% correction to the NLO result for $\eta_{cc}$ at the physical charm quark and the large $\mu_c$ dependence is not reduced at NNLO.  Aside from these ambiguities arising in the perturbative calculation, we do see a very large gap between the lattice and both the NLO results and the mass-dependent terms in our fit ansatz.  If we focus on the three heaviest charm quark masses shown in Fig.~\ref{fig:compare_with_nlo}, we see a 130-150$\%$ shift compared to the NLO results. This number is much larger than the 30$\%$ long distance contribution deduced at physical kinematics by comparing the NLO perturbative result with the experimental value for $\Delta M_K$ or the 11\% discrepancy seen at NNLO.  We view this as strong evidence that non-perturbative methods are needed to determine the contributions to $\Delta M_K$ from distances of $1/m_c$ and larger.  Of course, we must bear in mind the potentially large finite lattice spacing errors suggested by the values of $(m_c a)^2$ given in the first row of Table~\ref{tab:compare_with_nlo} which may be as large as 40-70\% for the two largest values of $m_c$.

\section{Conclusion and discussion}
\label{sec:conclusion}

We have addressed two objectives in this paper.  The first is the presentation of techniques needed to compute the $K_L$-$K_S$ mass difference with controlled systematic errors using the methods of lattice QCD.  We demonstrate that such a calculation should be possible if a lattice spacing is used which is sufficiently small that an explicit charm quark can be included in the calculation without introducing unacceptably large discretization errors.  Such a second-order weak calculation can be performed using effective, four-quark weak operators provided explicit subtractions are performed for specific exponentially growing terms that arise from intermediate states which are lighter than the kaon.  While we have been unable to study the contribution of two-pion states that are approximately degenerate with the kaon and the resulting finite volume effects, these can be controlled~\cite{Christ:2010zz,Christ:2012np} and will be the subject of future work.

Our second objective has been to demonstrate these methods by carrying out a first calculation of the long distance contributions to the  $K_L$-$K_S$ mass difference.  Although this calculation suffers from uncontrolled systematic errors caused by unphysical kinematics and a failure to include all relevant graphs, it is of some physical interest.   The experimental value of $K_L$-$K_S$ mass difference is $3.48\times 10^{-12}$\,MeV.  Our result ranges from $6.58(30)\times 10^{-12}$\,MeV to $11.89(81)\times 10^{-12}$\,MeV.  Since our kinematics are far from the physical and we have not included all possible diagrams, we are not able to compare with the experimental value directly. However, we do find that as the kaon mass decreases from $834$\,MeV to $563$\,MeV, the mass difference shrinks by nearly a factor of two. This implies that the result depends strongly on the kinematics. Currently we are using a pion mass of $421$\,MeV, which leaves open a possibly large decrease were we to use a physical $135$\,MeV pion mass.   Thus, our results are consistent with the conclusions drawn from the continuum, NNLO calculation of Brod and Gorban~\cite{Brod:2011ty} whose central value for the short distance part of $\Delta M_K$ accounts for 89\% of the experimental value.   

However, as Brod and Gorban point out, the apparent slow convergence of the perturbation series reflects large uncertainties in this application of perturbation theory at the energy scale of the charm quark mass, giving strong motivation for the sort of non-perturbative approach to the full calculation being developed here.  This conclusion is supported by the comparison between the NLO perturbative calculation of Herrlich and Nierste~\cite{Herrlich:1993yv} and our calculation of the same diagrams, a comparison that can be made for the same quark masses.  We find poor agreement, with the lattice result twice as large as those found from perturbation theory using a variety of prescriptions for treating the perturbative matching at the charm quark threshold.

We have neglected type 3 and type 4 diagrams in this calculation. Although this is a convenient choice in this first study of long distance effects, we must of course include these diagrams in a full calculation. We will need to calculate extra stochastic source propagators in order to evaluate type 3 and type 4 diagrams. This will make the calculation more expensive and noisier. More importantly, the type 4 disconnected diagrams pose the greatest challenge to a full calculation. Disconnected diagrams are extremely noisy and require very large statistics. Instead of analyzing more configurations, we can try to improve our method. In this work, we have fixed the spatial location of one operator and integrated only the position of the second operator over the whole spatial volume. To improve this, we can try to locate the first operator at more than one spatial point and then average the result over those added locations.  The all mode averaging technique~\cite{Blum:2012uh} may make it possible to include the calculation of these extra points with only a modest increase in computational cost.  This calculation is also a promising candidate for the use of all-to-all propagators which would make the integration over the entire space-time volume for both operators possible and also allow us to vary the source-sink separation, $t_f-t_i$ which was fixed at 27 in the present calculation.

Based on the results presented here, we have begun a more ambitious calculation with a lighter pion and larger volume which includes all diagrams.  This calculation should yield increased insight into the physics of the $K_L-K_S$ mass difference and a better understanding of the numerical requirements of a physical calculation.  If the substantially increased statistics in this next calculation, provided by the use of deflation and all mode averaging, are sufficient to give an accurate answer, then a follow-up calculation with physical kinematics should be possible.  However, substantially more computer resources than are presently available will be needed if we are to use a sufficiently small lattice spacing for the proper treatment of the charm quark.

\section*{Acknowledgments}
We thank Laurent Lellouch, Guido Martinelli and Steve Sharpe for very helpful discussions at the beginning of this work and our RBC and UKQCD colleagues for many valuable suggestions and encouragement.  We are particularly indebted to Christoph Lehner and Christian Sturm for providing us with their results for the one-loop, four-flavor RI/MOM to $\overline{{\rm MS}}$ conversion factors before publication.   These results were obtained using the DOE USQCD and RIKEN BNL Research Center QCDOC computers at the Brookhaven National Laboratory.  N.C. and J.Y. were supported in part by the US DOE grant DE-FG02-92ER40699, C.T.S. by STFC Grant ST/G000557/1, T.I. and A. S. by U.S. DOE contract DE-AC02-98CH10886 and T.I also by JSPS Grants  22540301 and 23105715.

\appendix

\section{Calculating $\Delta M_K$ and $\epsilon_K$ in the standard model}
\label{sec:epsilon}

In this appendix we review the various electro-weak diagrams that contribute to {$\Delta M_K$ and $\epsilon_K$ in the standard model and their expected sizes.   We then describe a framework for a combined perturbative and lattice QCD calculation of these quantities to sub-percent accuracy.  As in the body of this paper, we consider a four-flavor lattice calculation which includes explicit charm quark.  This allows a separation between the perturbative (short distance) and lattice QCD (long distance) parts of the calculation at a sufficiently large scale $\mu$ that both the lattice and perturbative errors can be controlled.  This discussion provides a basis for the complete calculation of $\Delta M_K$ begun in this paper and also identifies the ingredients that would be needed for a similar calculation of $\epsilon_K$~\cite{Christ:2012np} with accurate control of both long and short distance phenomena, addressing the issues raised in Ref.~\cite{Buras:2010pza}.

\begin{figure}[!htp]
\centering
\subfigure[]{\includegraphics[width=0.4\textwidth]{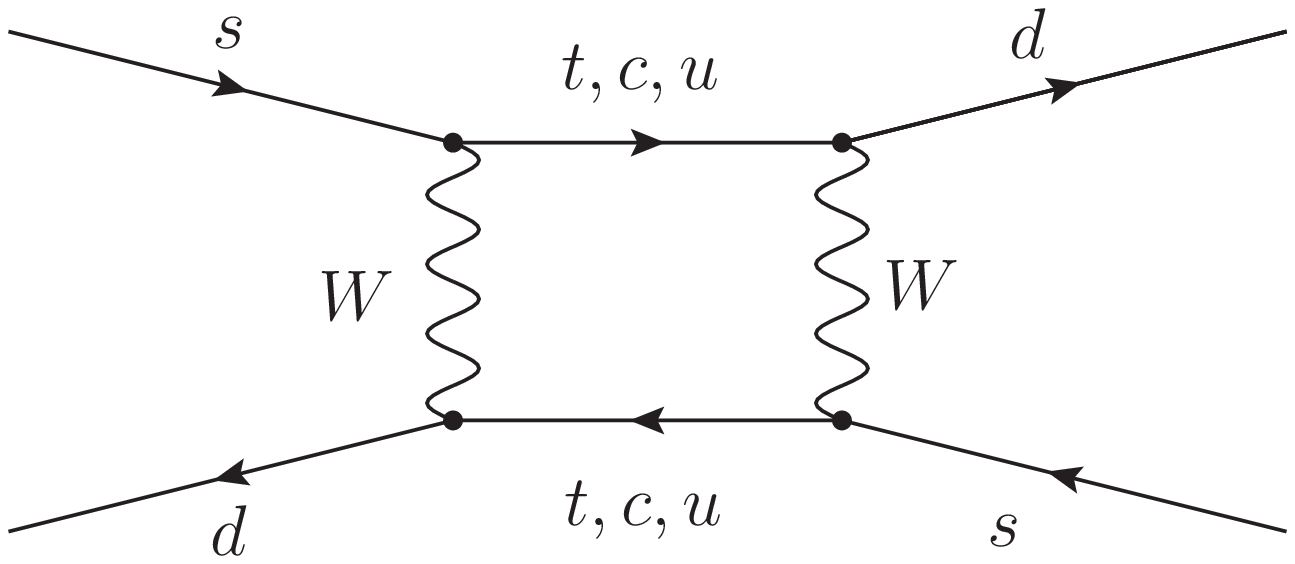}
\label{fig:topologies_a}} \hskip 0.5in
\subfigure[]{\includegraphics[width=0.4\textwidth]{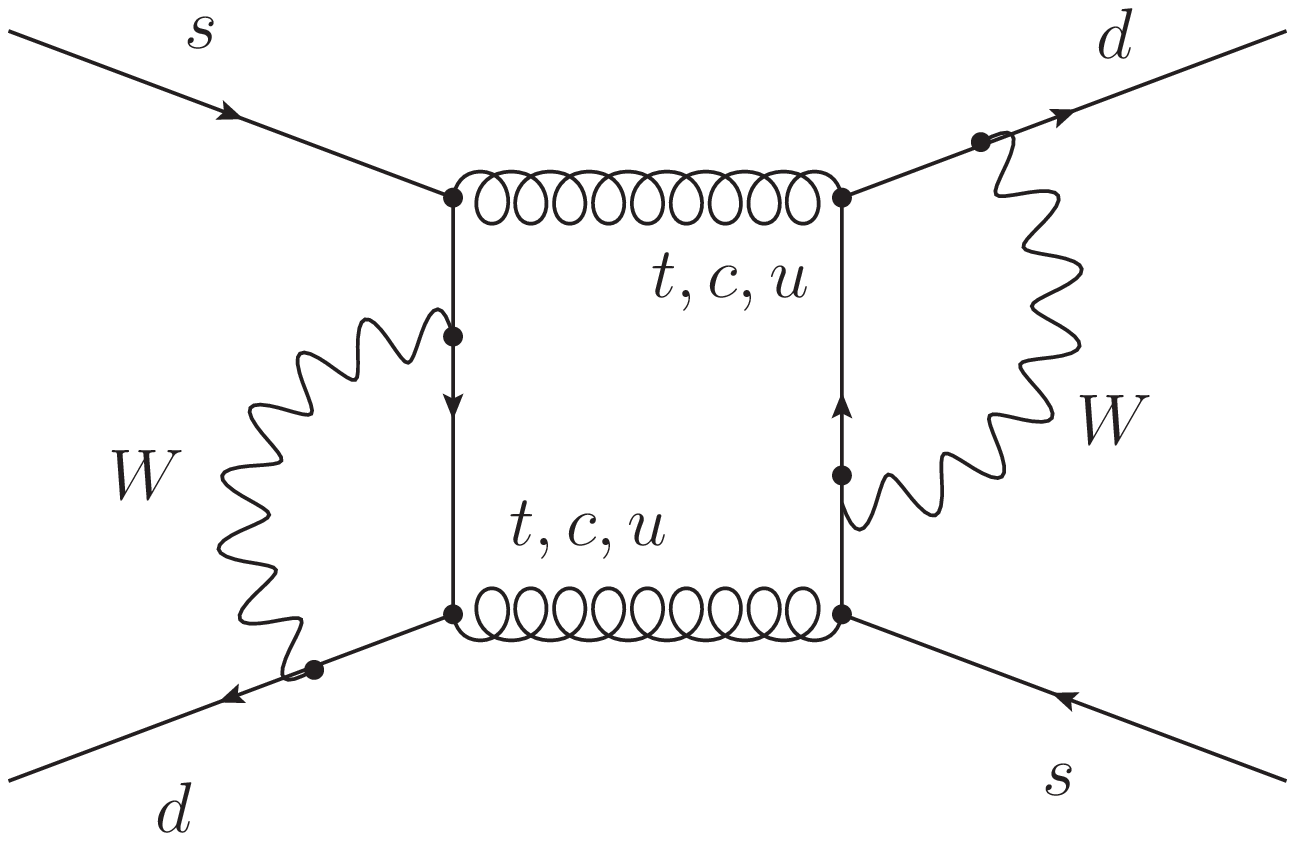}
\label{fig:topologies_b}}
\caption{Examples of the two types of Feynman diagrams that can contribute to $K^0 -\overline{K}^0$ mixing: a box graph (a) and a disconnected graph (b).  The box graph shows only the quarks and $W$'s with gluons to be included in all possible ways.  The disconnected graph shows a particular choice of gluon lines which will give a non-vanishing contribution.  Of course, many other arrangements of gluon lines are possible.}
\label{fig:topologies}
\end{figure}

Recall that $\Delta M_K$ and $\epsilon_K$ are derived from CP conserving and CP violating parts of the off diagonal $K^0-\overline{K^0}$ mixing matrix.  Here we focus on the so called ``dispersive part'' of the matrix which involves a sum over off-shell states such as found in Eq.~\eqref{eq:delta_mk_cont}.  We are not considering the on-shell ``absorptive part'' which can be determined directly from first-order, on-shell $K\to\pi\pi$ matrix elements that can be evaluated by more  familiar lattice methods.  To provide context for this discussion we show in Fig.~\ref{fig:topologies} the two types of graphs which contribute to $M_{\overline{0}0}$, the off-diagonal mass term  in the standard model, a box diagram on the left, \ref{fig:topologies_a}, and a disconnected graph on the right,  \ref{fig:topologies_b}.  In the former, each of the two exchanged $W$ bosons connects to both of the two quark lines carrying external flavor.  In the latter each $W$ boson is emitted and absorbed from a single quark line and only gluons join the two quark lines.  Diagrams of the sort which appear in Fig.~\ref{fig:topologies_a} will contribute to contractions of types 1 and 2 shown in Figs.~\ref{fig:type1} and \ref{fig:type2}, depending on how the external quark lines in Fig.~\ref{fig:topologies_a} are combined to form the kaon states.  Diagrams of the sort which appear in Fig.~\ref{fig:topologies_b} will contribute to contractions of types 3 and 4 shown in Figs.~\ref{fig:type3} and \ref{fig:type4}, depending on how the external quark lines in Fig.~\ref{fig:topologies_b} are arranged, making the resulting diagram disconnected in the $t$ or $s$ channel, respectively.  Throughout this paper a {\it disconnected} diagram is one which can be separated into two disjoint pieces by cutting only gluon lines.  In the case of the $t$-channel disconnected graphs entering the type 3 contractions of Fig.~\ref{fig:type3}, this must include cutting the gluon lines joining the $s$ and $d$ quarks making up the $K^0$ and $\overline{K^0}$ mesons.

The internal quark lines appearing in both types of diagrams between the weak vertices are of the up type, {\it i.e.} $u$, $c$ and $t$ and couple to the external $s$ and $d$ by the product of CKM matrix elements $\lambda_i = V_{id}^*V_{is}$ where $i=u$, $c$ or $t$.  Conventionally~\cite{Inami:1980fz} the $u$ quark coupling is eliminated by using the orthogonality of the first and second columns of the CKM matrix, allowing the sum over the three types of up quarks in each quark line of Fig.~\ref{fig:topologies_a} to be written in the GIM subtracted form:
\begin{equation}
\sum_{i=u,c,t}\frac{\lambda_i \slashed{p}}{p^2+m_i^2}
       = \lambda_c\left\{\frac{\slashed{p}}{p^2+m_c^2} - \frac{\slashed{p}}{p^2+m_u^2}\right\}
        + \lambda_t\left\{\frac{\slashed{p}}{p^2+m_t^2} - \frac{\slashed{p}}{p^2+m_u^2}\right\}
\label{eq:GIM_old}
\end{equation}
where we have dropped the usual mass term in the numerator because of the $V-A$ structure of the weak vertices.  Cancellation between the two propagators within the curly brackets reduces the degree of divergence of each subgraph containing such a GIM subtracted combination by two units.  (Note this reduction of the degree of divergence by two units for each GIM-subtracted propagator, is a consequence of chiral symmetry and holds more generally, even if additional intermediate gluon lines are coupled to that propagator.)  Using this scheme the resulting expression for the box diagram has been worked out at leading order (LO)~\cite{Inami:1980fz}, next-to-leading order (NLO)~\cite{Herrlich:1993yv} and next-to-next-to-leading order (NNLO)~\cite{Brod:2010mj,Brod:2011ty} in QCD perturbation theory and the results through NLO are summarized in Ref.~\cite{Buchalla:1995vs}.

For our calculation we find this standard approach unnecessarily awkward in two ways.  First, both $\lambda_t$ and $\lambda_c$ have imaginary parts, giving two partially related contributions to $\epsilon_K$.  Second, the term which involves a top-up subtraction combines short and long distance quantities in a fashion that is easily accommodated in neither a lattice nor a perturbative calculation.  The top quark is too massive to be treated directly using lattice methods while the up quark contribution can be sensitive to infrared effects that cannot be evaluated using perturbation theory. These two features are neatly avoided if CKM unitarity is instead used to eliminate $\lambda_c$:
\begin{equation}
\sum_{i=u,c,t}\frac{\lambda_i\slashed{p}}{p^2+m_i^2}
       = \lambda_u\left\{\frac{\slashed{p}}{p^2+m_u^2} - \frac{\slashed{p}}{p^2+m_c^2}\right\}
        + \lambda_t\left\{\frac{\slashed{p}}{p^2+m_t^2} - \frac{\slashed{p}}{p^2+m_c^2}\right\}.
\label{eq:GIM_new}
\end{equation}
With this approach $\lambda_u$ is real and all CP violation requires the presence of $\lambda_t$.  In addition, the propagator constructed from the difference of top and charm propagators will substantially suppress the contribution from low momentum, $p \ll m_c$ because of the relatively large mass of both the top and charm quarks.  This should make the perturbative calculation of amplitudes involving this difference more reliable and allow factors containing this difference to be treated as local in a lattice calculation with greater accuracy.
Since the real parts of $\lambda_u$ and $-\lambda_c$ are nearly identical, the structure of the $\Delta M_K$ calculation is affected very little by this change.   A similar construction with similar consequences can be performed separately for each of the two vertex subgraphs appearing in Fig.~\ref{fig:topologies_b}.

Using this framework $\Delta M_K$ and $\epsilon_K$, including both the perturbative calculation of short distance contributions and the non-perturbative calculation of long distance parts, can be naturally separated into six terms associated with the three factors $\lambda_u^2$, $\lambda_u\lambda_t$ and $\lambda_t^2$, (which we will refer to as $uu$, $ut$ and $tt$ respectively) for each of the box and disconnected topologies.  We will now discuss the techniques that can be used to compute each of these six terms and the size expected for each.  Before the factors of $\lambda_i\lambda_j$ have been applied, these six amplitudes, which must be determined by a combination of perturbative and lattice methods, are common to both $\Delta M_K$ and $\epsilon_K$.  Finally these six size estimates can be combined with the known values of the real and imaginary parts of the three $\lambda_i\lambda_j$ factors to reproduce the standard expectations for the relative contribution of each of these six terms to $\Delta M_K$ and $\epsilon_K$, allowing us to anticipate the precision that may be eventually achieved with the methods proposed here.  

Of course, these $\lambda_i\lambda_j$ factors play a large role in determining which of the $uu$, $ut$ and $tt$ pieces will be important and which can be neglected when computing $\Delta M_K$ and $\epsilon_K$.   The very different scale of the top and charm quark masses also affects the ultimate importance of these six amplitudes, where a factor of $(m_t/m_c)^2$ can produce a more than $10^4$ enhancement.  We will use the following values extracted from Ref.~\cite{Beringer:1900zz}:
\begin{alignat}{2}
\lambda_u                             &= 0.22 \\
\lambda_c                             &= -0.22           & +1.34\times 10^{-4} &i \\
\lambda_t                              &= 3.2\times 10^{-4} &  -1.34\times 10^{-4} &i \\
\left(\frac{m_t}{m_c}\right)^2 &= 2.4\times 10^4 
\label{eq:PDG_input}
\end{alignat}
where $\lambda_c$ is also given for completeness.

\textbf{Box topology, $\boldsymbol{uu}$:}
We first discuss the $uu$, $ut$ and $tt$ contributions coming from amplitudes with the box topology.  The $uu$ piece is simple to discuss.  The difference between up and charm quark propagators produces the GIM suppression needed for convergence for the box topology.  If the $W$ propagators are contracted to points, the resulting diagrams have a degree of divergence of $+2$.  For the $uu$ piece, as discussed above, the double $c-u$ difference reduces the degree of divergence to $-2$.  Thus, in a lattice calculation such as that undertaken in this paper, the $uu$ piece can be accurately determined from products of pairs of four-quark operators.  Since only momenta of the order of $m_c$ will enter, power counting implies that the $uu$ contribution will be of order $m_c^2/M_W^4 = x_c/M_W^2$, were we adopt the conventional notation $x_i=(m_i/M_W)^2$ for $i=u$, $c$ and $t$.

\textbf{Box topology, $\boldsymbol{ut}$:}
The $ut$ contribution to diagrams with the box topology is more complex.  Since only one quark line involves both up and charm quarks, we now have a single GIM subtraction and the diagram that results if the $W$ propagators are each contracted to a point will be logarithmically divergent.  Thus, if a lattice calculation is attempted including up and charm quarks using products of the two four quark operators given Eq.~\eqref{eq:operator}, there will be an incorrect, short distance contribution in which this logarithmic divergence is regulated not by the difference of the top and charm quark mass but by the lattice cutoff.  However, this difficulty can be accurately overcome by introducing the additional $\Delta S=2$ four-quark operator $O_{LL}$, defined in Eq.~\eqref{eq:OLL}, and adjusting its coefficient so that an appropriate off-shell, gauge-fixed four quark Green's function which includes both this new $O_{LL}$ insertion as well as the original operator product takes the value of the continuum amplitude instead of that given by the lattice cutoff~\cite{Christ:2010zz,Yu:2011gk}.  If the scale $\mu$ where this condition is imposed is sufficiently small ($\mu \ll 1/a$) that the lattice evaluation is accurate and sufficiently large ($\Lambda_{\rm QCD} \ll \mu$) that the perturbative value of the continuum amplitude can be reliably computed, then this $ut$ box contribution can be computed with controlled errors.  Since the most singular short distance behavior is logarithmic, power counting implies that the $ut$ box piece has the same $x_c/M_W^2$ scale as was found for the $uu$ contribution, up to logarithmic factors.

\textbf{Box topology, $\boldsymbol{tt}$:}
Next consider the $tt$ contribution to the box topology.  The natural momentum cutoff in this $t-c$, GIM subtracted amplitude is $m_t$ so this amplitude is dominated by short distances and has a natural size of $x_t/M_W^2$, more than $10^4$ larger than the $uu$ and $ut$ pieces just considered.  Thus, this piece can be computed at an ultimate accuracy of $10^{-4}$ by the standard evaluation of the product of a perturbative Wilson coefficient and a lattice matrix element of $O_{LL}$.

\textbf{Disconnected topology, $\boldsymbol{uu}$:}
The $uu$ contribution from the disconnected graphs is similar in structure to the box case discussed above.   The double GIM, $u-c$ subtractions make finite the diagrams resulting from reducing the $W$ propagators to points.  Thus, this piece also can be directly computed using lattice methods and the four-quark operators of Eq.~\eqref{eq:operator}.  However, the size of such a disconnected $uu$ piece is more difficult to estimate.  A phase space comparable to that present in the box contribution results if the momentum carried by the exchanged gluons in Fig.~\ref{fig:topologies_b} is on the order of $m_c$, giving the estimate $x_c/M_W^2$.  However, since momenta of the order of $m_c$ must be carried by gluons, these may be suppressed by $\alpha^2_s(m_c)$.  In contrast, the lower momentum region may receive some $I=0$ enhancement.  However, the smaller momentum suggests a $(m_K/m_c)^2$ phase space suppression factor.  As mentioned previously, such contributions are often described as Zweig suppressed.  Nevertheless, such a disconnected $uu$ contribution can be precisely defined in a lattice calculation which includes explicit charm quarks and its inclusion is the natural next step, following the calculation presented here.  Of course, such a lattice calculation would automatically include all diagrams with this disconnected topology and two, GIM-subtracted,  $u-c$ internal quark propagators, not only the specific diagram shown in Fig.~\ref{fig:topologies_b}.

\textbf{Disconnected topology, $\boldsymbol{ut}$:}
The disconnected $ut$ piece can also be included in a lattice calculation. Such a contribution can be viewed as arising from the product of two factors.  One factor is associated with $\lambda_u$ and involves the GIM-subtracted difference of up and charm quark lines connected to one of the current-current operators $Q^{\alpha\alpha}_i$ given in Eq.~\eqref{eq:operator} for $i=1,2$ and $\alpha=c,u$, obtained by shrinking to a point the $W$ propagator, shown for example in Fig.~\ref{fig:topologies_b}, which is associated with the factor $\lambda_u$.  The second factor, associated with $\lambda_t$, must be viewed as the sum of separate contributions of the top and charm quark propagators.  The large mass of the top quark implies that the dominant contribution involving the top quark will come from a gluonic vertex such as one of those shown in Fig.~\ref{fig:topologies_b}, with the vertices of the $W$ and top propagators treated as coincident.  This part can be accurately represented in a lattice calculation by the four standard, four-flavor ``QCD penguin'' operators
\begin{eqnarray}
P_3 &=& \sum_{q=u,d,c,s}(\overline{s_i}d_i)_{V-A}(\overline{q_j}q_j)_{V- A}, \quad
P_4 = \sum_{q=u,d,c,s}(\overline{s_i}d_j)_{V-A}(\overline{q_j}q_i)_{V-A}, \\
P_5 &=& \sum_{q=u,d,c,s}(\overline{s_i}d_i)_{V-A}(\overline{q_j}q_j)_{V+A} \quad
P_6 = \sum_{q=u,d,c,s}(\overline{s_i}d_j)_{V-A}(\overline{q_j}q_i)_{V+A},
\label{eq:QCD_penguin}
\end{eqnarray}
using the same notation as  in Eq.~\eqref{eq:operator}.
Such a dimension-six, QCD penguin operator will have a coefficient of order $1/M_W^2$ that can be computed in perturbation theory if that operator is renormalized at an appropriately large scale $\mu \gg \Lambda_\mathrm{QCD}$.  The combined $ut$ amplitude will be of order $x_c/M_W^2$ with the same uncertainties associated with scale of gluon momentum described above -- uncertainties that can be definitively resolved by a lattice calculation.

The charm quark part of this second factor requires more care.  While a portion can be reproduced by the lattice charm quark and a four quark vertex representing the $W$ exchange, lacking the GIM subtraction, this contribution will contain a logarithmically divergent gluon vertex subgraph which will also require the introduction of a QCD penguin subtraction.  As described in the discussion above of the $ut$ contribution to the box diagrams, this subtraction can be chosen to remove the short-distance artifacts introduced by this divergence and to replace them with the correct short distance part which can be computed in perturbation theory.  While somewhat involved, the necessary short-distance artifact subtraction can be determined non-perturbatively using Rome-Southampton methods and the perturbative short distance replacement included with accurately controlled errors.  

Thus, an accurate lattice representation for the two $W$ loops appearing in the disconnected diagrams, such as that of Fig.~\ref{fig:topologies_b} can be obtained for the $ut$ piece.  However, one final complexity arises because the product of the four-quark operators associated with each of the two $W$ exchanges will lead to a logarithmic singularity as their locations collide in the integral over their space-time positions.  (This singularity is not quadratic because of the GIM subtraction present in the factor associated with $\lambda_u$.)  This difficulty can also be handled by Rome-Southampton techniques and, as in the case of the $ut$ contribution to the box diagram, requires the introduction of the operator $O_{LL}$ to both remove the short-distance artifacts associated with this divergence in the product evaluated on the lattice and to provide the correct short distance contribution~\cite{Christ:2010zz,Yu:2011gk}.

To summarize, both the divergence associated with the charm quark loop appearing in the factor associated with $\lambda_t$ in this $ut$ product and the overall divergence resulting from the collision of the two four-quark operators representing the two exchanged $W$ bosons can be removed by imposing an off-shell, gauge-fixed, RI/MOM, Rome-Southampton condition on a four-quark Greens function containing the singular loop or product.  Such a condition is imposed at a scale $\mu$.  In each case a perturbative calculation of this same four-quark Green's function in the continuum theory, including any needed top-charm GIM cancellation must be carried out and the result at this same scale $\mu$ determined.  The required operator, either a combination of the QCD penguin operators in Eq.~\eqref{eq:QCD_penguin} or $O_{LL}$ must then be added to the lattice calculation to replace the lattice-regulated short distance part with the correct, continuum short distance contribution.

\textbf{Disconnected topology, $\boldsymbol{tt}$:}
Finally the disconnected $tt$ contribution are suppressed by two powers of $\lambda_t$.  As in the case of the box diagram, the GIM cancellation takes place at the scale of $m_t$ allowing the integration momentum to increase up to $p \approx m_t$ giving this disconnected $tt$ also an expected size of $x_t/M_W^2$, resulting in a contribution whose nominal size is the same as the $tt$ contribution to the box graph.  Of course, this is a short distance contribution and can be viewed as a NNLO correction to the usual $tt$ portion of the box diagram.

We conclude that the six $uu$, $ut$ and $tt$ contributions with box and disconnected topologies can be accurately determined by a combination of QCD and electroweak perturbation theory and four-flavor lattice calculations.  While the specific operators that must be included and subtractions required vary between the box and disconnected topologies and in some cases extra suppression factors of $(m_K/m_c)^2$ may arise, these have the relative nominal size of $x_c/M_W^2$, $x_c/M_W^2$ and $x_t/M_W^2$ respectively.

We can combine these estimates with the experimental values of $\lambda_u$, $\lambda_t$, $x_t/x_c$ given in Eq.~\eqref{eq:PDG_input} to determine which terms must be computed to obtain accurate results for $\Delta M_K$ and $\epsilon_K$.   The result is summarized in Tab.~\ref{tab:estimates}.  As described earlier in this paper, the largest contribution to $\Delta M_K$ comes from $uu$ pieces with box and disconnected topologies and can be computed directly in a four-quark theory such as that being used here.   The largest contribution likely comes from the $uu$ box piece which is the subject of the present calculation.  However, a disconnected $uu$ contribution should be included through the type 3 and 4 diagrams shown in Figs.~\ref{fig:type3} and \ref{fig:type4}.  The $tt$ contribution to $\Delta M_K$ is on the few percent level and can be determined from the usual product of a perturbatively determined Wilson coefficient times the matrix element  $\langle\overline{K^0}|O_{LL}|K^0\rangle$, analogous to the usual calculation of the dominant, short distance contribution to $\epsilon_K$.  The more challenging $ut$ piece, while accessible to lattice methods, is expected to contribute only at the fraction of a percent level.  

\begin{table}[th]
\centering
\begin{tabular}{lccc}
\hline\hline
 quarks &$M_{\overline{0}0}$
                                                             &Re$(M_{\overline{0}0})$ 
                                                                         &Im$(M_{\overline{0}0})$ \\
\hline
uu         & $\lambda_u^2 x_c$  & $1.1\times 10^{-5}$   & $0$  \\
tt           & $\lambda_t^2 x_t$    & $4.0\times 10^{-7}$   & $4.1\times 10^{-7}$  \\
ut          & $\lambda_u\lambda_t x_c$   
                                                 & $1.6\times 10^{-8}$   & $6.6\times 10^{-9}$ \\
\hline\hline
\end{tabular}
\caption{Explicit factors of CKM matrix elements and powers of quark masses which multiply the various terms considered here as they contribute to $\Delta M_K$ and $\epsilon_K$.  We adopt the standard notation, using $x_i=m_i^2/M_W^2$ for the ratio of the mass squared of the up-type quark $i=u$, $c$ and $t$ to the square of the mass of the $W$ boson.}
\label{tab:estimates}
\end{table}

For $\epsilon_K$ the situation is different.  As can be seen from Tab.~\ref{tab:estimates} the dominate piece comes from the familiar $tt$ contribution giving the usual product of $B_K$ and a Wilson coefficient.  The $ut$ term may contribute at the few percent level, involves long distance contributions and should be accessible to a combination of lattice and perturbative techniques.  The first piece to compute would likely be the box contributions which can be obtained in a four-flavor theory from a product of four-quark operators such as being studied here.  However, in contrast to the present calculation of $\Delta M_K$, an overall logarithmic divergence must be removed and a compensating short distance, $O_{LL}$ matrix element included.  A consistent calculation would also require the inclusion of disconnected graphs with the new QCD penguin operators and two levels of short distance corrections, discussed above.

In summary, this examination of the various terms that contribute to both $\Delta M_K$ and $\epsilon_K$ suggests that both quantities, including their long distance contributions, should be accessible to lattice methods with controlled systematic errors, ultimately at the sub-percent level.

It should be emphasized that the lattice calculation of the leading contributions to $\Delta M_K$ presented in this paper, and the extension of these methods to percent-level, sub-leading terms as discussed in this Appendix depend critically on the ability of lattice methods to describe accurately the four-flavor, effective theory of the standard model.  A calculation of these quantities with controlled systematic errors must seamlessly combine continuum perturbation theory with lattice calculation and it is essential that both approaches give the same result for kinematic regions in which they both apply.  This is made possible by the use of chiral lattice fermions which respect the full chiral symmetry of the effective four-flavor theory and by the Rome-Southampton renormalization methods which define a single set of renormalization conditions which can be applied using both methods. For example, the resulting close relation between continuum and lattice methods makes it entirely practical to introduce a subtraction which removes a $\ln(\mu a)$ term in a lattice calculation and to replace that subtracted term by the matrix element of a local operator whose coefficient is computed in perturbation theory giving, in the end, a physical result with controlled systematic errors.

\section{Second order energy shift from Euclidean four-point functions}
\label{sec:kinematics}

In Section~\eqref{sec:amplitude} we introduced an integrated, Euclidean four-point function, given in Eq.~\eqref{eq:integrated_correlator} and showed in Eq.~\eqref{eq:integration_result} how the finite-volume, second-order energy shift could be extracted from this integrated correlator.  In this Appendix, we provide added insight into this result by showing how the various terms in Eq.~\eqref{eq:integration_result} naturally arise from an application of standard perturbation theory to the time evolution operator generated by the sum of the QCD and weak interaction Hamiltonians, $H_\mathrm{QCD}+H_W$.  We will also discuss other constructions that could be used to achieve a similar result and explain why we chose to use the integrated correlator given in Eq.~\eqref{eq:integrated_correlator}.

The integrated correlator used in this paper to determine the finite volume, $K_L-K_S$ mass shift can be interpreted as a term in the expansion of the $K_{1/2} = (|K^0\rangle \mp |\overline{K^0}\rangle)/\sqrt{2}$ matrix elements of the following hybrid time evolution:
\begin{equation}
\mathscr{W}_\alpha = \langle 0|e^{-H_\mathrm{QCD}(T_\mathrm{tot}-t_f)} K_\alpha e^{-H_\mathrm{QCD}(t_f-t_b)} e^{-(H_\mathrm{QCD}+H_W)(t_b-t_a)}e^{-H_\mathrm{QCD}(t_a-t_i)}K_\alpha e^{-H_\mathrm{QCD}(t_i)}|0\rangle
\label{eq:hybrid_evolution_1}
\end{equation}
for $\alpha=1$ or 2.  In this appendix we use $T_\mathrm{tot}$ to represent the time extent of the lattice.  We can expand this matrix element in powers of $H_W$, and evaluate the second order term for $\alpha=1$ and 2.  The difference between the $CP$ even $\alpha=1$ and $CP$ odd $\alpha=2$ results is precisely the integrated correlator given in Eq.~\eqref{eq:integrated_correlator}.  The four terms appearing in the expression for this integrated correlator given in Eq.~\eqref{eq:integration_result} can then be easily understood by considering the all-orders time evolution given above in Eq.~\eqref{eq:hybrid_evolution_1} as follows. 

As in Eq.~\eqref{eq:unintegrated_correlator} we assume that the times $t_i$ and $T_\mathrm{tot}-t_f$ are sufficiently large that Euclidean time evolution with the QCD Hamiltonian $H_\mathrm{QCD}$ projects onto the vacuum state.  We also assume that the separations $t_a-t_i$ and $t_f-t_b$ are sufficiently large that only the QCD eigenstates $|K_\alpha\rangle$ propagate over this interval.  We can then use the sudden approximation to evaluate the expression in Eq.~\eqref{eq:hybrid_evolution_1} in terms of a sum over eigenstates $|\phi_n\rangle$ of the combined Hamiltonian $H_\mathrm{QCD}+H_W$ with eigenvalues $\mathscr{E}_n$:
\begin{equation}
\mathscr{W} = N_K^2 e^{-M_K(t_f-t_b)} \sum_n \langle K_\alpha|\phi_n\rangle e^{-\mathscr{E}_n(t_b-t_a)}\langle\phi_n|K_\alpha\rangle e^{-M_K(t_a-t_i)}.
\label{eq:hybrid_evolution_eval_1}
\end{equation}
The target of these considerations is the second-order, finite-volume energy shift $E_{K_\alpha}^{(2)}$ for the state $|K_\alpha\rangle$ given in standard perturbation theory by:
\begin{equation}
E_{K_\alpha}^{(2)} = \sum_{n\ne K_\alpha}\frac{\langle K_\alpha|H_W|n\rangle\langle n|H_W|K_\alpha\rangle}{M_K-E_n},
\end{equation}
where $|n\rangle$ and $E_n$ are the eigenstates and corresponding eigenvalues of the QCD Hamiltonian $H_\mathrm{QCD}$.  When Eq.~\eqref{eq:hybrid_evolution_eval_1} is expanded to second order in $H_W$, this second order energy shift will appear as the term:
\begin{equation}
N_K^2 e^{-M_K(t_f-t_i)} E_{K_\alpha}^{(2)}(t_b-t_a)
\end{equation}
which corresponds to the term proportional to $T$ in  Eq.~\eqref{eq:integration_result}.

A second type of term appearing in an expansion of Eq.~\eqref{eq:hybrid_evolution_eval_1} to second order in $H_W$ comes from expanding the $H_\mathrm{QCD}+H_W$ eigenstates $|\phi_n\rangle$ in terms of the eigenstates $|n\rangle$ of $H_W$.  To first order in $H_W$, a general state $|\phi_n\rangle$ will overlap with the state $|K_\alpha\rangle$ as given in standard first order perturbation theory:
\begin{equation}
\langle K_\alpha|\phi_n\rangle = \frac{\langle K_\alpha|H_W|\phi_n\rangle}{E_n-M_K}.
\end{equation}
This mixing with non-kaon states will introduce additional exponential time dependence in Eq.~\eqref{eq:hybrid_evolution_eval_1} through the terms:
\begin{equation}
N_K^2 e^{-M_K(t_f-t_i)} \sum_{n\ne K_\alpha} \frac{|\langle K_\alpha|H_W|\phi_n\rangle|^2}{(E_n-M_K)^2}e^{-(E_n-M_K)(t_b-t_a)}.
\label{eq:standard_lower_energy_states}
\end{equation}
This is the origin of the $e^{(M_K-E_n)}T$ term in  Eq.~\eqref{eq:integration_result}.  In a similar fashion, the $-1$ accompanying the $e^{(M_K-E_n)T_\mathrm{tot}}$ in that equation comes from the standard second order correction to the normalization of the state $|\phi_{K_\alpha}\rangle$:
\begin{equation}
\langle K_\alpha|\phi_{K_\alpha}\rangle = 1 - \frac{1}{2}\sum_{n\ne K_\alpha}\frac{|\langle n|H_W|K_\alpha\rangle|^2}{(M_K-E_n)^2},
\end{equation}
where $|\phi_{K_\alpha}\rangle$ is the eigenstate of $H_\mathrm{QCD}+H_W$ which is equal to $|K_\alpha\rangle$ to zeroth order in $H_W$.

Finally if the volume is chosen so that one of the $\pi\-\pi$ eigenstates of $H_\mathrm{QCD}$, $|n_0\rangle$ is degenerate with the kaon state, there will be a first order energy shift in the CP even state $|K_1\rangle$ given by standard degenerate perturbation theory as:
\begin{equation}
E_{K_1}^{(1)} = \pm \langle n_0|H_W|K_1\rangle.
\end{equation}
This energy shift will contribute a term proportional to $(t_b-t_a)^2$ when the expression in Eq.~\eqref{eq:hybrid_evolution_eval_1} is expanded to second order  in $H_W$.  This accounts for the final $T^2$ term in Eq.~\eqref{eq:integration_result}.

Given this straightforward interpretation of result in Eq.~\eqref{eq:integration_result} in terms of standard perturbation theory we can easily analyze alternative Green's functions that might be used to determine $\Delta M_K$.  For example, a simpler alternative integrates the two weak operators over the full time interval $[t_i,t_f]$ between the kaon source and sink instead of the restricted interval $[t_a,t_b]$ used here:
\begin{equation}
\mathscr{A}'=\frac{1}{2}\sum_{t_2=t_i}^{t_f}\sum_{t_1=t_i}^{t_f}\langle 0|T\left\{\overline{K^0}(t_f)H_W(t_2)H_W(t_1)\overline{K^0}(t_i)\right\}|0\rangle.
\label{eq:integrated_correlator_2}
\end{equation}
Following the above discussion, this Green's function can be recognized as a second order term in the $H_W$ expansion of the following matrix elements:
\begin{eqnarray}
\mathscr{W}'_\alpha &=& \langle 0|e^{-H_\mathrm{QCD}(T_\mathrm{tot}-t_f)} K_\alpha e^{-(H_\mathrm{QCD}+H_W)(t_f-t_i)}K_\alpha e^{-H_\mathrm{QCD}(t_i)}|0\rangle \\ 
&=& \sum_n \langle 0| K_\alpha|\phi_n\rangle e^{-\mathscr{E}_n(t_f-t_i)}\langle\phi_n|K_\alpha|0\rangle. 
\label{eq:hybrid_evolution_eval_2}
\end{eqnarray}
An expansion of this equation to second order in $H_W$ contains the term of interest, $N_K^2\Delta M_K (t_f-t_i)e^{-M_K(t_f-t_i)}/2$.  However, this expression has two disadvantages when compared to the quantity which we use.  The first is the need to vary the location of the source and sink positions if the linear dependence on $t_f-t_i$ is to be identified.  For the Green's functions which we consider we are able to work with fixed $t_f$ and $t_i$ and simply vary the interval $[t_a,t_b]$ over which the weak operator insertions are integrated.  Having fixed kaon source and sink locations reduces the number of propagators which must be evaluated in the calculation presented here. 

A second, far more serious difficulty with the expression in Eq.~\eqref{eq:hybrid_evolution_eval_2} arises from the analogue of the exponentially increasing terms given in Eq.~\eqref{eq:standard_lower_energy_states} for our method of choice.  In that previous case the coefficient of an exponentially increasing term coming from a QCD energy eigenstate $|n\rangle$ with energy $E_n$ lower than $M_K$ is a simple matrix element of $H_W$ between that state and a physical kaon state, a quantity easily determined in a separate lattice calculation.  However, for the matrix element $\mathscr{W}'_\alpha$ above these unwanted terms come with coefficients that are very difficult to determine and hence cannot be easily subtracted.  Specifically for $\mathscr{W}'_\alpha$ the term analogous to that in Eq.~\eqref{eq:standard_lower_energy_states}, involving energy eigenvalues of  $H_\mathrm{QCD}+H_W$  that are zeroth order in $H_W$ and energy eigenstates that are expanded to first order in $H_W$, gives the expression:
\begin{equation}
\sum_{n^{\prime\prime},n^\prime\ne n} \langle 0| K_\alpha|n^{\prime\prime}\rangle \frac{\langle n^{\prime\prime}|H_W|n\rangle}{E_n-E_{n^{\prime\prime}}}e^{-E_n(t_f-t_i)}\frac{\langle n|H_W|n^\prime\rangle}{E_n-E_{n^\prime}} \langle n^\prime|K_\alpha|0\rangle. 
\label{eq:hybrid_evolution_eval_2_failure}
\end{equation}
Here a term with energy $E_n < M_K$ which must be removed has a complicated coefficient given by a sum over matrix elements of $H_W$ between that state $|n\rangle$ and a series of excited states $|n'\rangle$,  a combination apparently inaccessible to a lattice QCD calculation.  Thus, a separate determination of the terms to be subtracted, such as was done for the pion state in the calculation presented here, appears very difficult.   Note this second difficulty only arises when there exist states of lower energy than that of the state being studied, in our case the kaon.  All of these unwanted terms with $E_n > m_K$ will not contribute for sufficiently large $t_f-t_i$.

Finally, a third alternative that we can examine integrates the product of the two weak operators $H_W(t_2)H_W(t_1)$ over the entire time interval $[0,T_\mathrm{tot}]$:
\begin{equation}
\mathscr{A}^{\prime\prime}=\frac{1}{2}\sum_{t_2=0}^{T_\mathrm{tot}}\sum_{t_1=0}^{T_\mathrm{tot}}\langle 0|T\left\{\overline{K^0}(t_f)H_W(t_2)H_W(t_1)\overline{K^0}(t_i)\right\}|0\rangle.
\label{eq:integrated_correlator_2}
\end{equation}
Since a strangeness change of two units, caused by the presence of $H_W(t_2)H_W(t_1)$ acting between the operators $\overline{K^0}(t_f)$ and $\overline{K^0}(t_i)$ is required to obtain a non-zero result, the amplitudes $\mathscr{A}'$ and $\mathscr{A}^{\prime\prime}$ must be equal and expanding the region of integration in this way should not affect the result.  Thus,  $\mathscr{A}'$ and $\mathscr{A}^{\prime\prime}$ will also suffer from the same shortcomings.

\bibliography{citation}

\begin{thebibliography}{32}%
\makeatletter
\providecommand \@ifxundefined [1]{%
 \@ifx{#1\undefined}
}%
\providecommand \@ifnum [1]{%
 \ifnum #1\expandafter \@firstoftwo
 \else \expandafter \@secondoftwo
 \fi
}%
\providecommand \@ifx [1]{%
 \ifx #1\expandafter \@firstoftwo
 \else \expandafter \@secondoftwo
 \fi
}%
\providecommand \natexlab [1]{#1}%
\providecommand \enquote  [1]{``#1''}%
\providecommand \bibnamefont  [1]{#1}%
\providecommand \bibfnamefont [1]{#1}%
\providecommand \citenamefont [1]{#1}%
\providecommand \href@noop [0]{\@secondoftwo}%
\providecommand \href [0]{\begingroup \@sanitize@url \@href}%
\providecommand \@href[1]{\@@startlink{#1}\@@href}%
\providecommand \@@href[1]{\endgroup#1\@@endlink}%
\providecommand \@sanitize@url [0]{\catcode `\\12\catcode `\$12\catcode
  `\&12\catcode `\#12\catcode `\^12\catcode `\_12\catcode `\%12\relax}%
\providecommand \@@startlink[1]{}%
\providecommand \@@endlink[0]{}%
\providecommand \url  [0]{\begingroup\@sanitize@url \@url }%
\providecommand \@url [1]{\endgroup\@href {#1}{\urlprefix }}%
\providecommand \urlprefix  [0]{URL }%
\providecommand \Eprint [0]{\href }%
\providecommand \doibase [0]{http://dx.doi.org/}%
\providecommand \selectlanguage [0]{\@gobble}%
\providecommand \bibinfo  [0]{\@secondoftwo}%
\providecommand \bibfield  [0]{\@secondoftwo}%
\providecommand \translation [1]{[#1]}%
\providecommand \BibitemOpen [0]{}%
\providecommand \bibitemStop [0]{}%
\providecommand \bibitemNoStop [0]{.\EOS\space}%
\providecommand \EOS [0]{\spacefactor3000\relax}%
\providecommand \BibitemShut  [1]{\csname bibitem#1\endcsname}%
\let\auto@bib@innerbib\@empty
\bibitem [{\citenamefont {Nakamura}\ \emph {et~al.}(2010)\citenamefont
  {Nakamura} \emph {et~al.}}]{Nakamura:2010zzi}%
  \BibitemOpen
  \bibfield  {author} {\bibinfo {author} {\bibfnamefont {K.}~\bibnamefont
  {Nakamura}} \emph {et~al.} (\bibinfo {collaboration} {Particle Data Group}),\
  }\href {\doibase 10.1088/0954-3899/37/7A/075021} {\bibfield  {journal}
  {\bibinfo  {journal} {J.Phys.G}\ }\textbf {\bibinfo {volume} {G37}},\
  \bibinfo {pages} {075021} (\bibinfo {year} {2010})}\BibitemShut {NoStop}%
\bibitem [{\citenamefont {Herrlich}\ and\ \citenamefont
  {Nierste}(1994)}]{Herrlich:1993yv}%
  \BibitemOpen
  \bibfield  {author} {\bibinfo {author} {\bibfnamefont {S.}~\bibnamefont
  {Herrlich}}\ and\ \bibinfo {author} {\bibfnamefont {U.}~\bibnamefont
  {Nierste}},\ }\href {\doibase 10.1016/0550-3213(94)90044-2} {\bibfield
  {journal} {\bibinfo  {journal} {Nucl. Phys.}\ }\textbf {\bibinfo {volume}
  {B419}},\ \bibinfo {pages} {292} (\bibinfo {year} {1994})},\ \Eprint
  {http://arxiv.org/abs/hep-ph/9310311} {arXiv:hep-ph/9310311} \BibitemShut
  {NoStop}%
\bibitem [{\citenamefont {Brod}\ and\ \citenamefont
  {Gorbahn}(2012)}]{Brod:2011ty}%
  \BibitemOpen
  \bibfield  {author} {\bibinfo {author} {\bibfnamefont {J.}~\bibnamefont
  {Brod}}\ and\ \bibinfo {author} {\bibfnamefont {M.}~\bibnamefont {Gorbahn}},\
  }\href {\doibase 10.1103/PhysRevLett.108.121801} {\bibfield  {journal}
  {\bibinfo  {journal} {Phys.Rev.Lett.}\ }\textbf {\bibinfo {volume} {108}},\
  \bibinfo {pages} {121801} (\bibinfo {year} {2012})},\ \Eprint
  {http://arxiv.org/abs/1108.2036} {arXiv:1108.2036 [hep-ph]} \BibitemShut
  {NoStop}%
\bibitem [{\citenamefont {Christ}(2010)}]{Christ:2010zz}%
  \BibitemOpen
  \bibfield  {author} {\bibinfo {author} {\bibfnamefont {N.~H.}\ \bibnamefont
  {Christ}} (\bibinfo {collaboration} {RBC and UKQCD Collaborations}),\
  }\href@noop {} {\bibfield  {journal} {\bibinfo  {journal} {PoS}\ }\textbf
  {\bibinfo {volume} {LATTICE2010}},\ \bibinfo {pages} {300} (\bibinfo {year}
  {2010})}\BibitemShut {NoStop}%
\bibitem [{\citenamefont {Glashow}\ \emph {et~al.}(1970)\citenamefont
  {Glashow}, \citenamefont {Iliopoulos},\ and\ \citenamefont
  {Maiani}}]{Glashow:1970gm}%
  \BibitemOpen
  \bibfield  {author} {\bibinfo {author} {\bibfnamefont {S.}~\bibnamefont
  {Glashow}}, \bibinfo {author} {\bibfnamefont {J.}~\bibnamefont {Iliopoulos}},
  \ and\ \bibinfo {author} {\bibfnamefont {L.}~\bibnamefont {Maiani}},\ }\href
  {\doibase 10.1103/PhysRevD.2.1285} {\bibfield  {journal} {\bibinfo  {journal}
  {Phys.Rev.}\ }\textbf {\bibinfo {volume} {D2}},\ \bibinfo {pages} {1285}
  (\bibinfo {year} {1970})}\BibitemShut {NoStop}%
\bibitem [{\citenamefont {Lellouch}\ and\ \citenamefont
  {Luscher}(2001)}]{Lellouch:2000pv}%
  \BibitemOpen
  \bibfield  {author} {\bibinfo {author} {\bibfnamefont {L.}~\bibnamefont
  {Lellouch}}\ and\ \bibinfo {author} {\bibfnamefont {M.}~\bibnamefont
  {Luscher}},\ }\href@noop {} {\bibfield  {journal} {\bibinfo  {journal}
  {Commun. Math. Phys.}\ }\textbf {\bibinfo {volume} {219}},\ \bibinfo {pages}
  {31} (\bibinfo {year} {2001})},\ \Eprint
  {http://arxiv.org/abs/hep-lat/0003023} {hep-lat/0003023} \BibitemShut
  {NoStop}%
\bibitem [{\citenamefont {Christ}(2011)}]{Christ:2012np}%
  \BibitemOpen
  \bibfield  {author} {\bibinfo {author} {\bibfnamefont {N.~H.}\ \bibnamefont
  {Christ}},\ }\href@noop {} {\bibfield  {journal} {\bibinfo  {journal} {PoS}\
  }\textbf {\bibinfo {volume} {LATTICE2011}},\ \bibinfo {pages} {277} (\bibinfo
  {year} {2011})},\ \Eprint {http://arxiv.org/abs/1201.2065} {arXiv:1201.2065
  [hep-lat]} \BibitemShut {NoStop}%
\bibitem [{\citenamefont {Blum}\ \emph {et~al.}(2011)\citenamefont {Blum},
  \citenamefont {Boyle}, \citenamefont {Christ}, \citenamefont {Garron},
  \citenamefont {Goode} \emph {et~al.}}]{Blum:2011pu}%
  \BibitemOpen
  \bibfield  {author} {\bibinfo {author} {\bibfnamefont {T.}~\bibnamefont
  {Blum}}, \bibinfo {author} {\bibfnamefont {P.}~\bibnamefont {Boyle}},
  \bibinfo {author} {\bibfnamefont {N.}~\bibnamefont {Christ}}, \bibinfo
  {author} {\bibfnamefont {N.}~\bibnamefont {Garron}}, \bibinfo {author}
  {\bibfnamefont {E.}~\bibnamefont {Goode}},  \emph {et~al.},\ }\href {\doibase
  10.1103/PhysRevD.84.114503} {\bibfield  {journal} {\bibinfo  {journal}
  {Phys.Rev.}\ }\textbf {\bibinfo {volume} {D84}},\ \bibinfo {pages} {114503}
  (\bibinfo {year} {2011})},\ \bibinfo {note} {40 pages, 12 figures},\ \Eprint
  {http://arxiv.org/abs/1106.2714} {arXiv:1106.2714 [hep-lat]} \BibitemShut
  {NoStop}%
\bibitem [{\citenamefont {Buras}\ \emph {et~al.}(2010)\citenamefont {Buras},
  \citenamefont {Guadagnoli},\ and\ \citenamefont {Isidori}}]{Buras:2010pza}%
  \BibitemOpen
  \bibfield  {author} {\bibinfo {author} {\bibfnamefont {A.~J.}\ \bibnamefont
  {Buras}}, \bibinfo {author} {\bibfnamefont {D.}~\bibnamefont {Guadagnoli}}, \
  and\ \bibinfo {author} {\bibfnamefont {G.}~\bibnamefont {Isidori}},\ }\href
  {\doibase 10.1016/j.physletb.2010.04.017} {\bibfield  {journal} {\bibinfo
  {journal} {Phys. Lett.}\ }\textbf {\bibinfo {volume} {B688}},\ \bibinfo
  {pages} {309} (\bibinfo {year} {2010})},\ \Eprint
  {http://arxiv.org/abs/1002.3612} {arXiv:1002.3612 [hep-ph]} \BibitemShut
  {NoStop}%
\bibitem [{\citenamefont {Isidori}\ \emph {et~al.}(2006)\citenamefont
  {Isidori}, \citenamefont {Martinelli},\ and\ \citenamefont
  {Turchetti}}]{Isidori:2005tv}%
  \BibitemOpen
  \bibfield  {author} {\bibinfo {author} {\bibfnamefont {G.}~\bibnamefont
  {Isidori}}, \bibinfo {author} {\bibfnamefont {G.}~\bibnamefont {Martinelli}},
  \ and\ \bibinfo {author} {\bibfnamefont {P.}~\bibnamefont {Turchetti}},\
  }\href {\doibase 10.1016/j.physletb.2005.11.044} {\bibfield  {journal}
  {\bibinfo  {journal} {Phys.Lett.}\ }\textbf {\bibinfo {volume} {B633}},\
  \bibinfo {pages} {75} (\bibinfo {year} {2006})},\ \Eprint
  {http://arxiv.org/abs/hep-lat/0506026} {arXiv:hep-lat/0506026 [hep-lat]}
  \BibitemShut {NoStop}%
\bibitem [{\citenamefont {Yu}(2011)}]{Yu:2011gk}%
  \BibitemOpen
  \bibfield  {author} {\bibinfo {author} {\bibfnamefont {J.}~\bibnamefont
  {Yu}},\ }\href@noop {} {\bibfield  {journal} {\bibinfo  {journal} {PoS}\
  }\textbf {\bibinfo {volume} {LATTICE2011}},\ \bibinfo {pages} {297} (\bibinfo
  {year} {2011})},\ \Eprint {http://arxiv.org/abs/1111.6953} {arXiv:1111.6953
  [hep-lat]} \BibitemShut {NoStop}%
\bibitem [{\citenamefont {Buchalla}\ \emph {et~al.}(1996)\citenamefont
  {Buchalla}, \citenamefont {Buras},\ and\ \citenamefont
  {Lautenbacher}}]{Buchalla:1995vs}%
  \BibitemOpen
  \bibfield  {author} {\bibinfo {author} {\bibfnamefont {G.}~\bibnamefont
  {Buchalla}}, \bibinfo {author} {\bibfnamefont {A.~J.}\ \bibnamefont {Buras}},
  \ and\ \bibinfo {author} {\bibfnamefont {M.~E.}\ \bibnamefont
  {Lautenbacher}},\ }\href {\doibase 10.1103/RevModPhys.68.1125} {\bibfield
  {journal} {\bibinfo  {journal} {Rev.Mod.Phys.}\ }\textbf {\bibinfo {volume}
  {68}},\ \bibinfo {pages} {1125} (\bibinfo {year} {1996})},\ \Eprint
  {http://arxiv.org/abs/hep-ph/9512380} {arXiv:hep-ph/9512380 [hep-ph]}
  \BibitemShut {NoStop}%
\bibitem [{Note1()}]{Note1}%
  \BibitemOpen
  \bibinfo {note} {Together with G.~Martinelli, we are also exploring the
  possibility of controlling finite volume effects without requiring a two-pion
  state to be degenerate with the kaon, generalizing the approach to
  finite-volume effects in $K\to \pi \pi $ decays developed in \cite
  {Lin:2001ek,Kim:2005gf}.}\BibitemShut {Stop}%
\bibitem [{\citenamefont {Allton}\ \emph {et~al.}(2007)\citenamefont {Allton}
  \emph {et~al.}}]{Allton:2007hx}%
  \BibitemOpen
  \bibfield  {author} {\bibinfo {author} {\bibfnamefont {C.}~\bibnamefont
  {Allton}} \emph {et~al.} (\bibinfo {collaboration} {RBC and UKQCD}),\
  }\href@noop {} {\bibfield  {journal} {\bibinfo  {journal} {Phys. Rev.}\
  }\textbf {\bibinfo {volume} {D76}},\ \bibinfo {pages} {014504} (\bibinfo
  {year} {2007})},\ \Eprint {http://arxiv.org/abs/hep-lat/0701013}
  {hep-lat/0701013} \BibitemShut {NoStop}%
\bibitem [{\citenamefont {Stathopoulos}\ and\ \citenamefont
  {Orginos}(2010)}]{Stathopoulos:2007zi}%
  \BibitemOpen
  \bibfield  {author} {\bibinfo {author} {\bibfnamefont {A.}~\bibnamefont
  {Stathopoulos}}\ and\ \bibinfo {author} {\bibfnamefont {K.}~\bibnamefont
  {Orginos}},\ }\href@noop {} {\bibfield  {journal} {\bibinfo  {journal} {SIAM
  J.Sci.Comput.}\ }\textbf {\bibinfo {volume} {32}},\ \bibinfo {pages} {439}
  (\bibinfo {year} {2010})},\ \Eprint {http://arxiv.org/abs/0707.0131}
  {arXiv:0707.0131 [hep-lat]} \BibitemShut {NoStop}%
\bibitem [{\citenamefont {Liu}(2011)}]{Liu:2011jp}%
  \BibitemOpen
  \bibfield  {author} {\bibinfo {author} {\bibfnamefont {Q.}~\bibnamefont
  {Liu}},\ }\href@noop {} {\  (\bibinfo {year} {2011})},\ \Eprint
  {http://arxiv.org/abs/1110.2143} {arXiv:1110.2143 [hep-lat]} \BibitemShut
  {NoStop}%
\bibitem [{\citenamefont {Aoki}\ \emph {et~al.}(2011)\citenamefont {Aoki} \emph
  {et~al.}}]{Aoki:2010dy}%
  \BibitemOpen
  \bibfield  {author} {\bibinfo {author} {\bibfnamefont {Y.}~\bibnamefont
  {Aoki}} \emph {et~al.} (\bibinfo {collaboration} {RBC Collaboration, UKQCD
  Collaboration}),\ }\href {\doibase 10.1103/PhysRevD.83.074508} {\bibfield
  {journal} {\bibinfo  {journal} {Phys.Rev.}\ }\textbf {\bibinfo {volume}
  {D83}},\ \bibinfo {pages} {074508} (\bibinfo {year} {2011})},\ \Eprint
  {http://arxiv.org/abs/1011.0892} {arXiv:1011.0892 [hep-lat]} \BibitemShut
  {NoStop}%
\bibitem [{\citenamefont {Zweig}(1964)}]{Zweig:1964jf}%
  \BibitemOpen
  \bibfield  {author} {\bibinfo {author} {\bibfnamefont {G.}~\bibnamefont
  {Zweig}},\ }\href@noop {} {\  (\bibinfo {year} {1964})}\BibitemShut {NoStop}%
\bibitem [{\citenamefont {Okubo}(1963)}]{Okubo:1963fa}%
  \BibitemOpen
  \bibfield  {author} {\bibinfo {author} {\bibfnamefont {S.}~\bibnamefont
  {Okubo}},\ }\href {\doibase 10.1016/S0375-9601(63)92548-9} {\bibfield
  {journal} {\bibinfo  {journal} {Phys.Lett.}\ }\textbf {\bibinfo {volume}
  {5}},\ \bibinfo {pages} {165} (\bibinfo {year} {1963})}\BibitemShut {NoStop}%
\bibitem [{\citenamefont {Iizuka}(1966)}]{Iizuka:1966fk}%
  \BibitemOpen
  \bibfield  {author} {\bibinfo {author} {\bibfnamefont {J.}~\bibnamefont
  {Iizuka}},\ }\href {\doibase 10.1143/PTPS.37.21} {\bibfield  {journal}
  {\bibinfo  {journal} {Prog.Theor.Phys.Suppl.}\ }\textbf {\bibinfo {volume}
  {37}},\ \bibinfo {pages} {21} (\bibinfo {year} {1966})}\BibitemShut {NoStop}%
\bibitem [{\citenamefont {Martinelli}\ \emph {et~al.}(1995)\citenamefont
  {Martinelli}, \citenamefont {Pittori}, \citenamefont {Sachrajda},
  \citenamefont {Testa},\ and\ \citenamefont {Vladikas}}]{Martinelli:1995ty}%
  \BibitemOpen
  \bibfield  {author} {\bibinfo {author} {\bibfnamefont {G.}~\bibnamefont
  {Martinelli}}, \bibinfo {author} {\bibfnamefont {C.}~\bibnamefont {Pittori}},
  \bibinfo {author} {\bibfnamefont {C.~T.}\ \bibnamefont {Sachrajda}}, \bibinfo
  {author} {\bibfnamefont {M.}~\bibnamefont {Testa}}, \ and\ \bibinfo {author}
  {\bibfnamefont {A.}~\bibnamefont {Vladikas}},\ }\href@noop {} {\bibfield
  {journal} {\bibinfo  {journal} {Nucl. Phys.}\ }\textbf {\bibinfo {volume}
  {B445}},\ \bibinfo {pages} {81} (\bibinfo {year} {1995})},\ \Eprint
  {http://arxiv.org/abs/hep-lat/9411010} {hep-lat/9411010} \BibitemShut
  {NoStop}%
\bibitem [{\citenamefont {Blum}\ \emph {et~al.}(2003)\citenamefont {Blum} \emph
  {et~al.}}]{Blum:2001xb}%
  \BibitemOpen
  \bibfield  {author} {\bibinfo {author} {\bibfnamefont {T.}~\bibnamefont
  {Blum}} \emph {et~al.} (\bibinfo {collaboration} {RBC Collaboration}),\
  }\href {\doibase 10.1103/PhysRevD.68.114506} {\bibfield  {journal} {\bibinfo
  {journal} {Phys.Rev.}\ }\textbf {\bibinfo {volume} {D68}},\ \bibinfo {pages}
  {114506} (\bibinfo {year} {2003})},\ \Eprint
  {http://arxiv.org/abs/hep-lat/0110075} {arXiv:hep-lat/0110075 [hep-lat]}
  \BibitemShut {NoStop}%
\bibitem [{\citenamefont {Lehner}\ and\ \citenamefont
  {Sturm}(2011)}]{Lehner:2011fz}%
  \BibitemOpen
  \bibfield  {author} {\bibinfo {author} {\bibfnamefont {C.}~\bibnamefont
  {Lehner}}\ and\ \bibinfo {author} {\bibfnamefont {C.}~\bibnamefont {Sturm}},\
  }\href {\doibase 10.1103/PhysRevD.84.014001} {\bibfield  {journal} {\bibinfo
  {journal} {Phys.Rev.}\ }\textbf {\bibinfo {volume} {D84}},\ \bibinfo {pages}
  {014001} (\bibinfo {year} {2011})},\ \Eprint {http://arxiv.org/abs/1104.4948}
  {arXiv:1104.4948 [hep-ph]} \BibitemShut {NoStop}%
\bibitem [{\citenamefont {Giusti}\ \emph {et~al.}(2004)\citenamefont {Giusti},
  \citenamefont {Hernandez}, \citenamefont {Laine}, \citenamefont {Weisz},\
  and\ \citenamefont {Wittig}}]{Giusti:2004an}%
  \BibitemOpen
  \bibfield  {author} {\bibinfo {author} {\bibfnamefont {L.}~\bibnamefont
  {Giusti}}, \bibinfo {author} {\bibfnamefont {P.}~\bibnamefont {Hernandez}},
  \bibinfo {author} {\bibfnamefont {M.}~\bibnamefont {Laine}}, \bibinfo
  {author} {\bibfnamefont {P.}~\bibnamefont {Weisz}}, \ and\ \bibinfo {author}
  {\bibfnamefont {H.}~\bibnamefont {Wittig}},\ }\href {\doibase
  10.1088/1126-6708/2004/11/016} {\bibfield  {journal} {\bibinfo  {journal}
  {JHEP}\ }\textbf {\bibinfo {volume} {0411}},\ \bibinfo {pages} {016}
  (\bibinfo {year} {2004})},\ \Eprint {http://arxiv.org/abs/hep-lat/0407007}
  {arXiv:hep-lat/0407007 [hep-lat]} \BibitemShut {NoStop}%
\bibitem [{\citenamefont {Sturm}\ \emph {et~al.}(2009)\citenamefont {Sturm},
  \citenamefont {Aoki}, \citenamefont {Christ}, \citenamefont {Izubuchi},
  \citenamefont {Sachrajda} \emph {et~al.}}]{Sturm:2009kb}%
  \BibitemOpen
  \bibfield  {author} {\bibinfo {author} {\bibfnamefont {C.}~\bibnamefont
  {Sturm}}, \bibinfo {author} {\bibfnamefont {Y.}~\bibnamefont {Aoki}},
  \bibinfo {author} {\bibfnamefont {N.}~\bibnamefont {Christ}}, \bibinfo
  {author} {\bibfnamefont {T.}~\bibnamefont {Izubuchi}}, \bibinfo {author}
  {\bibfnamefont {C.}~\bibnamefont {Sachrajda}},  \emph {et~al.},\ }\href
  {\doibase 10.1103/PhysRevD.80.014501} {\bibfield  {journal} {\bibinfo
  {journal} {Phys.Rev.}\ }\textbf {\bibinfo {volume} {D80}},\ \bibinfo {pages}
  {014501} (\bibinfo {year} {2009})},\ \Eprint {http://arxiv.org/abs/0901.2599}
  {arXiv:0901.2599 [hep-ph]} \BibitemShut {NoStop}%
\bibitem [{\citenamefont {Allton}\ \emph {et~al.}(2008)\citenamefont {Allton}
  \emph {et~al.}}]{Allton:2008pn}%
  \BibitemOpen
  \bibfield  {author} {\bibinfo {author} {\bibfnamefont {C.}~\bibnamefont
  {Allton}} \emph {et~al.} (\bibinfo {collaboration} {RBC-UKQCD
  Collaboration}),\ }\href {\doibase 10.1103/PhysRevD.78.114509} {\bibfield
  {journal} {\bibinfo  {journal} {Phys.Rev.}\ }\textbf {\bibinfo {volume}
  {D78}},\ \bibinfo {pages} {114509} (\bibinfo {year} {2008})},\ \Eprint
  {http://arxiv.org/abs/0804.0473} {arXiv:0804.0473 [hep-lat]} \BibitemShut
  {NoStop}%
\bibitem [{\citenamefont {Blum}\ \emph {et~al.}(2012)\citenamefont {Blum},
  \citenamefont {Izubuchi},\ and\ \citenamefont {Shintani}}]{Blum:2012uh}%
  \BibitemOpen
  \bibfield  {author} {\bibinfo {author} {\bibfnamefont {T.}~\bibnamefont
  {Blum}}, \bibinfo {author} {\bibfnamefont {T.}~\bibnamefont {Izubuchi}}, \
  and\ \bibinfo {author} {\bibfnamefont {E.}~\bibnamefont {Shintani}},\
  }\href@noop {} {\  (\bibinfo {year} {2012})},\ \Eprint
  {http://arxiv.org/abs/1208.4349} {arXiv:1208.4349 [hep-lat]} \BibitemShut
  {NoStop}%
\bibitem [{\citenamefont {Inami}\ and\ \citenamefont
  {Lim}(1981)}]{Inami:1980fz}%
  \BibitemOpen
  \bibfield  {author} {\bibinfo {author} {\bibfnamefont {T.}~\bibnamefont
  {Inami}}\ and\ \bibinfo {author} {\bibfnamefont {C.}~\bibnamefont {Lim}},\
  }\href {\doibase 10.1143/PTP.65.297} {\bibfield  {journal} {\bibinfo
  {journal} {Prog.Theor.Phys.}\ }\textbf {\bibinfo {volume} {65}},\ \bibinfo
  {pages} {297} (\bibinfo {year} {1981})}\BibitemShut {NoStop}%
\bibitem [{\citenamefont {Brod}\ and\ \citenamefont
  {Gorbahn}(2010)}]{Brod:2010mj}%
  \BibitemOpen
  \bibfield  {author} {\bibinfo {author} {\bibfnamefont {J.}~\bibnamefont
  {Brod}}\ and\ \bibinfo {author} {\bibfnamefont {M.}~\bibnamefont {Gorbahn}},\
  }\href {\doibase 10.1103/PhysRevD.82.094026} {\bibfield  {journal} {\bibinfo
  {journal} {Phys.Rev.}\ }\textbf {\bibinfo {volume} {D82}},\ \bibinfo {pages}
  {094026} (\bibinfo {year} {2010})},\ \Eprint {http://arxiv.org/abs/1007.0684}
  {arXiv:1007.0684 [hep-ph]} \BibitemShut {NoStop}%
\bibitem [{\citenamefont {Beringer}\ \emph {et~al.}(2012)\citenamefont
  {Beringer} \emph {et~al.}}]{Beringer:1900zz}%
  \BibitemOpen
  \bibfield  {author} {\bibinfo {author} {\bibfnamefont {J.}~\bibnamefont
  {Beringer}} \emph {et~al.} (\bibinfo {collaboration} {Particle Data Group}),\
  }\href {\doibase 10.1103/PhysRevD.86.010001} {\bibfield  {journal} {\bibinfo
  {journal} {Phys.Rev.}\ }\textbf {\bibinfo {volume} {D86}},\ \bibinfo {pages}
  {010001} (\bibinfo {year} {2012})}\BibitemShut {NoStop}%
\bibitem [{\citenamefont {Lin}\ \emph {et~al.}(2001)\citenamefont {Lin},
  \citenamefont {Martinelli}, \citenamefont {Sachrajda},\ and\ \citenamefont
  {Testa}}]{Lin:2001ek}%
  \BibitemOpen
  \bibfield  {author} {\bibinfo {author} {\bibfnamefont {C.~J.~D.}\
  \bibnamefont {Lin}}, \bibinfo {author} {\bibfnamefont {G.}~\bibnamefont
  {Martinelli}}, \bibinfo {author} {\bibfnamefont {C.~T.}\ \bibnamefont
  {Sachrajda}}, \ and\ \bibinfo {author} {\bibfnamefont {M.}~\bibnamefont
  {Testa}},\ }\href@noop {} {\bibfield  {journal} {\bibinfo  {journal} {Nucl.
  Phys.}\ }\textbf {\bibinfo {volume} {B619}},\ \bibinfo {pages} {467}
  (\bibinfo {year} {2001})},\ \Eprint {http://arxiv.org/abs/hep-lat/0104006}
  {hep-lat/0104006} \BibitemShut {NoStop}%
\bibitem [{\citenamefont {Kim}\ \emph {et~al.}(2005)\citenamefont {Kim},
  \citenamefont {Sachrajda},\ and\ \citenamefont {Sharpe}}]{Kim:2005gf}%
  \BibitemOpen
  \bibfield  {author} {\bibinfo {author} {\bibfnamefont {C.~h.}\ \bibnamefont
  {Kim}}, \bibinfo {author} {\bibfnamefont {C.~T.}\ \bibnamefont {Sachrajda}},
  \ and\ \bibinfo {author} {\bibfnamefont {S.~R.}\ \bibnamefont {Sharpe}},\
  }\href@noop {} {\bibfield  {journal} {\bibinfo  {journal} {Nucl. Phys.}\
  }\textbf {\bibinfo {volume} {B727}},\ \bibinfo {pages} {218} (\bibinfo {year}
  {2005})},\ \Eprint {http://arxiv.org/abs/hep-lat/0507006} {hep-lat/0507006}
  \BibitemShut {NoStop}%
\end{thebibliography}%

\end{document}